\documentclass[letterpaper]{article}
\pdfoutput=1
\usepackage{jheppub}
\usepackage{color,latexsym,array,multirow,verbatim}
\usepackage[config,singlelinecheck=true]{caption}
\usepackage{url}
\usepackage{subfigure}
\usepackage{xcolor}
\usepackage{xspace}
\usepackage{tikz}

\preprint{}

\newcommand{\GeV}{\text{ GeV}}
\newcommand{\TeV}{\text{ TeV}}
\newcommand{\ecm}{E_\text{CM}}

\newcommand{\tc}{\tau_2^{\text{cut}}}
\newcommand{\Tcut}{\cT_{\text{cut}}}

\newcommand{\muM}{\mu_R}%\text{MG}}
\newcommand{\Htree}{H_\text{tree}}
\newcommand{\colorindex}{{IJ}}
\newcommand{\Fcorr}{F_c}
\newcommand{\abs}[1]{\left| #1 \right|}
\newcommand{\Gc}{\Gamma}
\newcommand{\cO}{\mathcal{O}}
\DeclareMathOperator{\Tr}{Tr}

\newcommand{\Pythia}{\textsc{Pythia}\xspace}
\newcommand{\MadGraph}{\textsc{MadGraph}\xspace}
\newcommand{\Mathematica}{\textsc{Mathematica}\xspace}
\newcommand{\Alpgen}{\textsc{Alpgen}\xspace}
\newcommand{\Sherpa}{\textsc{Sherpa}\xspace}
\newcommand{\Caesar}{\textsc{Caesar}\xspace}
\newcommand{\Geneva}{\textsc{Geneva}\xspace}

\newcommand{\obs}{\ensuremath{\mathcal{O}}\xspace}
\newcommand{\cT}{\ensuremath{\mathcal{T}}}
\def\be{\begin{equation}}
\def\ee{\end{equation}}
\newcommand{\Eq}[1]{Eq.~\eqref{eq:#1}}
\newcommand{\Fig}[1]{Fig.~\ref{fig:#1}}
\newcommand{\Sec}[1]{Sec.~\ref{sec:#1}}

\newcommand{\App}[1]{App.~\ref{app:#1}}

\begin{document}

\title{Streamlining resummed QCD calculations using Monte Carlo integration}

\author{David Farhi, Ilya Feige, Marat Freytsis, Matthew D. Schwartz}

\affiliation{Center for the Fundamental Laws of Nature, Harvard University, Cambridge, MA 02138, USA}

\emailAdd{farhi@physics.harvard.edu}
\emailAdd{feige@physics.harvard.edu}
\emailAdd{freytsis@physics.harvard.edu}
\emailAdd{schwartz@physics.harvard.edu}

\abstract{
Some of the most arduous and error-prone aspects of precision resummed calculations are related to the partonic hard process, having nothing to do with the resummation. In particular, interfacing to parton-distribution functions, combining various channels, and performing the phase space integration can be limiting factors in completing calculations. Conveniently, however, most of these tasks are already automated in many Monte Carlo programs, such as \MadGraph, \Alpgen or \Sherpa. In this paper, we show how such programs can be used to produce distributions of partonic kinematics with associated color structures representing the hard factor in a resummed distribution. These distributions can then be used to weight convolutions of jet, soft and beam functions producing a complete resummed calculation. In fact, only around 1000 unweighted events are necessary to produce precise distributions. A number of examples and checks are 
provided, including $e^+e^-$ two- and four-jet event shapes, $n$-jettiness and jet-mass related observables at hadron colliders. Attached code can be used to modify \MadGraph to export the relevant leading-order hard functions and color structures for arbitrary processes.}

\maketitle

%%%%%%%%%%%%%%%%%%%%%%%%%%%%%%%%%%
%%%%%%%%%%%%%%%%%%%%%%%%%%%%%%%%%%
\section{Introduction}
\label{sec:intro}
%%%%%%%%%%%%%%%%%%%%%%%%%%%%%%%%%%
%%%%%%%%%%%%%%%%%%%%%%%%%%%%%%%%%%
One of the main goals of collider experiments like the Large Hadron Collider (LHC) is to detect or bound phenomena beyond the Standard Model. Of course, to find deviations from the Standard Model, we must understand the predictions of the Standard Model itself. As the LHC analyses exhaust the limits of what can be learned from clean events with leptons and photons, the need for precise predictions of 
more sophisticated observables involving jets and hadronic event and jet shapes  grows.

The usual approach in processes too complicated for analytic predictions is to use Monte Carlo (MC) simulations. However, these MC simulations are generally limited to leading-log resummation. Soft Collinear Effective Theory (SCET)~\cite{Bauer:2000ew, Bauer:2000yr,Beneke:2002ph,Beneke:2002ni} provides a systematically improvable framework for resumming the soft and collinear logarithms in QCD. It has been used to compute cross sections for observables to next-to-leading logarithmic order (NLL), NNLL and even NNNLL. SCET has been applied to a variety of jet-related observables in different processes at colliders, from thrust~\cite{Schwartz:2007ib,Becher:2008cf} to jet mass~\cite{Chien:2012ur,Jouttenus:2013hs} or the Higgs cross section with a jet veto~\cite{Berger:2010xi,Becher:2012qa}.

Resummed computations have been performed for processes with 2~jets (as in $e^+e^-$ colliders), 3~jets (or 2~incoming beams and 1 outgoing jet)~\cite{Chien:2012ur,Jouttenus:2013hs,Becher:2009th}, or 4~jets in special cases~\cite{Dasgupta:2012hg,Kelley:2010fn,Kelley:2010qs}. For processes with many jets, the phase space can become prohibitively difficult to integrate. For instance, in \cite{Dasgupta:2012hg}, the authors compute a distribution for a 4-parton observable, but only at a single phase space point. Many interesting observables have 4, 5, or even more jet or beam directions. For example, jet mass in dijet events has been measured~\cite{ATLAS:2012am} but not yet computed. Another example is 2-subjettiness~\cite{Thaler:2010tr}, useful for telling boosted $W$ or Higgs bosons from QCD background. The signal distribution has been computed at NNNLL level~\cite{Feige:2012vc}, but the background, which involves $pp\to X+ \text{2 collinear jets}$, has eluded calculation so far.

There are basically three steps in computing a resummed distribution. First, one needs a factorization formula indicating the relevant objects (usually hard, jet, soft and beam functions) which dominate the distribution in a certain threshold limit. Second, to get to a particular accuracy, one needs the fixed-order computation of these objects and the relevant anomalous dimensions. These first two steps comprise the intellectual meat of the prediction. The third step is to put it all together and get numbers out. Unfortunately, this last step can take years. Anyone who has worked on precision calculations knows the pain of getting factors of two right, hammering out the small coding errors, and cross-checking against other calculations. Although the past decade has seen the complete automation of the numerical step for fixed-order calculations with public codes, resummed calculations are still often done with Mathematica on a user's personal computer. In this paper, we show that much of the tedium of the numerical step can be automated in resummed calculations as well. We propose a way to combine analytic resummation with a numerical computation of the leading-order hard function and phase space integrals using existing MC generators.

Distributions of observables in SCET are given by the product of a hard function with a convolution of jet, soft and beam functions, integrated over phase space. We can write this as
\begin{equation}
  \label{eq:main}
  \frac{d\sigma}{d\obs} = \sum_{\text{channels}}\ \sum_{\text{colors}~I,J}\
                          \sigma_0 \int d\Phi\, H^\colorindex(\Phi)F^\colorindex(\Phi, \obs)
\end{equation}
Where $\Phi$ is the phase space, $H$ is the hard function, and $F$ is the combination of jet, soft and beam functions. $IJ$ indexes the matrix structure of the objects in color space. The basic idea is that the phase space integral $\int d\Phi$ and the hard function (which at leading order is just the tree-level squared matrix element for partonic scattering) are computed numerically with a MC generator. The rest is computed analytically. Since MC generators sample phase space proportionally to squared matrix-elements, roughly speaking we simply make a MC sample, weight each point $\Phi$ by $F^\colorindex(\Phi, \obs)$, and sum them to get the final differential cross section $\frac{d\sigma}{d\obs}(\obs)$. For hadronic collisions, parton-distribution functions are included in the $H^\colorindex$, since they are handled efficiently by the MC. Then a correction factor is added to change the scale at which the PDF is evaluated or to turn the PDF into a beam function.

In addition to computing the difficult integrals, this method is nicely extensible and automatable. The resummed function $F^\colorindex(\Phi,\obs)$ is universal in the sense that it does not depend on the parton-level process, only the number and type of the jets. This allows the whole computation to be mostly automated. After the anomalous dimensions have been calculated, $F^\colorindex(\Phi,\obs)$ can be calculated once and then the distribution can be generated automatically for any partonic process. For example, the calculation of jet mass in dijet events or dijet-plus-photon events involves the same function $F^\colorindex(\Phi,\obs)$.

The method we propose is distinct from previous approaches to semi-automated resummation. The \Caesar framework performs resummation of hadronic event shapes at NLL in an entirely automated fashion~\cite{Banfi:2004yd,Banfi:2012yh}. \Geneva~\cite{Bauer:2008qj,Bauer:2008qh,Alioli:2012fc} attempts to produce fully differential distributions which reproduce certain observables at a given order. The connection between resummation in event generators and SCET has been explored in~\cite{Bauer:2006mk,Bauer:2006qp}. Recently, distributions for processes with a jet veto have been computed  by interfacing with \MadGraph at NLO~\cite{Becher:2014aya}. Some differences with our work are that Ref.~\cite{Becher:2014aya} exploited \MadGraph to compute the NLO hard function, allowing for NNLL resummation, and that Ref.~\cite{Becher:2014aya} only considered observables for which there was no color evolution. With color singlet final states Ref.~\cite{Becher:2014aya} did not have to modify \MadGraph to provide the color structures necessary to the evolution. In addition to showing that \MadGraph can be used to provide color structures, we also demonstrate averaging distributions for each hard phase point (like $\frac{d \sigma}{d \cT_4}$) rather than using only single values of an observable at each point (like the $W^+W^-$ transverse momentum).

To demonstrate the power of the method, in this paper we provide the first calculation of 2-jettiness in $pp\to 2j + \gamma$ at NLL. We also explore various features and checks of related distributions such as thrust and 4-jettiness in $e^+e^-$ collisions.  All the results in this paper at at NLL. Going beyond NLL is straightforward using this method, and discussed in the conclusion.

%%%%%%%%%%%%%%%%%%%%%%%%%%%%%%%%%%
%%%%%%%%%%%%%%%%%%%%%%%%%%%%%%%%%%
\section{Overview}
  \label{sec:overview}
%%%%%%%%%%%%%%%%%%%%%%%%%%%%%%%%%%
%%%%%%%%%%%%%%%%%%%%%%%%%%%%%%%%%%
The focus of this paper is the resummation of jet-based observables to higher order than that provided by current parton-shower based Monte Carlo event generators. The observables $\obs$ for which resummation is useful typically have singular distributions at fixed order in perturbation theory. For example, in perturbation theory, the distribution of jet mass blows up for small mass. The singularities appear as large logarithms arise in the cumulative cross section typically of the form $\alpha_s^n \ln^m \obs$, with $m \le 2n$. Formally, Monte Carlo event generators get only the leading singularities, with $m=2n$ correct. We are interested in extending the resummation of logarithms for such observables to higher order in a systematically improvable way using the framework of SCET.

Calculations in SCET proceed through the separation of a full calculation for a given process by way of a factorization theorem into contributions from sectors that are separately calculated and then assembled to provide a  final result. A general framework for $n$-jet events schematically takes the form~\cite{Bauer:2008jx}
\begin{equation}
\label{eq:Njetfactor}
  d\sigma \sim  \sigma_0\,   d\Phi_{N+X} H^{IJ}\otimes
                J_1 \otimes \dotsm \otimes J_N \otimes S^{IJ} \otimes B_a \otimes B_b ,
\end{equation}
Here, $\sigma_0$ is the total leading-order cross section, $d\Phi$ is the differential phase space for a final state of $N$ hard partons along with any other uncolored final-state particles $X$, $H^{IJ}$ is the hard function, encoding the Wilson coefficients providing information about the short-distance scattering process. $J_i$ and $B_a$ are jet and beam functions which describe the collinear evolution of final and initial state partons respectively, and $S^{IJ}$ is the soft function describing the crosstalk of soft QCD radiation between these collinear sectors. $I$ and $J$ are color-structure indices. All the terms involving beam functions in this and subsequent equations will be absent in the case of leptonic initial states.

The differential distribution of a given observable is produced by integrating the differential cross section against a measurement function.  In order to effect the resummation of large logarithms, it is necessary to evolve each of the pieces in \Eq{Njetfactor} from their natural scales to a common scale using the renormalization evolution equation of SCET. Thus the resummed distribution for the relevant observables takes the following form.
\begin{multline}
\label{eq:resumfact}
  \frac{d\sigma}{d\obs} = \sigma_0 \int d\Phi H^{IJ}(\Phi,\mu_h) U_H(\mu,\mu_h)
     \left(\bigotimes_i  U_{J_i}(\mu,\mu_j) J_i(m_i, \mu_h) \right) \\
    \otimes U_S(\mu,\mu_s) S(k_i,\mu_s) \otimes U_{B_a}(\mu,\mu_B) B_a(m_a, \mu_B) \otimes  U_{B_b}(\mu,\mu_B) B_b(m_b, \mu_B) \delta(\obs - f_\obs(m_i,k_i))
\end{multline}
where the $U_i(\mu,\nu)$ are evolution kernels encoding the resummation of large logarithms. (For a recent review with more details, see Ref.~\cite{Becher:2014oda}.) For the observables of interest, everything in this formula after $H^{IJ}(\Phi,\mu_h)$ can be computed analytically. The important point for this paper is that the hard function $H^{IJ}(\Phi,\mu_h)$ and hard-parton phase-space integrals $\int d\Phi$ are exactly what many existing MC codes are designed to calculate. We can thereby combine MC calculations with analytic resummation to efficiently compute the differential distribution. Compressing the notation even more results in the heuristic formula in \Eq{main}, where $F^\colorindex(\Phi,\obs)$ represents everything after $H^{IJ}(\Phi,\mu_h)$ in \Eq{resumfact}.

The procedure we  propose is then
\begin{enumerate}
\item  Use \MadGraph to produce events with a probability proportional to $H d\Phi$.
\item Compute the function $F^\colorindex(\Phi,\obs)$ analytically as a function of the phase space point $\Phi$ (it is a matrix in color-structure space, indexed by $IJ$). At each point $F^\colorindex(\Phi,\obs)$ is a distribution in $\obs$.
\item Average  the $F^\colorindex(\Phi,\obs)$ distributions over all events.
\end{enumerate}
The result is the prediction for the resummed distribution.

The function $F^\colorindex(\Phi,\obs)$ is both analytically computable and general in the sense that it depends only on the number and type (quark vs gluon) of the jets. Thus, for instance, it needs to be computed once for the 4-jettiness observable, and then can be applied to 4-jettiness in $e^+e^-\to 4j$, $e^+e^-\to \gamma+4j$, $e^+e^-\to Z+4j$ and any other 4-jet process, even when the phase space is very different. With suitable universal modifications, crossing 2 partons to the initial state allows the same results to be used to compute $pp$ observables with 2 jets in the final state.

Normally, MCs produce only distributions of momenta. To perform the resummation, we also need to know the color structure of the hard colored particles. This information is already computed by the MCs, so extracting it is just a matter of modifying the code to include more information in the Les Houches Event Format (LHEF) file~\cite{Boos:2001cv,Alwall:2006yp}. We do this by adding a comment to the event. For example an $e^+e^- \to u\bar{d} \bar{u} d$ event might look like this:
\begin{verbatim}
<event>
 6   0  0.2609800E-07  0.50,00000E+03  0.7546771E-02  0.9399810E-01
   -11  -1   0   0    0    0  0.0000E+00  0.0000E+00  0.2500E+03  0.2500E+03  0.0E+00 0. -1.
    11  -1   0   0    0    0 -0.0000E+00 -0.0000E+00 -0.2500E+03  0.2500E+03  0.0E+00 0.  1.
     1   1   1   2  501    0 -0.1007E+03 -0.2251E+02 -0.1204E+02  0.1039E+03  0.0E+00 0.  1.
     2   1   1   2  502    0 -0.5886E+02  0.1209E+03 -0.1251E+02  0.1351E+03  0.0E+00 0.  1.
    -1   1   1   2    0  502  0.1321E+03  0.1112E+02 -0.2925E+02  0.1358E+03  0.0E+00 0. -1.
    -2   1   1   2    0  501  0.2749E+02 -0.1095E+03  0.5380E+02  0.1251E+03  0.0E+00 0. -1.
<mgrwt>
  ...MadGraph5 scale information for reweighting here...
</mgrwt>
<colordecomposition>
0.11342428E+15 -.34027285E+15 -.34027285E+15 0.10208186E+16
0.90000000E+01 0.30000000E+01 0.90000000E+01 0.90000000E+01
</colordecomposition>
</event>
\end{verbatim}
The middle six lines give the particle momenta, which we might draw as \raisebox{-0.5\height}{\includegraphics[width=1.25in,trim=50 20 50 20, clip]{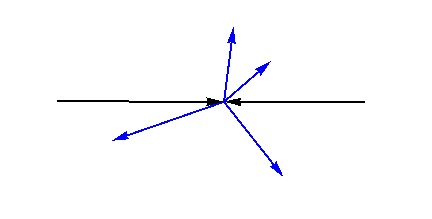}}. The \texttt{colordecomposition} block at the end of the event specifies the entries of the hard function and the color structure, respectively, discussed in more detail in \Sec{hard}.

For each event, we then use the momenta to determine the directions of the Wilson lines in the soft function, and the kinematic variables on which the RGE evolution kernel for the hard function depends. With the directions fixed, we can then compute the differential distribution of the observable at that phase space point. Heuristically, 
\begin{align}
  \begin{array}{c}
    \texttt{<event>...</event>} \\
    \parbox{1.5in}{\includegraphics[width=1.5in]{Plots/SchematicOverview/ExampleEventDoohickey1}}
  \end{array}
  \hspace{0.3in} &\longrightarrow \hspace{0.7in}
  \parbox{2in}{\includegraphics[width=2in]{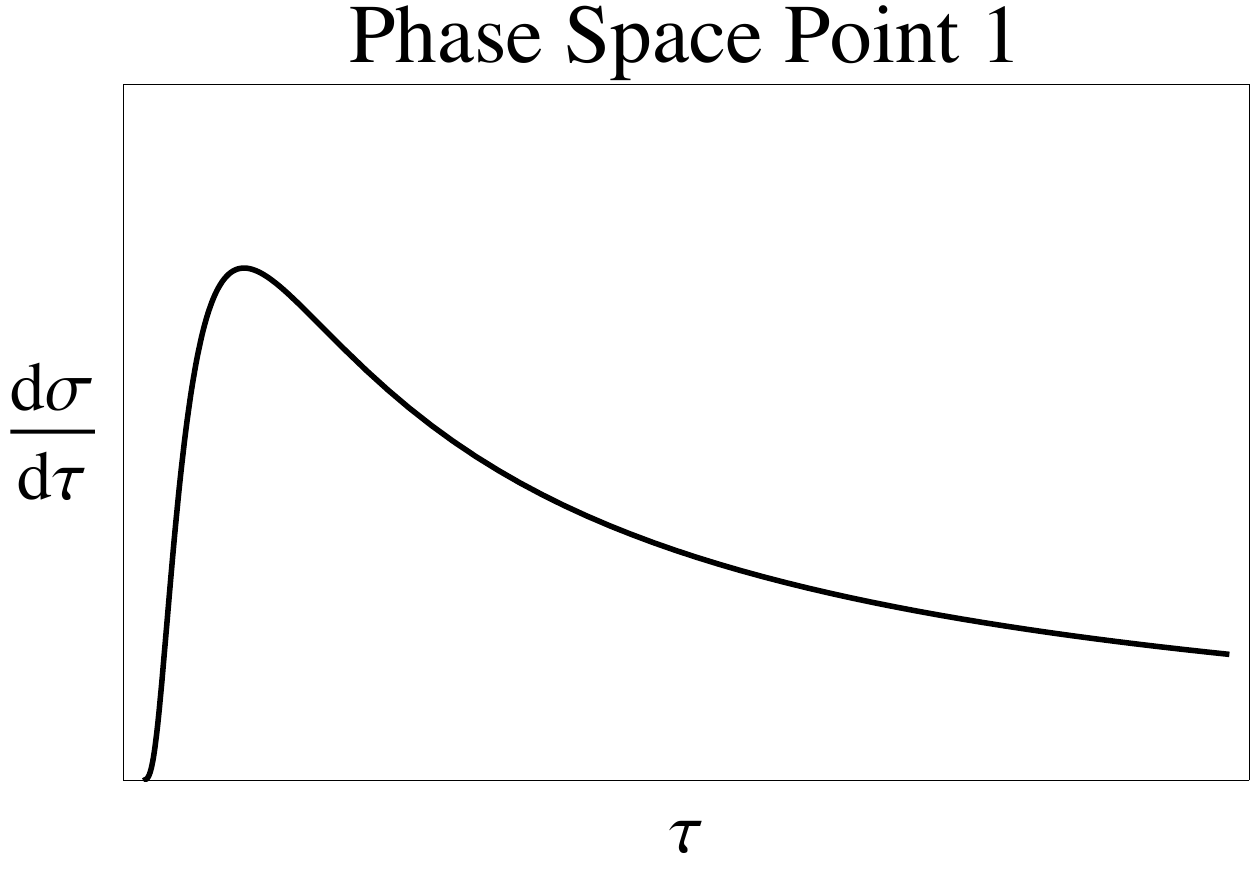}} \\  
\intertext{We then do this for all the phase space points computed by \MadGraph:}
  \begin{array}{c}
    \texttt{<event>...</event>} \\
    \parbox{1.5in}{\includegraphics[width=1.5in]{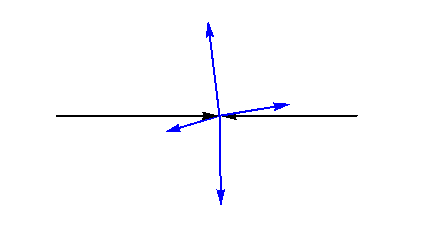}}
  \end{array}
  \hspace{0.3in} &\longrightarrow \hspace{0.7in}
  \parbox{2in}{\includegraphics[width=2in]{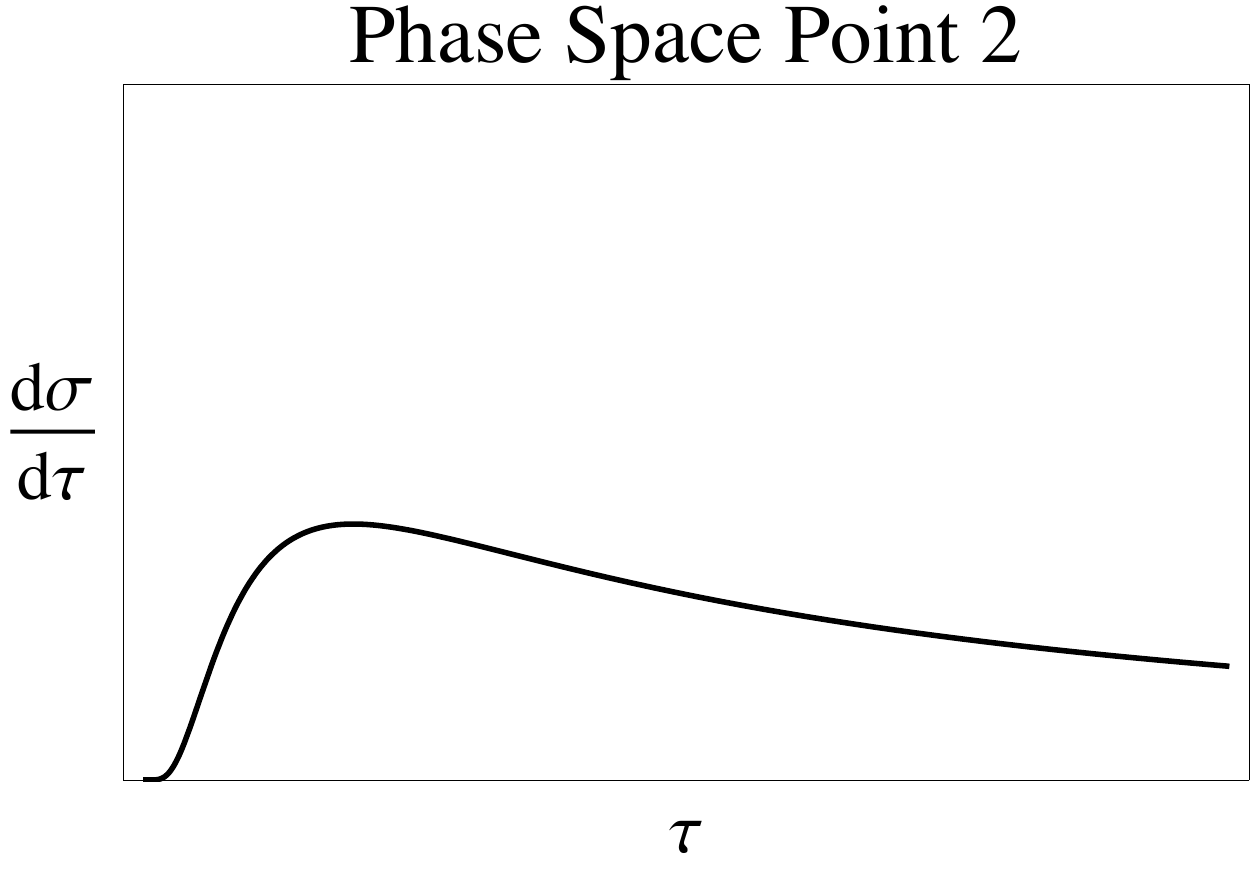}} \\
  \begin{array}{c}
    \texttt{<event>...</event>} \\
    \parbox{1.5in}{\includegraphics[width=1.5in]{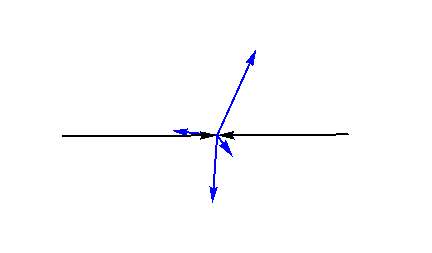}}
  \end{array}
  \hspace{0.3in} &\longrightarrow \hspace{0.7in}
  \parbox{2in}{\includegraphics[width=2in]{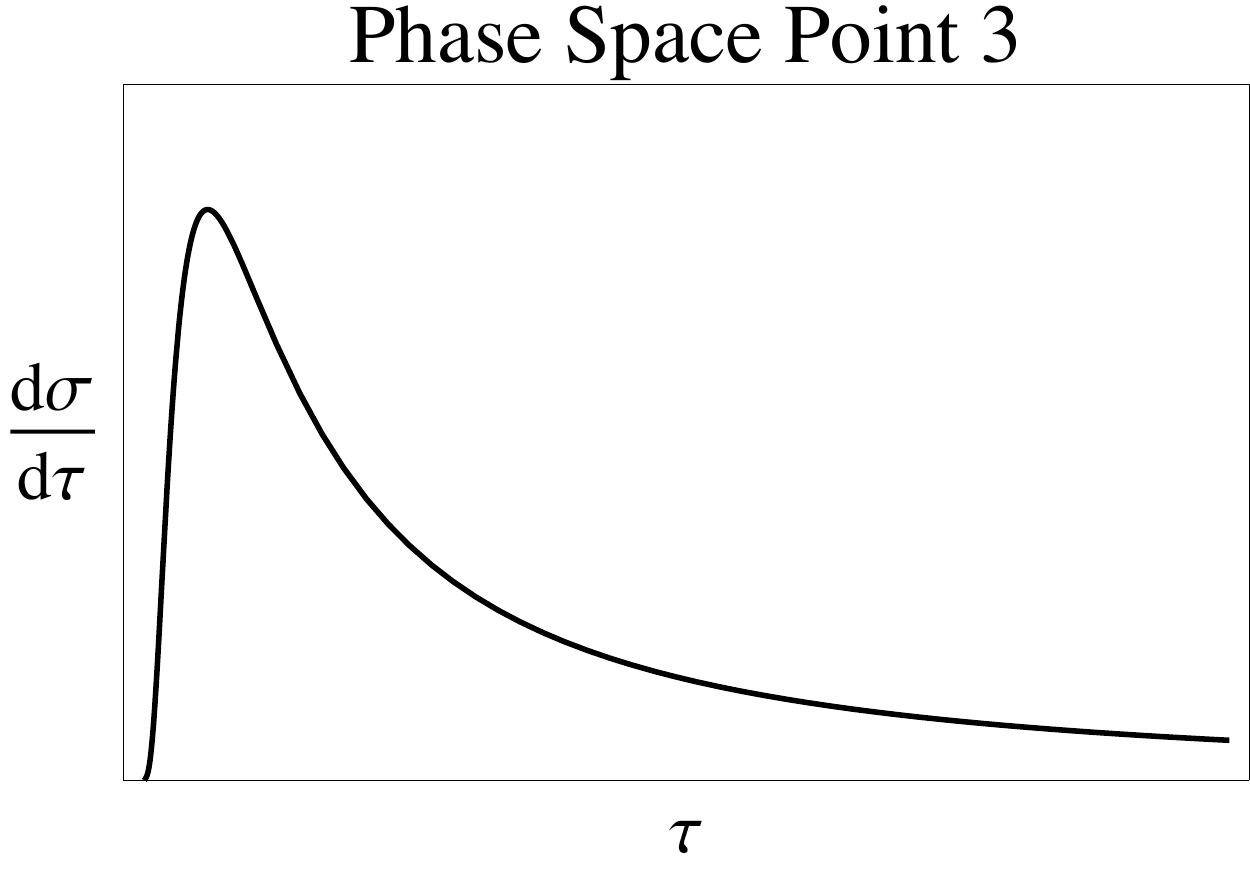}} \\
  \vdots \qquad &\vdots \qquad \vdots
\end{align}
Then all these distributions are averaged to produce the final distribution
\begin{equation}
  \frac{1}{N_\text{events}}
    \left( \parbox{1in}{\includegraphics[width=1in]{Plots/SchematicOverview/PS1Distribution}} +
           \parbox{1in}{\includegraphics[width=1in]{Plots/SchematicOverview/PS2Distribution}} +
           \parbox{1in}{\includegraphics[width=1in]{Plots/SchematicOverview/PS3Distribution}} + \dotsb
    \right)
    = \parbox{1in}{\includegraphics[width=1in]{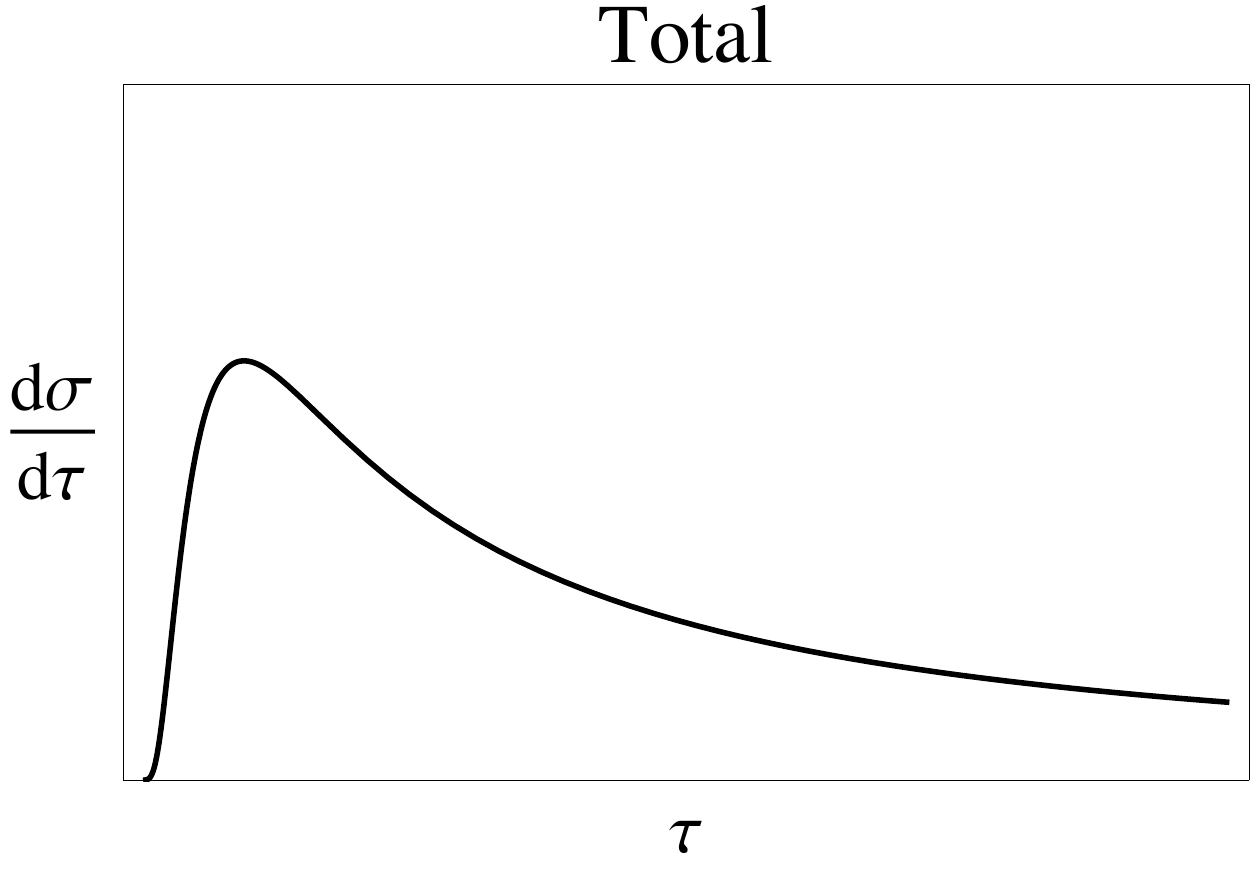}}
\end{equation}
For an observable like thrust in $e^+e^-$, the distributions for each phase space point are the same (since each point has back-to-back quarks). More generally, the averaging
is nontrivial and important.

%%%%%%%%%%%%%%%%%%%%%%%%%%%%%%%%%%
%%%%%%%%%%%%%%%%%%%%%%%%%%%%%%%%%%
\section{The hard function numerically from \MadGraph}
  \label{sec:hard}
%%%%%%%%%%%%%%%%%%%%%%%%%%%%%%%%%%
%%%%%%%%%%%%%%%%%%%%%%%%%%%%%%%%%%

In order to compute
%observables with NLL precision, 
resummed observables the hard and soft functions need to be evolved using the RG evolution equations. In a given channel, if multiple color structures are allowed the evolution of the Wilson coefficients will mix these structures as a function of scale. In certain cases, such as electroweak production with a jet veto~\cite{Becher:2014aya}, only a single color structure contributes at Born level, and evolution can be effected by simply reweighting the tree-level matrix element evaluated by a MC generator.

In contrast, for hadronic collisions with jets in the final state the color structures are nontrivial. Even for a channel like $qq' \to qq'$, described by a single Feynman diagram at tree level, multiple (in this case two) color structures mix during RG evolution. All the necessary ingredients for computing the function $F^\colorindex(\Phi,\obs)$ are often available in the literature. As an example, for all fully-hadronic $2 \to 2$ processes these were presented in Ref.~\cite{Kelley:2010fn}. Unfortunately, the most natural color basis for presenting these results and carrying out the resummation is often not the basis in which MC generators carry out their calculations. Moreover, even if this were not the case, the details of the calculation are intentionally hidden from the end user. \MadGraph by default returns only events from a sampling of the process phase space, and by the time these events are generated, the internal code itself only has access to the spin- and color-averaged squared matrix element, so that all color information is obscured.

We have implemented a modification of the \MadGraph evaluation of matrix elements that allows us to retain this information. \MadGraph evaluates squared matrix elements by picking a point in phase space, and evaluating all wavefunctions, propagators, and vertices for the given kinematics and every choice of color structure. This matrix element is then contracted with an appropriate color matrix, yielding a single number. The contributions from all helicity configurations are then added together. Schematically, for phase space point $\Phi$, this corresponds to
\begin{equation}
  \label{eq:maddefault}
  |\mathcal{M}(\Phi)|^2 = \sum_{h,IJ} \mathcal{M}(\Phi)_h^{I*} S_{IJ}^\text{tree} \mathcal{M}(\Phi)_h^J,
\end{equation}
where $I, J$ index the color structures of the process, and we write the matrix encoding the color contraction as $S_{IJ}^\text{tree}$ since, up to a rescaling, it in fact corresponds to the tree-level soft function for some choice of basis in color space. For our purposes, the shortcoming of this procedure is that all color information is traced over before spin states are summed over, so that there is no way to extract the value of the hard function for each color structure.

The solution to this is to simply change the order of summation, so that \MadGraph sums over helicity states before contracting the colors. This requires the introduction of an auxiliary matrix,
\begin{equation}
  \label{eq:madhard}
  H^\colorindex_\text{tree}(\Phi) = \sum_h \mathcal{M}(\Phi)_h^{I*} \mathcal{M}(\Phi)_h^J.
\end{equation}
The squared matrix element is then simply 
\begin{equation}
  \label{eq:MSH}
  |\mathcal{M}|^2 = \sum_{IJ} H^\colorindex_\text{tree} S_{IJ}^\text{tree}
\end{equation}
while $H^\colorindex_\text{tree}$ (and $S_{IJ}^\text{tree}$), passed along with the rest of the event information, contains the necessary information to implement the resummation and then contract the color structures in order to implement the necessary event reweighting. Sometimes, passing the full matrix $H^\colorindex_\text{tree}$ can be unnecessary. If only a single helicity structure contributes to a process, $H^\colorindex_\text{tree}$ can be trivially decomposed as $C^{I\dagger} C^J$. This can sometimes be done even when multiple helicity combinations are non-zero, due to various identities making helicities of a given color ordering linearly dependent at tree-level~\cite{Kleiss:1988ne,DelDuca:1999rs,Bern:2008qj,Melia:2013bta,Johansson:2015oia}. In these cases $C$ also acts as the Wilson coefficient matching the SCET operators onto full QCD at leading order in $\cO(\alpha_s)$. Since this is not possible for all processes, however, a fully general code should not implement such a decomposition. After modifying \MadGraph's matrix element calculation procedure as described above, as events are generated $H^\colorindex_\text{tree}$ and $S_{IJ}^\text{tree}$ are also appended to the normal data stored for each event in the generated LHEF files. This information is then readily accessible in order to implement the resummation.

Beyond extracting the color information from \MadGraph, it is necessary to understand the basis in which this information is presented, so that the values of the right operators are evolved by RG evolution at every phase space point. In this section, the structures needed for the observables we compute in this paper are described. Details of how \MadGraph organizes the color information for arbitrary processes
are provided in \App{MGappendix}.

Four-quark matrix elements are computed in \MadGraph decomposed in the color dipole basis. For concreteness, we discuss the $qq' \bar{q}\bar{q}'$ operator, for which \MadGraph uses the basis
\begin{align}
  \mathcal{O}_1 &= \bar{q}q\, \bar{q}'q'   , \\
  \mathcal{O}_2 &= \bar{q}{q}' \bar{q}'q \,.
\end{align}
Because the tree level process is proportional to $\bar{q}T^aq\, \bar{q}' T^a q' = -\frac{1}{6}\mathcal{O}_1 + \frac{1}{2}\mathcal{O}_2$, the Wilson coefficients are proportional to $\left(-\frac{1}{6}, \frac{1}{2}\right)$. (A more complicated channel would have more complicated, momentum-dependent values).

The basis in which the SCET resummation formulas we use in $F^\colorindex(\Phi,\obs)$ have been computed is different from the basis above. For example, for the $qq'\bar q\bar q'$ channel, Ref.~\cite{Kelley:2010fn} uses the singlet/octet basis:
\begin{align}
  \mathcal{O}_O &= \bar{q}T^a q\, \bar{q}'T^a q'   , \\
  \mathcal{O}_S &= \bar{q}q    \, \bar{q}' q'    \,.
\end{align}
This is related to the \MadGraph basis by Fierz identities, which are straightforward work out. Since we have thus far restricted ourselves to processes with at most four colored states in the matrix elements at a time, we have computed these transformations explicitly, but they can be easily automated for higher-multiplicity states. For this channel, for instance, we see that
\begin{align}
 \begin{pmatrix} C_1 \\ C_2 \end{pmatrix}_\text{MG}
 = R \begin{pmatrix} C_O \\ C_S \end{pmatrix}_\text{F}
 = \begin{pmatrix} -\frac{1}{6} & 1 \\ \frac{1}{2} & 0 \end{pmatrix}
   \begin{pmatrix} C_O \\ C_S \end{pmatrix}_\text{F}.
\end{align}
The fixed-order fully-averaged matrix element must be the same in every basis, and so we must transform $H^\colorindex_\text{tree}$ and $S_{IJ}^\text{tree}$ with opposite senses in order to maintain this property. Additionally, this provides a sanity check since then $S_{IJ}^\text{tree}$ must then correspond to the tree-level soft function of the SCET calculations we use as inputs. The hard function can then be transformed to the preferred basis for resummation by acting on the matrix output provided by \MadGraph as $H_\text{F} = R\, H_\text{MG}\, R^\dagger$, while the color matrix provided by \MadGraph transforms into the soft function in the same basis as $S_\text{F} = (R^{\dagger})^{-1}\, S_\text{MG}\, R^{-1}$. It is worth noting that this basis transformation is non-unitary, so we must be careful about when to use its inverse and when to use its adjoint.

The situation with $q\bar{q}gg$ operators is simpler, since both \MadGraph and our formulas use the same basis,
\begin{align}
  \mathcal{O}_1^{\mu\nu} &= \bar{q}T^a T^b q\, A_a^\mu A_b^\nu \,, \\
  \mathcal{O}_2^{\mu\nu} &= \bar{q}T^b T^a q\, A_a^\mu A_b^\nu \,,
\end{align}
so no rotation is necessary. However, an additional operator appears in the RG evolution through mixing with operators generated at tree-level. We thus need to embed the hard function returned by \MadGraph as a submatrix into one also including the operator 
\begin{equation}
  \mathcal{O}_3^{\mu\nu} = \bar{q} q\, A_a^\mu A_a^\nu
\end{equation}
in its basis for the resummation.

So in practice we take a matrix from \MadGraph, computed from the decomposition of the matrix element into its preferred color basis, convert it to the SCET basis using a rotation matrix which is computed from the color basis information for the given process provided by \MadGraph, and possibly supplement the tree-level basis with additional operators involved in the resummation. We also keep track of the color matrix \MadGraph generates associated with the hard function entries to weight individual phase space points properly.

A note on the normalization of the MC sum is in order. \MadGraph produced events $\Phi_i$ with weights $w_i$ (which might be the same if we are using unweighted events, or different if we are using weighted events). It produces them such that they approximate the integral over phase space for any function $f$:
\begin{equation}
  \label{eq:summing}
  \int d\Phi |\mathcal{M}(\Phi)|^2 f(\Phi) =
    \int d\Phi H^\colorindex_\text{tree}(\Phi)S^\text{tree}_{IJ} f(\Phi) = \sum_i w_i f(\Phi_i) 
\end{equation}
using \Eq{MSH}. The SCET calculation (\Eq{resumfact}) needs us to calculate:
\begin{equation}
  \frac{d\sigma}{d\mathcal{O}} = \int d\Phi  H^\colorindex_\text{tree}(\Phi)F_{IJ}(\Phi,\obs)
\end{equation}
Thus, according to \Eq{summing}, the function $f$ which we need to evaluate at each phase space point (and then sum according to the weights) is:
\begin{equation}
  \label{eq:withnorm}
  \frac{d\sigma}{d\mathcal{O}} = \sum_i w_i \frac{H^\colorindex_\text{tree}(\Phi)F_{IJ}(\Phi,\obs)}
                                                 {H^\colorindex_\text{tree}(\Phi)S^{\text{tree}}_\colorindex}
\end{equation}

\section{Everything else analytically with SCET}
\label{sec:SCET}
We have discussed how to use \MadGraph to produce the function $H^\colorindex(\Phi)$ in \Eq{main} numerically. This is to be combined with analytical calculations of the function $F^\colorindex(\Phi,\obs)$ using SCET.

\subsection{$e^+e^-$ observables}
To begin, consider the observable thrust in $e^+e^-$ collisions. There, the phase space $\Phi$ is trivial (all tree-level quark pairs are back-to-back with half the center-of-mass energy), and there is only one color structure so $F^\colorindex(\Phi,\obs)=F(\tau)$. For thrust, we  have an  analytical closed form expression for $F(\tau)$~\cite{Becher:2008cf}:
\begin{multline}
  \label{eq:complicated}
  F(\tau) %= \frac{1}{\sigma_0} \frac{{\rm d} \sigma_2}{{\rm d} \tau}
           = H(Q^2,\mu_h) \exp \left[ 4 S (\mu_h, \mu) - 2 A_H (\mu_h, \mu) - 8 S (\mu_j, \mu)
             + 4 A_J (\mu_j, \mu) + 4 S (\mu_s, \mu) + 2 A_S (\mu_s, \mu) \right] \\
  \times \left(\frac{Q^2}{\mu_h^2}\right)^{-2A_\Gamma(\mu_h,\mu)} 
 \left[ \widetilde{j} (\partial_{2 \eta_j},\mu_j)\right]^2 \widetilde{s}_T (\partial_{\eta_s},\mu_s)
   \left[ \frac{1}{\tau} \left( \frac{\tau Q^2}{\mu_j^2} \right)^{2 \eta_j}
     \left( \frac{\tau Q}{\mu_s} \right)^{\eta_s}
     \frac{e^{- \gamma_E (2 \eta_j + \eta_s)}}{\Gamma (2 \eta_j + \eta_s)} \right] \,.
\end{multline}
See~\cite{Becher:2008cf} for the definitions and perturbative expression for  $S(\nu,\mu)$, $A(\nu,\mu)$ and $\eta_i$ as well as precise expressions for the fixed order jet- and soft-functions $\tilde j$ and $\tilde s_T$.\footnote{The conventions used for $S$ and $A$ in \Eq{complicated} and Ref.~\cite{Becher:2008cf} include a factor of $C_F$ in the cusp anomalous dimension $\Gamma$. In our calculations of more complicated variables, where the presence of gluons means that the cusp anomalous dimension appears with $C_F$ or $C_A$, we do not include either factor in $\Gamma$. This is why the factors of $S$ and $A$ in \Eq{t4} have extra factors of $C_F$ and $C_A$ as compared to \Eq{complicated}.}

Taking a step backwards, this $F(\tau)$ is calculated as the product of evolution kernels and a convolution of fixed-order jet and soft functions. Let us write it as
\begin{equation}
  \label{eq:Ft}
  F(\tau) = \Fcorr \, U_H(\mu_h,\mu) U_J(\mu_j,\mu) U_J(\mu_j,\mu) U_S(\mu_s,\mu)
            J(\mu,\tau) \otimes J(\mu,\tau) \otimes S_T(\mu,\tau)
\end{equation}
Here the $U_i(\nu,\mu)$ are RG evolution kernels. For example, the hard evolution kernel is 
\begin{equation}
  U(\mu_h,\mu) = \exp\left[ 4S(\mu_h,\mu) - 2A_\Gamma(\mu_h,\mu) \ln\frac{Q^2}{\mu_h^2}
                            - 2A_H(\mu_h,\mu)\right]
\end{equation}
In \Eq{Ft}, $J(\mu,\tau)$ is a jet function and $S_T(\mu,\tau)$ is the thrust soft function, both of which are included a fixed order.

The factor $\Fcorr$ in \Eq{Ft} is a correction factor, compensating for the parts of the hard function that \MadGraph does not compute. One contribution to $\Fcorr$ comes from the fact that \MadGraph is tree-level, but for NLL resummation, we need the large-logarithmic parts of the 1-loop hard function.  The hard function for thrust can be written as
\begin{equation}
  H(Q,\mu) = H(Q,\mu_h) U(\mu_h,\mu) =  \Htree(Q,\mu_h)\frac{H(Q,\mu_h)}{\Htree(Q,\mu_h)} U(\mu_h,\mu)
\end{equation}
The tree-level hard function, $\Htree$ is computed by \MadGraph. The rest has to be included in $\Fcorr$. For thrust, $\Htree=1$ and
\begin{equation}
  \Fcorr = \frac{H(Q,\mu_h)}{\Htree(Q,\mu_h)}
         = 1 + C_F\frac{\alpha_s(\mu_h)}{4\pi}
               \left( -2\ln^2\frac{Q^2}{\mu_h^2} - 6\ln\frac{Q^2}{\mu_h^2} \right)
\end{equation}
More generally, for NLL, only the 1-loop logarithms are included in $H(Q,\mu_h)$ and in $F_c$.

Thrust in $e^+e^-$ events has  $\Htree=1$. For other process, the hard function at tree-level is non-trivial. For example, for 3-jet event shapes (such as 3-jettiness) based on $e^+e^- \to q \bar{q} g$,  the hard function is the tree-level differential cross-section:
\begin{equation}
  \Htree(s,t,u,\mu_h)d\Phi = C_F\frac{\alpha_s(\mu_h)}{2\pi}\frac{s^2+t^2+2 u}{s t} ds d t
\end{equation}
Rather than using the analytic form for this hard function, we use \MadGraph to sample phase space according to it. At NLL, the 1-loop corrections to $H$ are proportional to $\Htree$, so the correction factor $\frac{H}{\Htree} = 1 + \alpha_s (\cdots)$.

For more complicated $e^+e^-$ processes, closed form expressions for the cross sections are neither available nor necessary. For a generic process, the leading order cross sections will be proportional to  $\alpha_s^n(\mu_h)$, where $n$ is the number of factors of $g_s$ in the tree-level Feynman diagrams. Because $n$ is fixed, we do not have to vary $\mu_h$ within \MadGraph. Instead, we can run \MadGraph with a fixed scale $\muM$ and compensate for this scale in $F_c$. Thus, for $e^+e^-$ observables, we take
\begin{equation}
  \Fcorr = \left(\frac{\alpha_s(\mu_h)}{\alpha_s(\muM)}\right)^n
             \frac{H(\{p_i\},\mu_h)}{\Htree( \{p_i\},\mu_h)}
\end{equation}
This first factor in $\Fcorr$ is how we include hard scale variation in our examples. Alternatively, one could run \MadGraph separately at $\muM$, $2\muM$ and $\frac{1}{2}\muM$, but varying the scales through $\Fcorr$ is easier since \MadGraph only has to be run once. Note that this correction factor is in addition to the evolution kernel $U(\mu_h,\mu)$ which is always included as part of $F(\Phi)$.

\subsection{Hadron collisions}
For hadron collisions, there can be two additional features. First, in \MadGraph, the incoming parton momenta are not fixed, but chosen according to parton distribution functions. Second, factorizable observables in hadron collisions often involve beam functions. We will now explain how to handle these two new features.

The inclusion of parton distributions in the hard function is straightforward. Although the incoming parton momenta are not the same for each event, we can still use the hadron collision events from \MadGraph exactly like $e^+e^-$ collision events. The only difference is that if we want
to change the scale from the factorization scale $\muM$ chosen by \MadGraph, we must multiply and divide by PDFs. So the correction factor becomes
\be
\Fcorr
= \left(\frac{\alpha_s(\mu_h)}{\alpha_s(\muM)}\right)^n\frac{H(\{p_i\},\mu_h)}{\Htree( \{p_i\},\mu_h)}
 \frac{f_a(x_1,\mu_f)}{f_a(x_1,\muM)}
 \frac{f_b(x_2,\mu_f)}{f_b(x_2,\muM)}
\ee
Note that we know $x_1$ and $x_2$ from the hard kinematics on an event-by-event basis, so this factor is just a number at each phase space point (not a convolution). Also, note that although it is standard to choose the renormalization and factorization scales equal in \MadGraph, we can separate them afterwards for the analytical part of the calculation.

For certain observables, such as boson spectra at high $p_T$ or threshold hadronic event shapes, there is no collinear radiation in the direction of the beams by construction. Alternative observables, like 2-jettiness, do not require a threshold expansion and control the collinear radiation in the beam direction through beam functions. Heuristically, with beam functions,

\begin{equation}
  \label{eq:BeamFunctionReplacement}
  (H\otimes f_a\otimes f_b)\otimes (J_1\otimes J_2\otimes S) \to
  H\otimes B_a\otimes B_b\otimes J_1\otimes J_2\otimes S
\end{equation}
This seems dangerous, since beam functions mix up information about the PDFs (which \MadGraph was handling before) and jet-related information (which SCET was handling before). However beam functions are a simple convolution of these two pieces of information, so they can still be separated.

As explained in \cite{Jouttenus:2011wh}, beam functions can be written as convolutions of PDFs and perturbatively calculable functions ${\mathcal{I}}_{ij}$:
\begin{equation}
B_i(\tau, x, \mu) = \sum_j \int_x^1 \frac{d\xi}{\xi}\mathcal{I}_{ij}(\tau, \frac{x}{\xi}, \mu)f_j(\xi, \mu)
\end{equation}
Because of this, the PDFs must be evaluated at scales $\xi$ different from the scales $x_1$ and $x_2$ where \MadGraph evaluates them. This is a complication, but not an obstruction. We simply calculate the beam functions numerically, then divide by the PDFs. 

To be concrete, for $n$-jettiness, which involves two beam functions and $n$ jets, we have
\begin{multline}
F(\Phi,\tau) = \Fcorr \,U_H(\mu_h,\mu) U_B(\mu_b,\mu) U_B(\mu_b,\mu) U_J(\mu_j,\mu) \cdots U_J(\mu_j,\mu) U_S(\mu_s,\mu)  \\[2mm]
\times B_1(x_1,\mu,\tau)\otimes B_2(x_2,\mu,\tau)\otimes J(\mu,\tau) \otimes \cdots \otimes J(\mu,\tau) \otimes S_T(\mu,\tau) \label{eq:Ftnjet}
\end{multline}
with a correction factor
\be
\Fcorr = \left(\frac{\alpha_s(\mu_h)}{\alpha_s(\muM)}\right)^n\frac{H(\{p_i\},\mu_h)}{\Htree( \{p_i\},\mu_h)}
 \frac{1}{f_a(x_1,\muM)}  \frac{1}{f_b(x_2,\muM)}
\ee
Note that for each event in \MadGraph, the energy fractions $x_1$ and $x_2$ are included in the event record. These same $x_i$ are used in the correction factor, to remove the PDF, and in the beam functions.

%%%%%%%%%%%%%%%%%%%%%%%%%%%%%%%%%%
%%%%%%%%%%%%%%%%%%%%%%%%%%%%%%%%%%
\section{Applications}
%%%%%%%%%%%%%%%%%%%%%%%%%%%%%%%%%%
%%%%%%%%%%%%%%%%%%%%%%%%%%%%%%%%%%

Next we discuss some applications. We begin with the $e^+e^-$ event shapes thrust and 4-jettiness, then look at $pp$ observables.

An obvious leading question is what do we compare to, and how do we know we are doing the calculation correctly? Comparing to data might be great, if there were appropriate data, and obviously that is the ultimate goal. However, data is complicated by many experimental issues and contributions from backgrounds and the underlying event, which we do not include. Instead, we can compare to Monte Carlo event generators like \Pythia, which are known to agree pretty well with data in most cases. We can also compare to a full analytical calculation (without the \MadGraph matching step) of an observable if that calculation is available. 

An advantage of this framework, besides its efficient exploitation of the matrix element computation and sampling, is that one can easily look at the same observable computed using different phase space points. In an inclusive calculation, these points are integrated over, but using our method, we can explore how the shape of the observable changes point-to-point. For example, when all the partons have roughly the same energy, we expect resummation to work well. However, when there are very different scales involved at the parton level, there may be additional large logarithms which we do not resum, but which are resummed at the leading-log level by \Pythia. We will discuss these issues as we go through some examples. 

In our comparisons, we have switched off \Pythia's hadronization and multiple-parton interaction models. We have also set the value of $\alpha_s$ in the parton shower to the same value we use in our calculations, and run the coupling at two-loops, as in our analytic calculation.  With these features off, for $e^+e^- \to $ 2 jet events, \Pythia is reduced to a final-state, timelike-branching parton shower. It is performing essentially the same calculation as SCET and we expect \Pythia to correspond to some particular scale choice of SCET. We can look at what this scale is and compare to the scale where the scale uncertainty is smallest in the SCET calculation.

When the processes are complex enough to involve multiple color structures, we expect to find bigger differences between \Pythia and SCET, even in the leptonic collisions. For example, in $pp \to 2j + \gamma$, in the all-quark channel $qq\to qq + \gamma$, there are two color structures (the first incoming quark can be color-connected to the first outgoing or to the second outgoing). In \MadGraph/\Pythia, one of the color connections is chosen at random by \MadGraph proportional to the relative cross sections. That color connection is then sent to \Pythia for showering. In \MadGraph/SCET, the entire vector of amplitudes is kept ($H^\colorindex(\Phi)$ is a matrix, made up of the outer product of those vectors), and contracted with the SCET function $F^\colorindex(\Phi,\obs)$) . If the RG running of the hard and soft functions for the different color structures had no cross terms, \emph{i.e.}, $H^\colorindex(\Phi)$ were diagonal, then we would expect that \Pythia's random sampling and shower (up to a rescaling of the coupling strength~\cite{Catani:1990rr}) to be equivalent to SCET's averaging even at NLL. However, when the resummation includes off-diagonal terms, then choosing one of the structures according to the diagonal entries loses some information. Thus we expect that in these more complex channels, \MadGraph/SCET will be sensitive to the off-diagonal running that \MadGraph/\Pythia ignores. As this effect first appears at NLL, we do not expect it to be numerically large. It would certainly be interesting to find evidence for this effect in theoretical calculations, and ultimately in data.

One additional complication is that for $pp$ collisions \Pythia does not treat the final-state timelike-branching and initial-state spacelike-branching parton showers simultaneously. \Pythia implements a parton shower for the initial and final states separately while imposing phase space restrictions to minimize overlap and dead zones in the final multi-body phase space. This is an imperfect procedure, but the imperfections are partially compensated for in hadronized events through the tuning of hadronization parameters. Because of the greater sensitivity to tuning, we correspondingly expect the agreement between \MadGraph/SCET and \Pythia at the parton level to be worse for hadron collisions than for $e^+e^-$ collisions.

\subsection{\texorpdfstring{$e^+e^- \to \text{2 jets}$}{2 jets at lepton colliders}} 

\begin{figure}[t]
  \begin{center}
\includegraphics[width=0.6\textwidth]{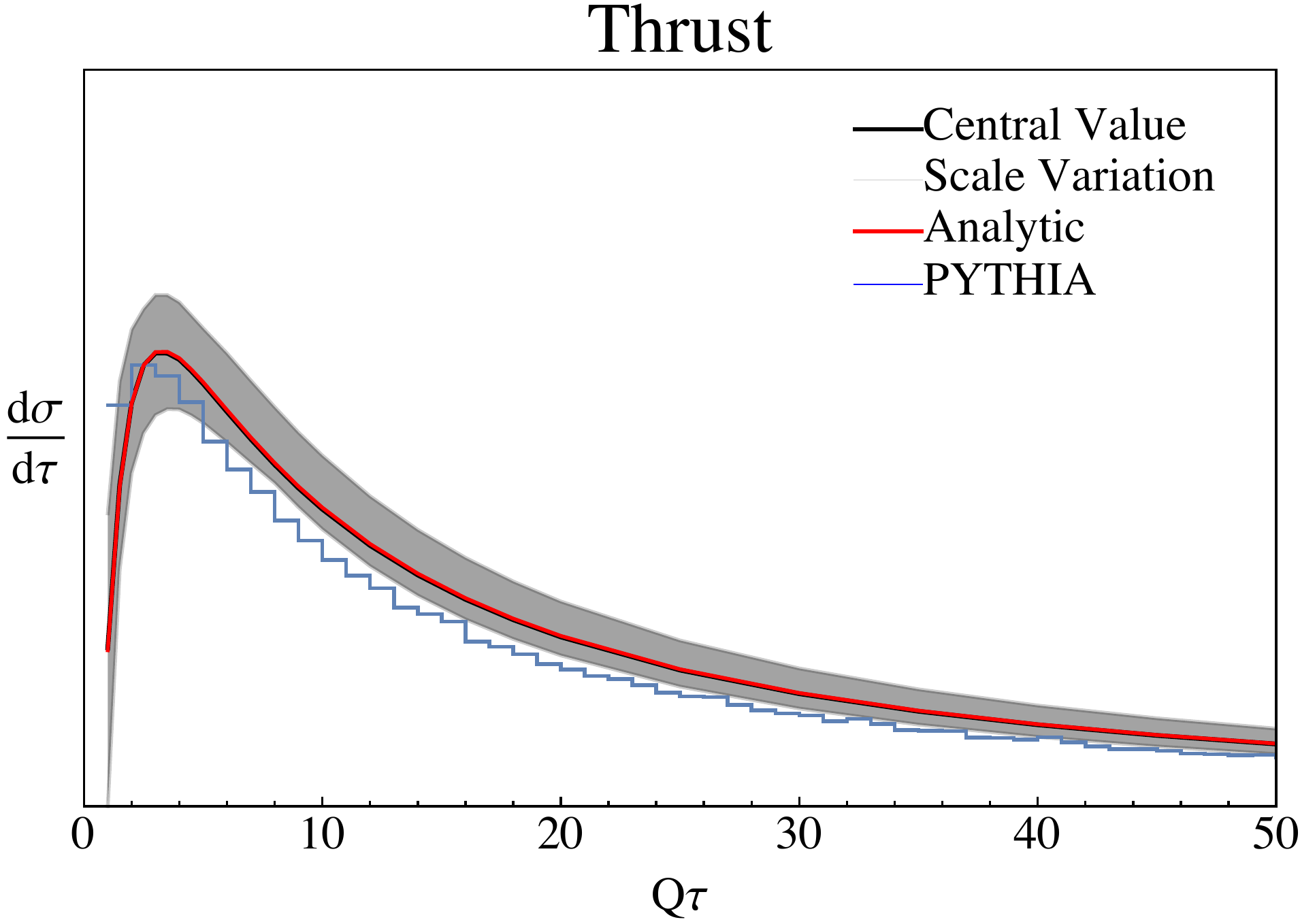}
\caption{Computation of thrust in $e^+e^-$ collisions at center of mass energy of 500 GeV. The histogram is from {\sc \MadGraph} matched to \Pythia, with hadronization off; the red curve is a purely analytic calculation~\cite{Feige:2012vc}. The blue is from \MadGraph matched to SCET, as discussed here. Not surprisingly, the central value of the blue band agrees exactly the red curve, as the \MadGraph part is trivial. }
\label{fig:eejj}
\end{center}
\end{figure}

The \MadGraph matching step is trivial for $e^+e^-\to 2j$ event shapes like thrust. For these cases, the soft function does not depend on the phase space point, and thus $F^\colorindex(\Phi, \obs)$ does not depend on $\Phi$. This means that the \MadGraph sum is trivial, and we should reproduce the analytic thrust results exactly. 

For scale choices, we use the natural scales
\begin{align}
\mu_h &= Q \\
\mu_j &= \sqrt{\tau} Q\\
\mu_s &= \tau Q
\end{align}
with $Q=500$ GeV the center of mass energy. One could use more sophisticated profile functions for scale choices~\cite{Abbate:2010xh}, but we use the natural scales for simplicity.

In \Fig{eejj} we show the output of our general framework for thrust. We generate $e^+e^- \to \bar{q} q$ events with \MadGraph. The \Pythia histogram is generated by then showering those events through \Pythia. For the central value curve, we use the same events and compute thrust for each event with SCET. Varying the scales by a factor of 2 produces the band. The red curve is the purely analytical calculation~\cite{Schwartz:2007ib,Becher:2008cf} at NLL (without using the \MadGraph events). It agrees exactly with the central value of the blue band. Note that the \MadGraph/\Pythia histogram is slightly off the analytic result. Some discrepancy is not surprising, as we are just running \Pythia out of the box and do not try to match scales or optimize the tuning.

\subsection{\texorpdfstring{$e^+e^- \to \text{4 jets}$}{4 jets at lepton colliders}} 
Next, we consider 4-jettiness in $e^+e^-$ events. This process is much more complicated than thrust, since it involves multiple color structures, multiple channels and the phase space is too complicated to integrate analytically. Indeed, this is the first NLL resummation of 4-jettiness.

To define 4-jettiness~\cite{Stewart:2010tn,Jouttenus:2011wh}, we need to partition phase space into 4 sectors $\mathcal{J}_i$ and assign a lightlike direction $n_i^\mu$ to each sector. Then, 4-jettiness is
\begin{equation}
  \label{eq:ee4jetdef}
  \cT_4 = \sum_{i=1}^4\cT^i
    \qquad \text{with} \qquad
  \cT^i \equiv \sum_{p_k \in \text{Jet}_i} n_i \cdot p_k = \frac{m_i^2}{E_i} + \dotsb
\end{equation}
While the original definition defined the $n_i$ with a minimization formula, the partitioning and assignment of $n_i$ can be done in different ways that are equivalent at leading power. Since we are interfacing to \MadGraph, we take the $n_i$ as the parton directions on an event-by-event basis. Note that, unlike thrust, $\cT_4$ is conventionally defined to have dimensions of energy. 

For $e^+e^- \to$ 4 jets, there are two partonic channels relevant to the order we work: $e^+e^-\to q\bar{q}gg$ and $e^+e^-\to qq\bar{q}\bar{q}$. Thus we need two different functions $F^\colorindex(\Phi, \cT_4)$, which we call $F^\colorindex_{qqqq}(\Phi, \cT_4)$ and $F^\colorindex_{qqgg}(\Phi, \cT_4)$. Details of these functions are given in \App{maincalculationappendix}. This process also has non-trivial color structure; the $qq\bar{q}\bar{q}$ channel has two color structures and the $q\bar{q}gg$ channel has three. Thus $F^\colorindex_{qqqq}(\Phi, \cT_4)$ is a $2\times2$ matrix, and $F^\colorindex_{qqgg}(\Phi, \cT_4)$ a $3\times3$ matrix.

For scale choices, we take
\begin{align}
\mu_h &= x \,(E_1 E_2 E_3 E_4)^{1/4}\\
\mu_j^i &= E_i \sqrt{\cT_4}\\
\mu_s &= \frac{1}{x} \cT_4
\end{align}
where $E_i$ are the energy of the partons and $x$ is an adjustment factor. These are natural in the sense that the soft scale only depends on the observable, each jet scale only depends on the energy of the relevant jet, and the hard scale satisfies $\mu_h \mu_s = \bar{\mu}_J^2$, with $\bar\mu_j = \sqrt{\mu_j^1 \mu_j^2 \mu_j^3 \mu_j^4}$ the geometric mean of the jet scales. We have included a numerical adjustment factor $x$ which should be of order 1. For example, reducing to thrust where $\mu_h$ is the center of mass energy, $x=2$. Varying $x$ is equivalent to varying the hard scale keeping $\mu_s \mu_h = \bar{\mu}_J^2$ (the {\it anti-correlated} scale variation from~\cite{Becher:2008cf}).

Because our method involves summing up distributions at each phase space point, we can compare the distributions from \MadGraph/SCET to \MadGraph/\Pythia at each phase space point individually. To do so we take a single \MadGraph event and shower it several thousand times with \Pythia. Recall that \Pythia showers the event assuming a particular color structure, chosen randomly once and for all at the start. So for purposes of comparing to \Pythia in these single phase-space-point plots, we ignore the color structure of the hard function in the SCET calculation and instead use the same definite color structure as \Pythia.

We consider two phase space points. First, a highly \textbf{symmetric event}:
\begin{equation}
\label{eq:symm}
  {\parbox{3cm}{\begin{tikzpicture}[thick,scale=2]
                \tikzset{>=latex};
                \draw[->, line width=2,opacity=1] (0.05,0) -- node[below]{$p_1$} (1,0) ;
                \draw[->, line width=2,opacity=1,rotate=90] (0.05,0) -- node[left]{$p_2$}  (1,0) ;
                \draw[->, line width=2,opacity=1] (-0.05,0) -- node[below]{$p_3$}  (-1,0) ;
                \draw[->, line width=2,opacity=1,rotate=90] (-0.05,0) --  node[right]{$p_4$} (-1,0) ;
                \end{tikzpicture}}}
  \hspace{30mm} 
  \begin{array}{l}
    p_1 = (800, 800, 0, 0) \GeV \\
    p_2 = (800, 0, 800, 0) \GeV \\
    p_3 = (800, -800, 0 , 0) \GeV \\
    p_4 = (800, 0, -800, 0) \GeV
  \end{array}
\end{equation}
Of course, taking a tetrahedron structure rather than a square would be more symmetric, but having two back-to-back partons facilitations the comparison with $\cT_2$ in $pp$ collisions, where the same phase space point can be used. Second, we consider a {\bf angled event} with outgoing partons separated by 30 degrees
\be
{
\parbox{3cm}
{
\begin{tikzpicture}
[thick,scale=2]
\tikzset{>=latex};
\draw[->, line width=2,opacity=1] (0.05,0) -- node[below]{$p_1$} (1,0) ;
\draw[->, line width=2,opacity=1,rotate=30] (0.05,0) -- node[above]{$p_2$}  (1,0) ;
\draw[->, line width=2,opacity=1] (-0.05,0) -- node[above]{$p_3$}  (-1,0) ;
\draw[->, line width=2,opacity=1,rotate=30] (-0.05,0) --  node[below]{$p_4$} (-1,0) ;
\end{tikzpicture}
}
}
\hspace{30mm} 
\begin{array}{l}
p_1 = (800, 800, 0, 0) \GeV \\
p_2 = (800, 692, 400, 0) \GeV \\
p_3 = (800, -800, 0 , 0) \GeV \\
p_4 = (800, -692, -400, 0) \GeV
\end{array}
\label{eq:scissor}
\ee
The center-of-mass energy for this collision is $\ecm = 3200$ GeV. 

For these two phase space points, where all partons have $E_i = 800$ GeV, the scales are
\begin{equation}
  \label{eq:mhvary}
  \begin{split}
    \mu_h   &= x\,  800 \GeV \\
    \mu_j^i &= \sqrt{800 \GeV \cT_4} \\
    \mu_s   &= \frac{1}{x} \cT_4
  \end{split}
\end{equation}
we will show plots for different $x$, or equivalently different choices of $\mu_h$.

We can make qualitative observations about the SCET calculation at a given phase space point in two ways: using scale variations, and comparing to \Pythia. Comparing to \Pythia is dangerous, of course, as SCET/\MadGraph and \Pythia are not expected to agree, and \Pythia with hadronization and underlying event turned off is IR unsafe and tuning-dependent. On the other hand, since \Pythia resums all logarithms at the LL level, while SCET resums some large logarithms (those of $\cT_4$ to NLL) and others not even to LL (\emph{e.g.} logs of ratios of hard kinematic invariants, like $\ln\frac{s}{t}$), the comparison at different phase space points can illustrate the relevant tradeoffs.

\begin{figure}[t]
  \begin{center}
    \begin{tabular}{lcccc}
      {} & $\mu_H = 400\GeV$ & $\mu_H = 600\GeV$ & $\mu_H = 800\GeV$ & $\mu_H = 1600\GeV$ \\
      \tikz{\node [rotate=90] (text) {$e^+e^- \to u\bar{u}u\bar{u}$};} &
      \includegraphics[width=0.22\textwidth]{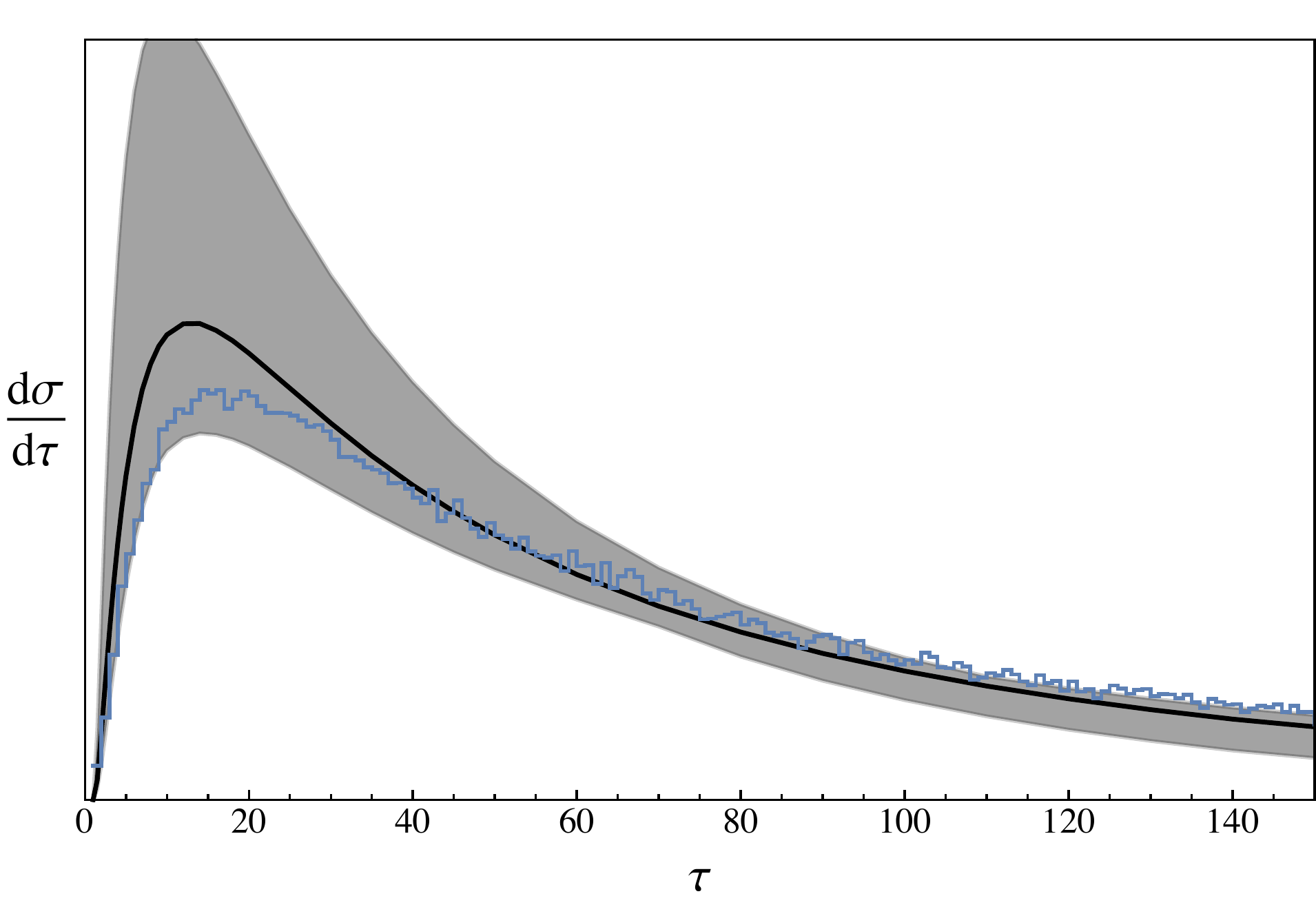} &
      \includegraphics[width=0.22\textwidth]{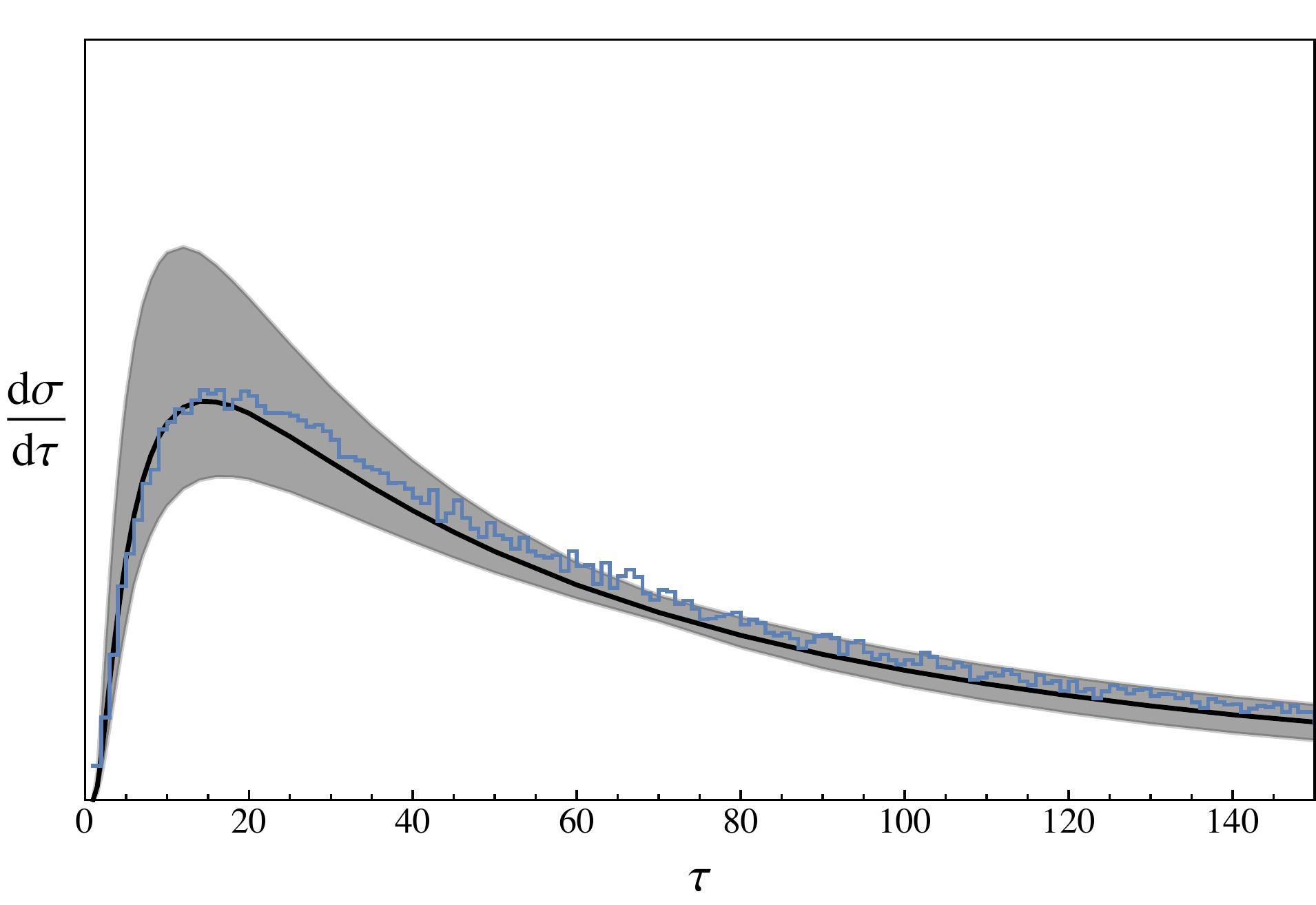} &
      \includegraphics[width=0.22\textwidth]{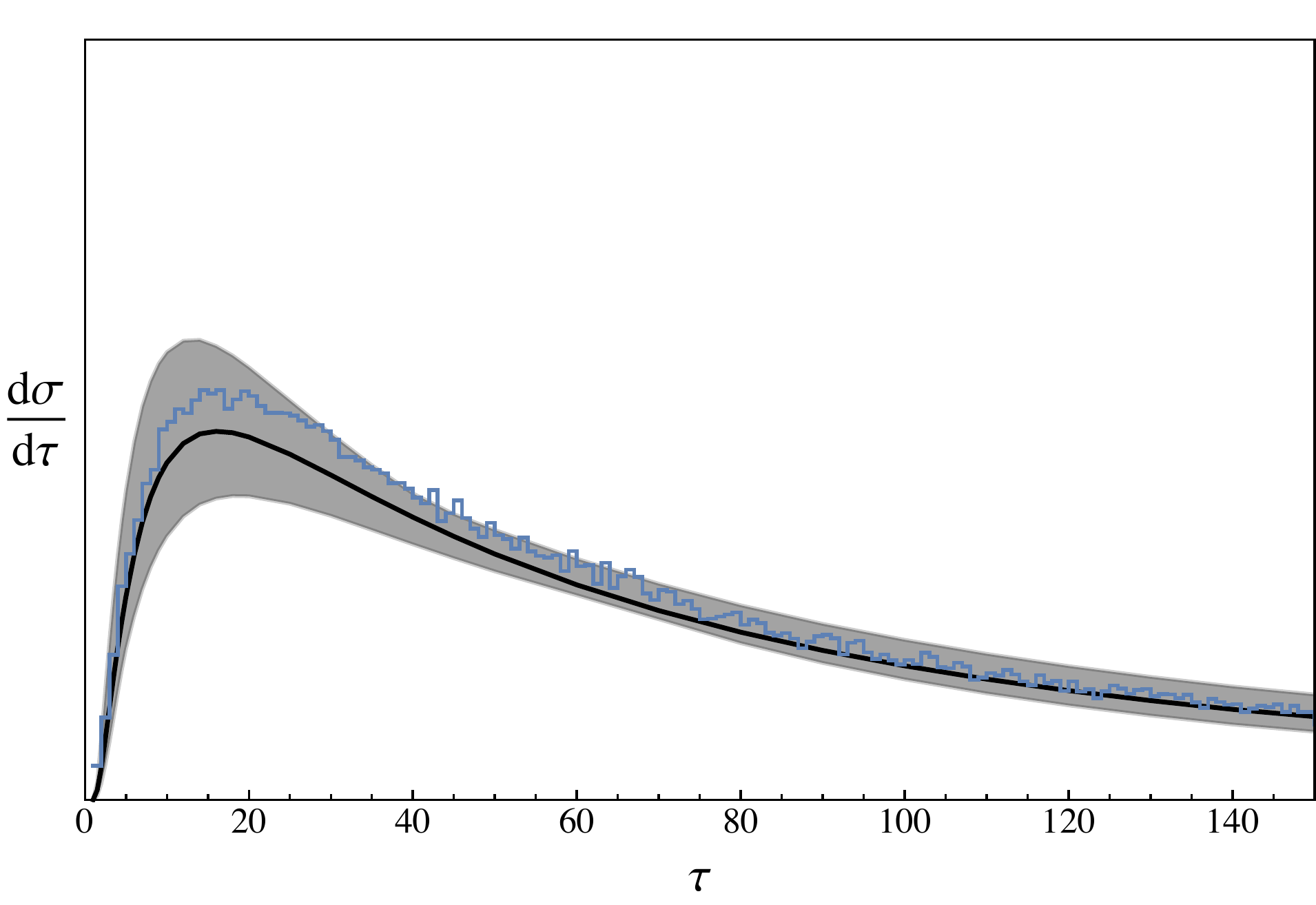} &
      \includegraphics[width=0.22\textwidth]{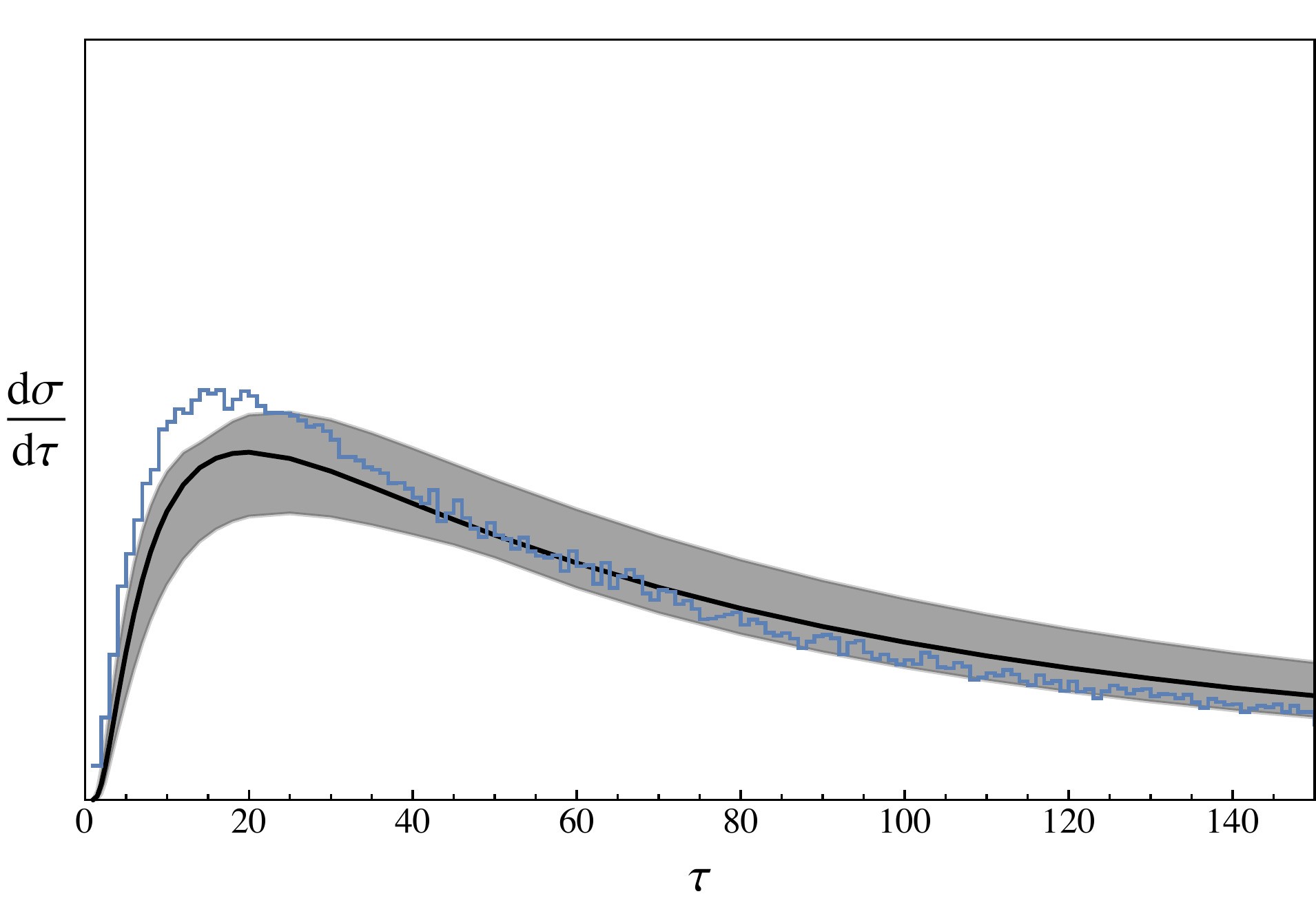} \\ 
      \tikz{\node [rotate=90] (text) {$e^+e^- \to u\bar{u}gg$};} &
      \includegraphics[width=0.22\textwidth]{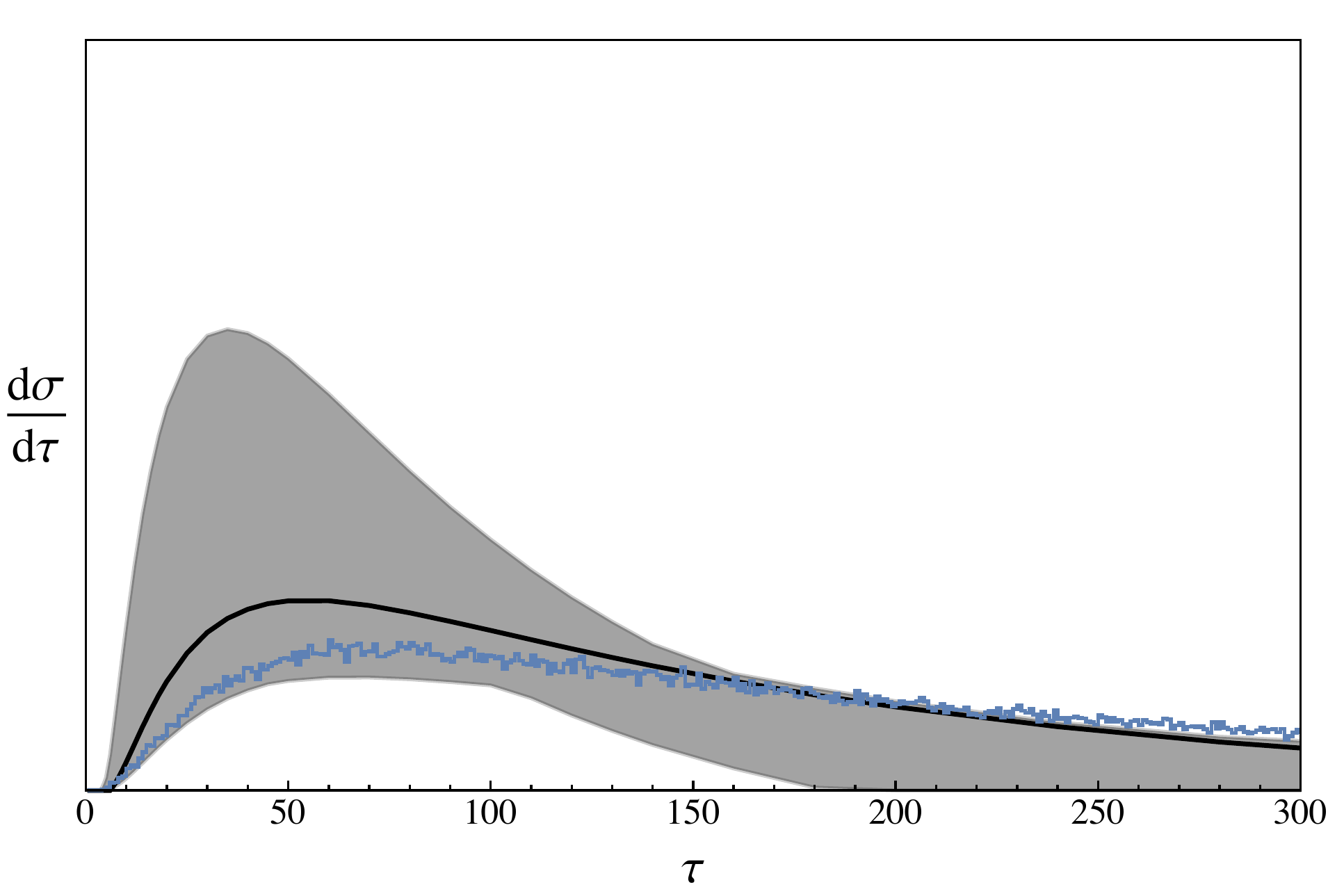} &
      \includegraphics[width=0.22\textwidth]{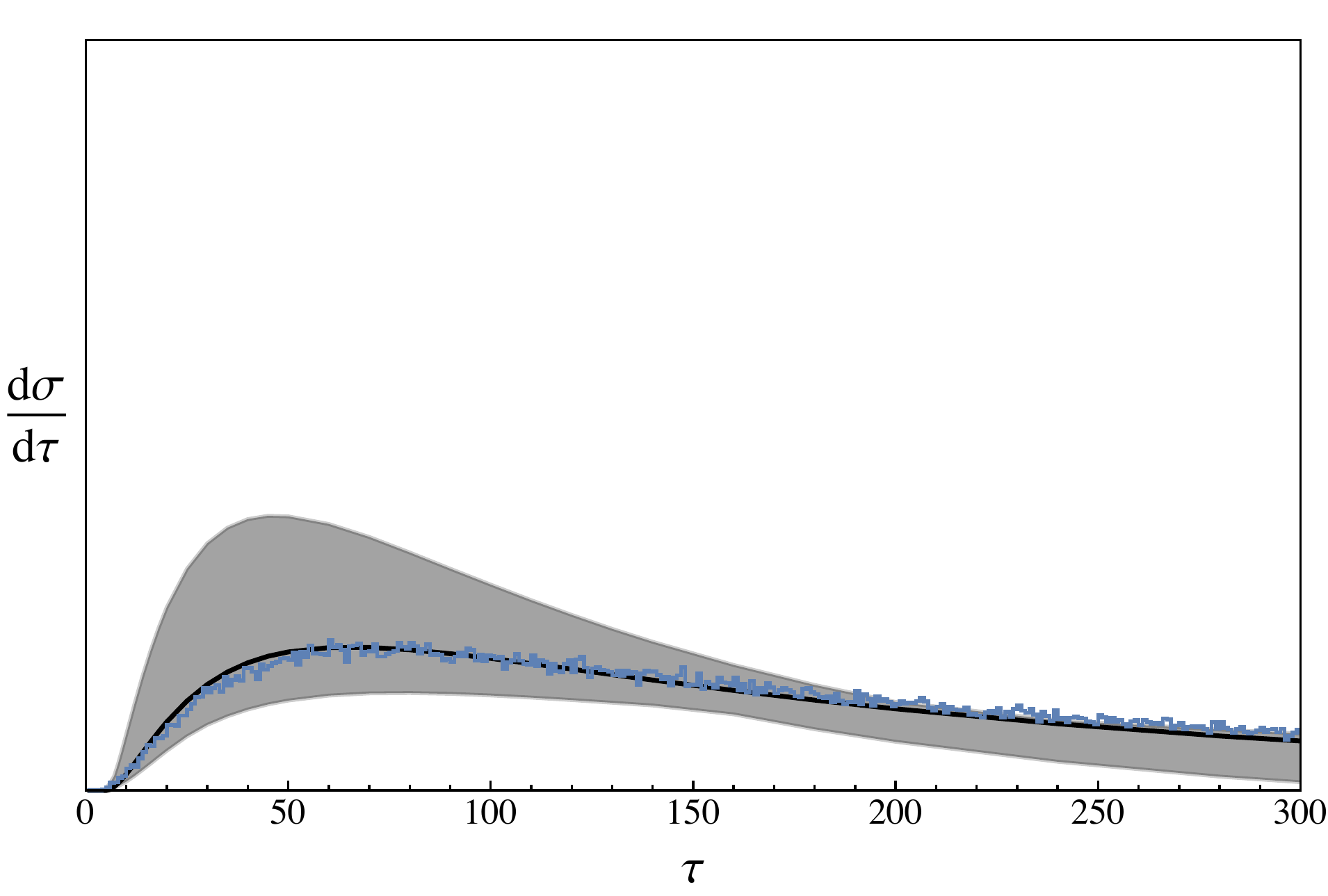} &
      \includegraphics[width=0.22\textwidth]{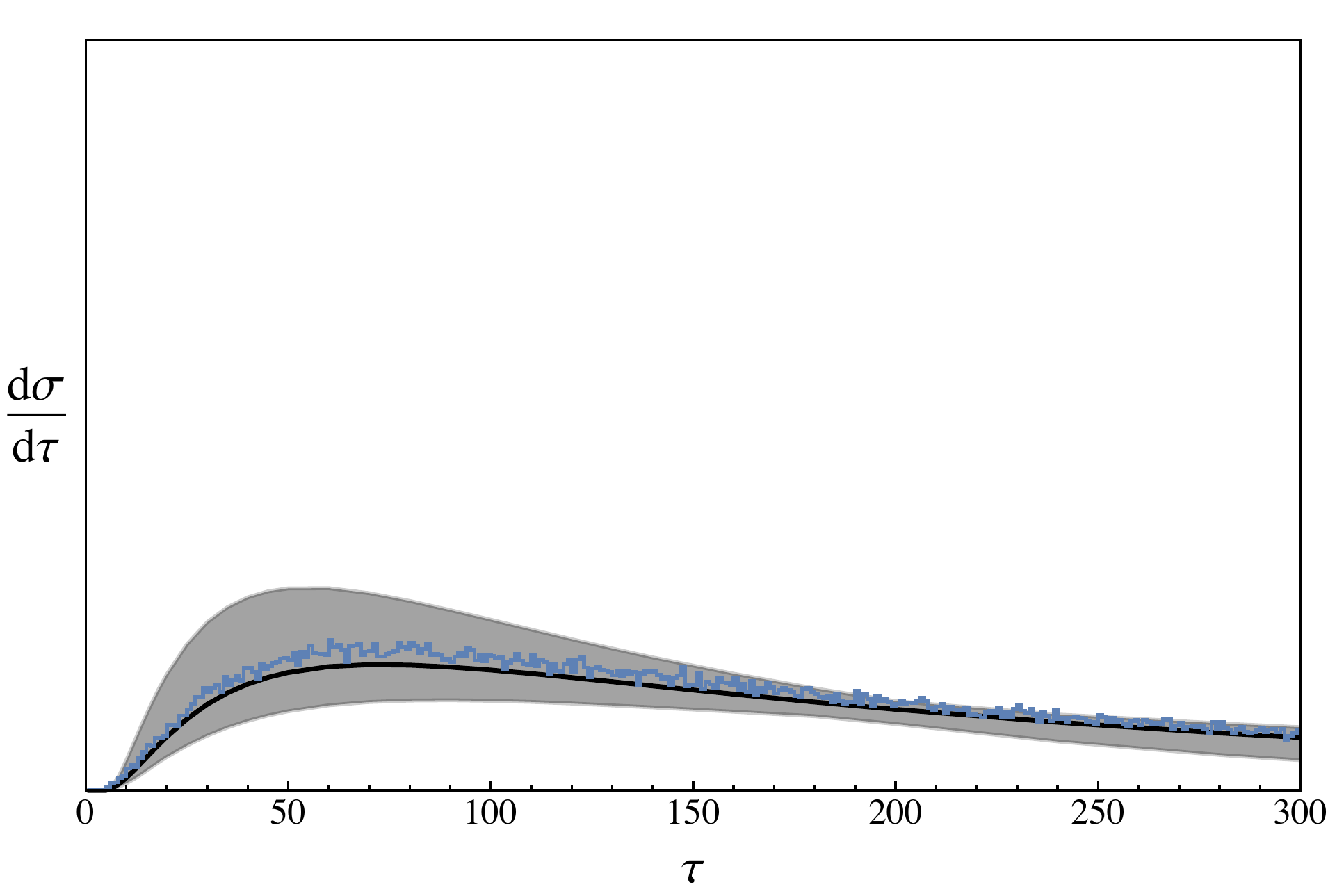} &
      \includegraphics[width=0.22\textwidth]{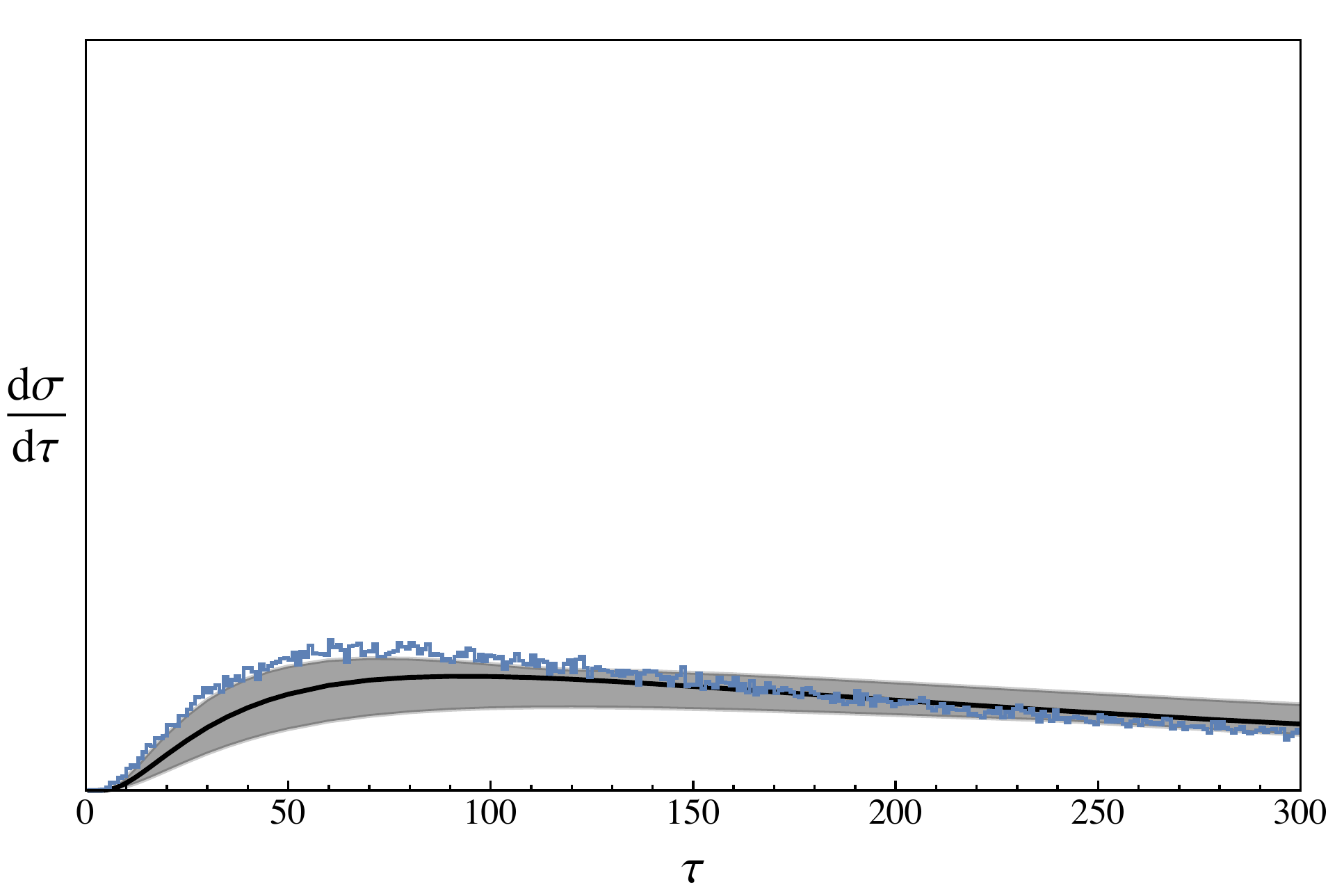} \\
      {} & \multicolumn{4}{c}{\includegraphics[width=0.66\textwidth]{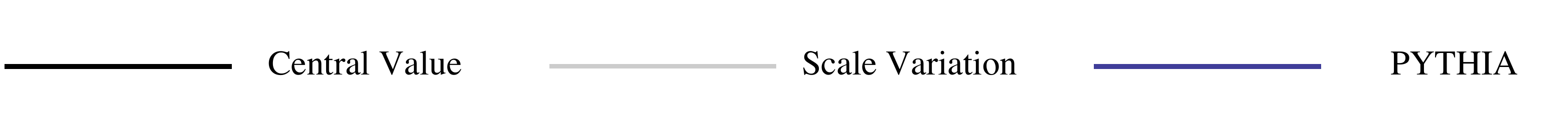}}
    \end{tabular}
  \end{center}
  \caption{Distributions of $\cT_4$ from the  $e^+e^- \to u \bar{u} u \bar{u}$ partonic channel (top) and  $e^+e^- \to u \bar{u} g g$  (bottom) for the symmetric event phase space point in \Eq{symm}. \Pythia is the blue histograms. Black is the central value for \MadGraph/SCET (at this phase space point only) with different choices of $\mu_h$ according to \Eq{mhvary}. Gray bands are the union of hard, jet and soft scale variations by factors of 2 around the central value.}
  \label{fig:ee4jSquareEvent}
\end{figure}

\Fig{ee4jSquareEvent} shows the results for the symmetric phase space point with $e^+e^- \to u\bar{u} u\bar{u}$ at the parton level (top) and $e^+e^- \to u\bar{u} gg$ (bottom). We see that the scale variation is minimized for $\mu_h \sim 800~\GeV ~(x \sim 1)$. There is good agreement between \Pythia and SCET for $ x \lesssim 1$ as well. We conclude that for symmetric phase points, which have only one scale, the resummation of $\cT_4$ using SCET is as effective as the resummation of thrust ($\cT_2$). This is not surprising, yet reassuring.

For the symmetric phase space point there is really only one scale in the hard function, and so we expect the resummation to miss no large logs. Therefore it is not shocking that there is a scale at which SCET and \Pythia agree; \Pythia's shower is equivalent to some particular SCET scale choice (around 600 GeV). However, there is no reason to expect this scale choice to be the same as the natural SCET scale choice (the one where the scale variations are minimized). This difference will be more pronounced in $pp$ scattering later (Fig.~\ref{fig:ppSquare}), but already here we see a slight difference: the SCET error bands are minimized closer to 800 GeV or higher, where the disagreement with \Pythia is slightly worse.

For the angled phase space point, the scales are all roughly the same, but there are some (small) logarithms which can be relevant. For example, $\frac{p_1 \cdot p_3}{p_1 \cdot p_2}=3.8$. Because of the lack of symmetry, we will consider two different color configurations:
\be
\parbox{3cm}{
  \begin{tikzpicture}[thick,scale=2]
  \tikzset{>=latex};
  \draw[->, line width=2,opacity=1] (0.05,0) -- node[below]{$u$} (1,0) ;
  \draw[->, line width=2,opacity=1,rotate=30] (0.05,0) -- node[above]{$\bar{u}$}  (1,0) ;
  \draw[->, line width=2,opacity=1] (-0.05,0) -- node[above]{$u$}  (-1,0) ;
  \draw[->, line width=2,opacity=1,rotate=30] (-0.05,0) --  node[below]{$\bar{u}$} (-1,0) ;
  \draw[-,line width=1,red]   (-0.8,0.2) to[out=-10,in=-140] (0.6, 0.6);
  \draw[-,line width=1,blue]   (-0.6,-0.6) to[out=40,in=170] (0.8, -0.2);
  \end{tikzpicture}
}
\hspace{20mm}
\parbox{3cm}{
  \begin{tikzpicture}[thick,scale=2]
  \tikzset{>=latex};
  \draw[->, line width=2,opacity=1] (0.05,0) -- node[below]{$u$} (1,0) ;
  \draw[->, line width=2,opacity=1,rotate=30] (0.05,0) -- node[above]{$\bar{u}$}  (1,0) ;
  \draw[->, line width=2,opacity=1] (-0.05,0) -- node[above]{$u$}  (-1,0) ;
  \draw[->, line width=2,opacity=1,rotate=30] (-0.05,0) --  node[below]{$\bar{u}$} (-1,0) ;
  \draw[-,line width=1,red]   (-0.8,-0.1) to[out=5,in=25,distance = 10] (-0.7, -0.3);
  \draw[-,line width=1,blue]   (0.8,0.1) to[out=185,in=205,distance = 10] (0.7, 0.3);
\end{tikzpicture}
}
\ee

\begin{figure}[t]
  \begin{center}
    \begin{tabular}{lccc}
      {} & $\mu_H = 400\GeV$ & $\mu_H = 800\GeV$ & $\mu_H = 1600\GeV$ \\
      \begin{tikzpicture}[thick,scale=1] %, every node/.style={scale=1}]
        \useasboundingbox (1,-1) rectangle (0,1);
        \tikzset{>=latex};
        \draw[->, line width=1,opacity=1] (0.05,0) -- node[below]{} (1,0) ;
        \draw[->, line width=1,opacity=1,rotate=30] (0.05,0) -- node[above]{}  (1,0) ;
        \draw[->, line width=1,opacity=1] (-0.05,0) -- node[above]{}  (-1,0) ;
        \draw[->, line width=1,opacity=1,rotate=30] (-0.05,0) --  node[below]{} (-1,0) ;
        \draw[-,line width=1,red]   (-0.8,-0.1) to[out=5,in=25,distance = 10] (-0.7, -0.3);
        \draw[-,line width=1,blue]   (0.8,0.1) to[out=185,in=205,distance = 10] (0.7, 0.3);
      \end{tikzpicture} &
      \includegraphics[width=0.22\textwidth]{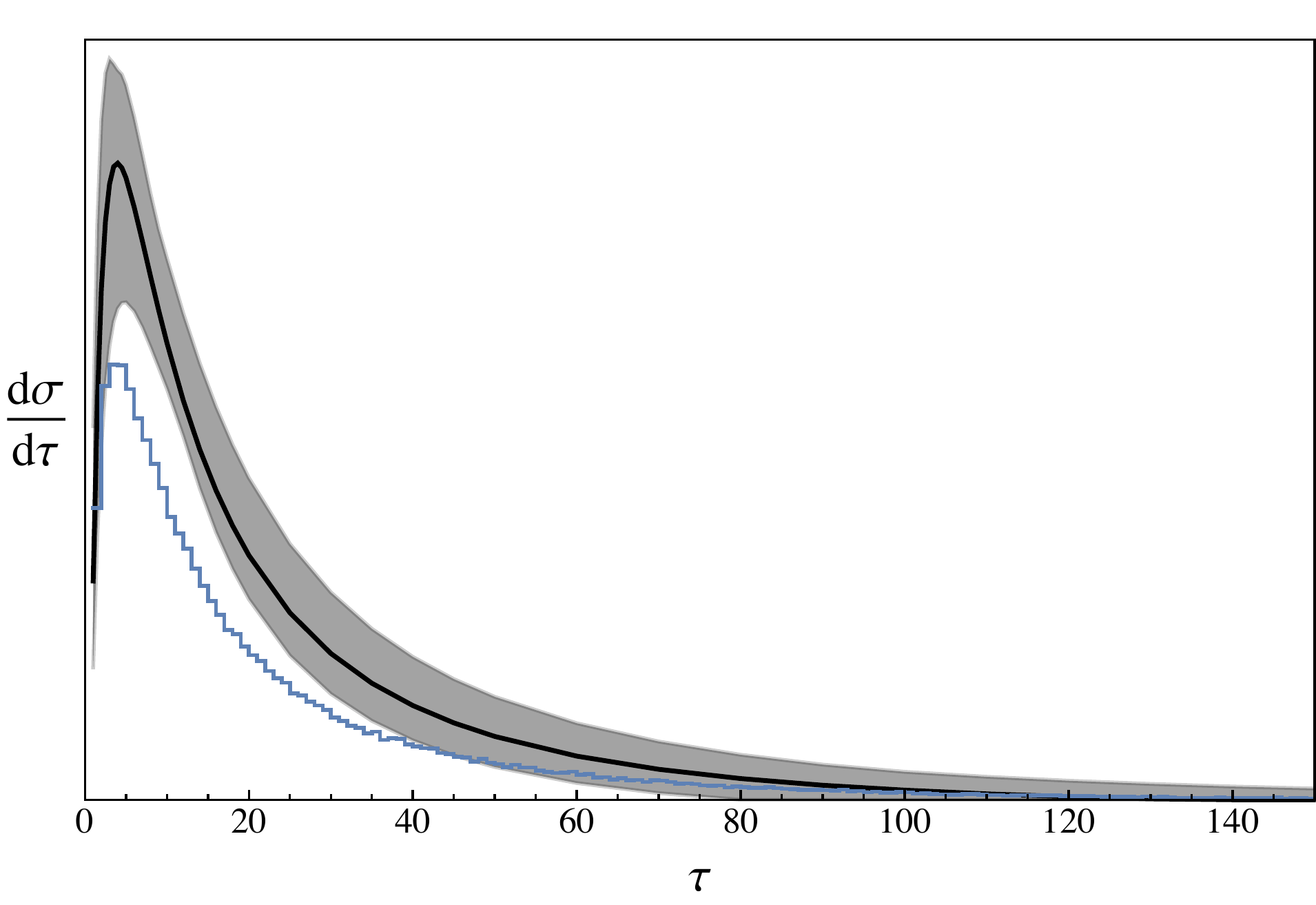} &
      \includegraphics[width=0.22\textwidth]{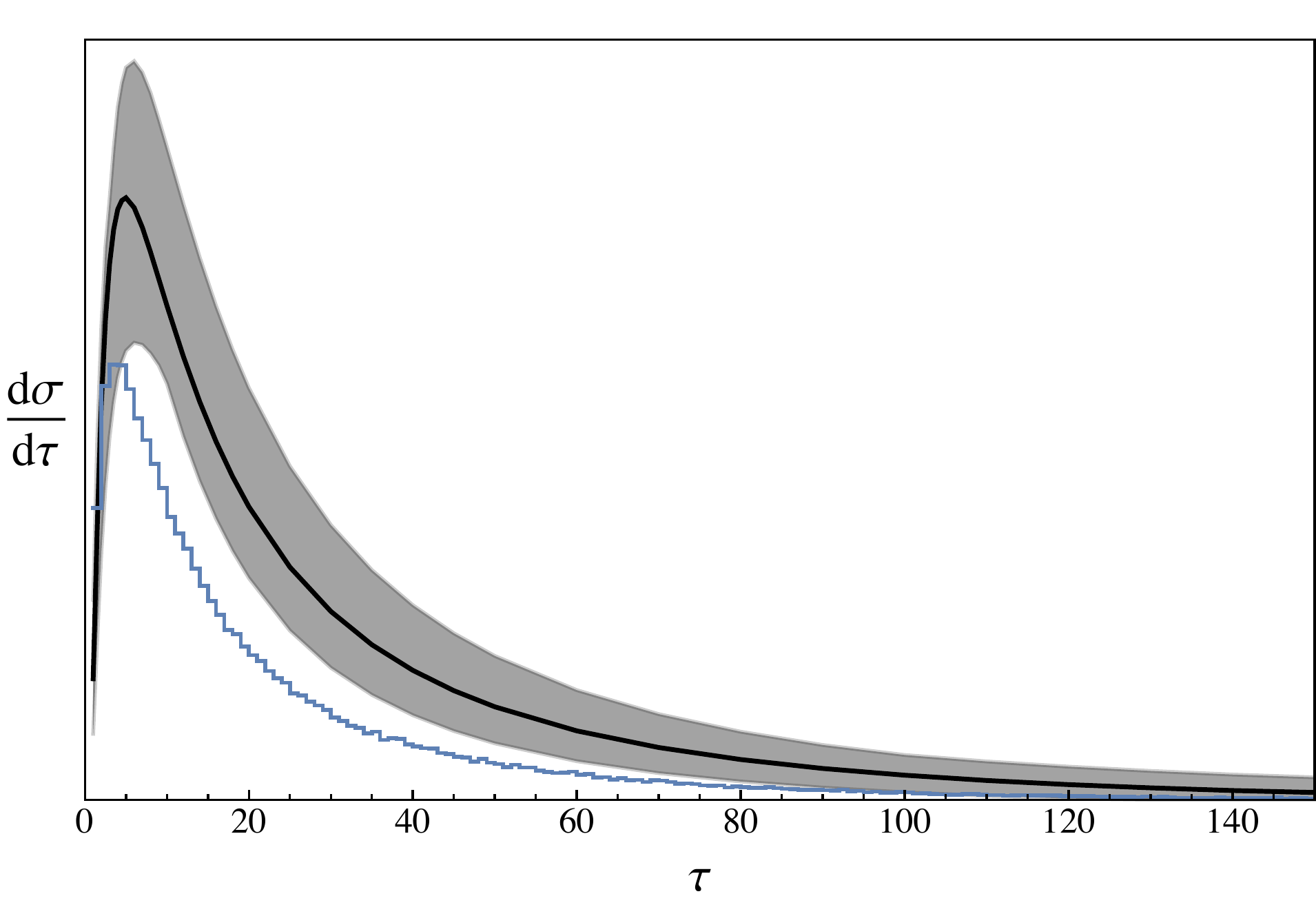} &
      \includegraphics[width=0.22\textwidth]{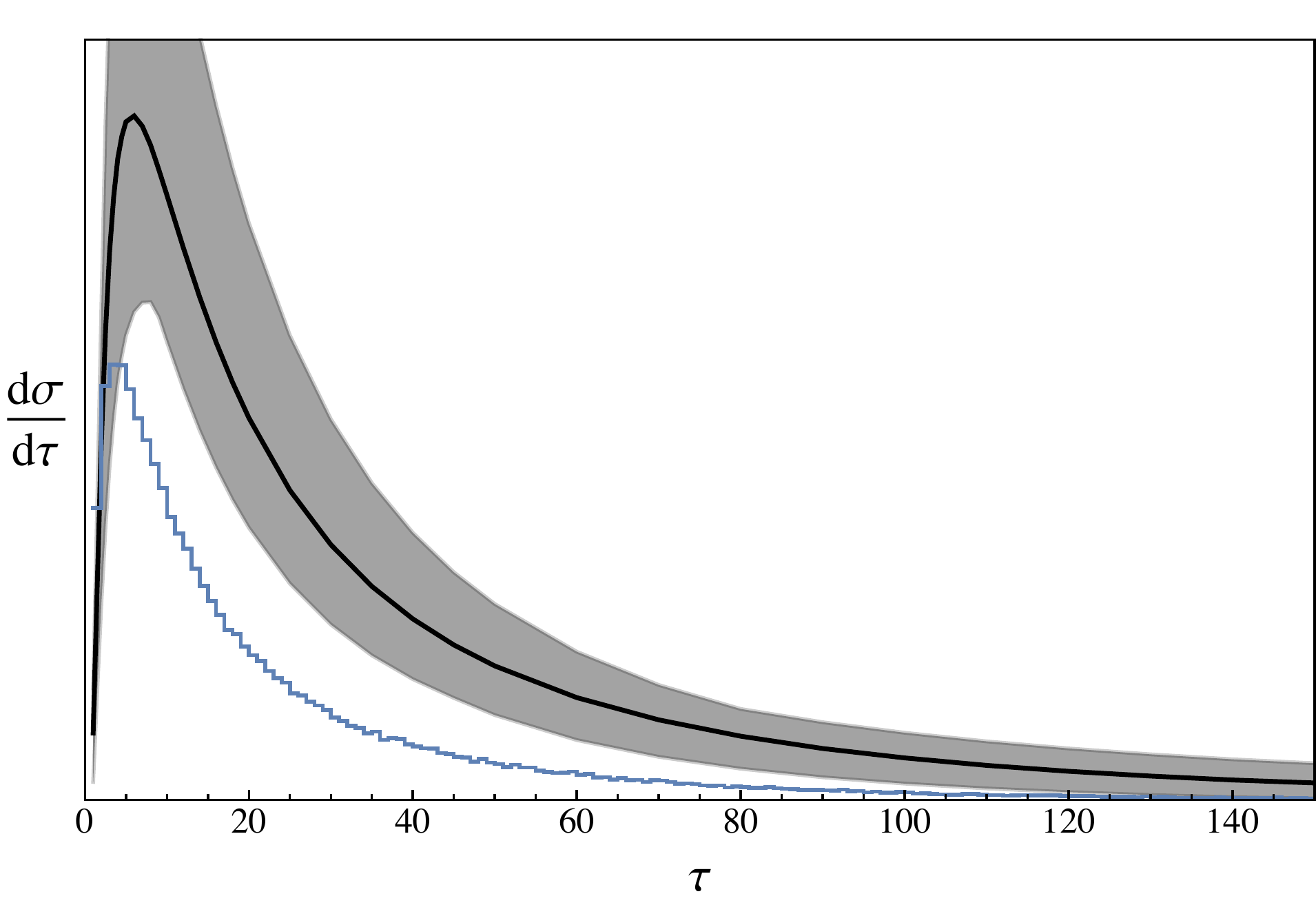} \\
      \begin{tikzpicture}[thick,scale=1]
        \useasboundingbox (1,-1) rectangle (0,1);
        \tikzset{>=latex};
        \draw[->, line width=1,opacity=1] (0.05,0) -- node[below]{} (1,0) ;
        \draw[->, line width=1,opacity=1,rotate=30] (0.05,0) -- node[above]{}  (1,0) ;
        \draw[->, line width=1,opacity=1] (-0.05,0) -- node[above]{}  (-1,0) ;
        \draw[->, line width=1,opacity=1,rotate=30] (-0.05,0) --  node[below]{} (-1,0) ;
        \draw[-,line width=1,red]   (-0.8,0.2) to[out=-10,in=-140] (0.6, 0.6);
        \draw[-,line width=1,blue]   (-0.6,-0.6) to[out=40,in=170] (0.8, -0.2);
      \end{tikzpicture} &
      \includegraphics[width=0.22\textwidth]{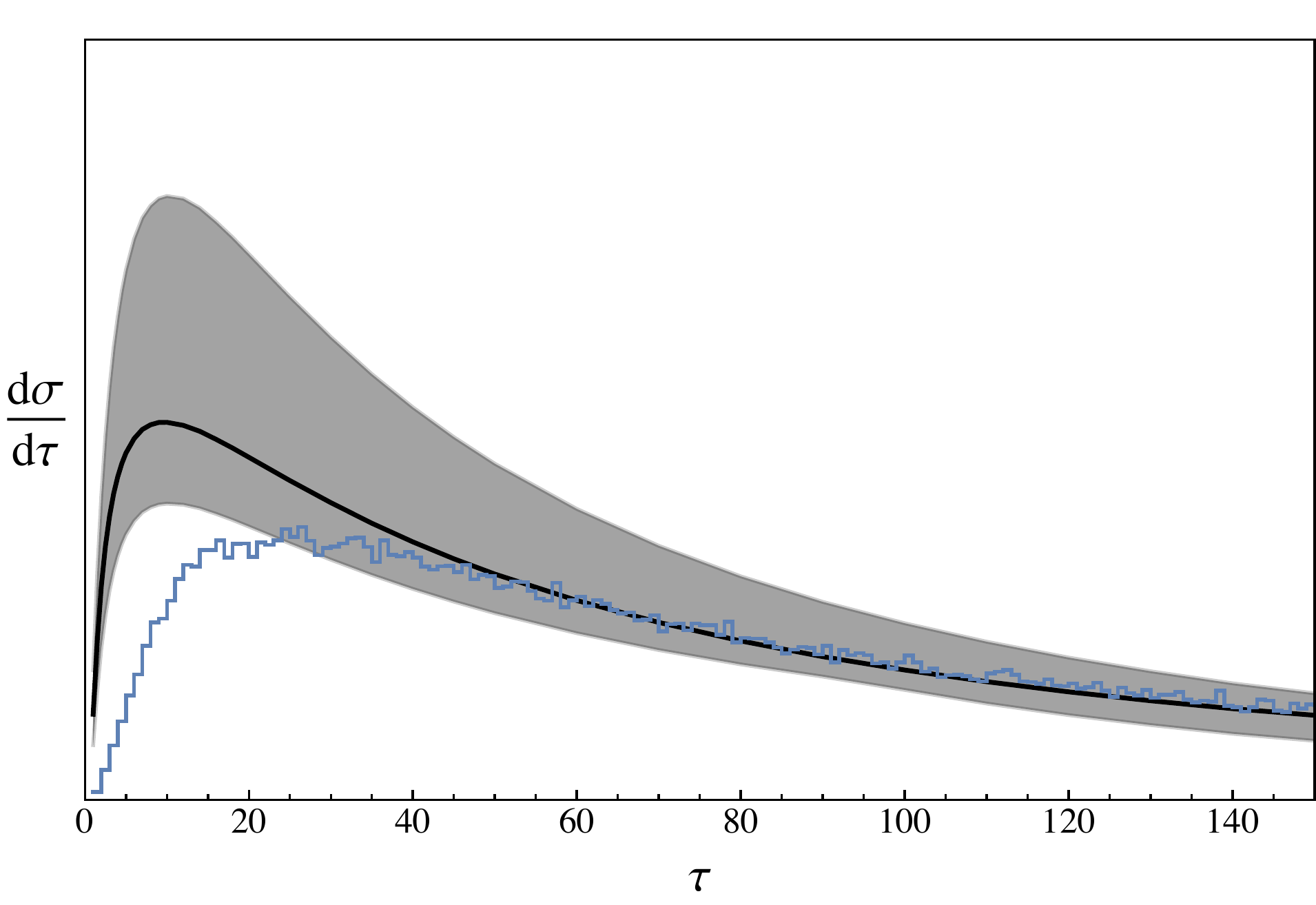} &
      \includegraphics[width=0.22\textwidth]{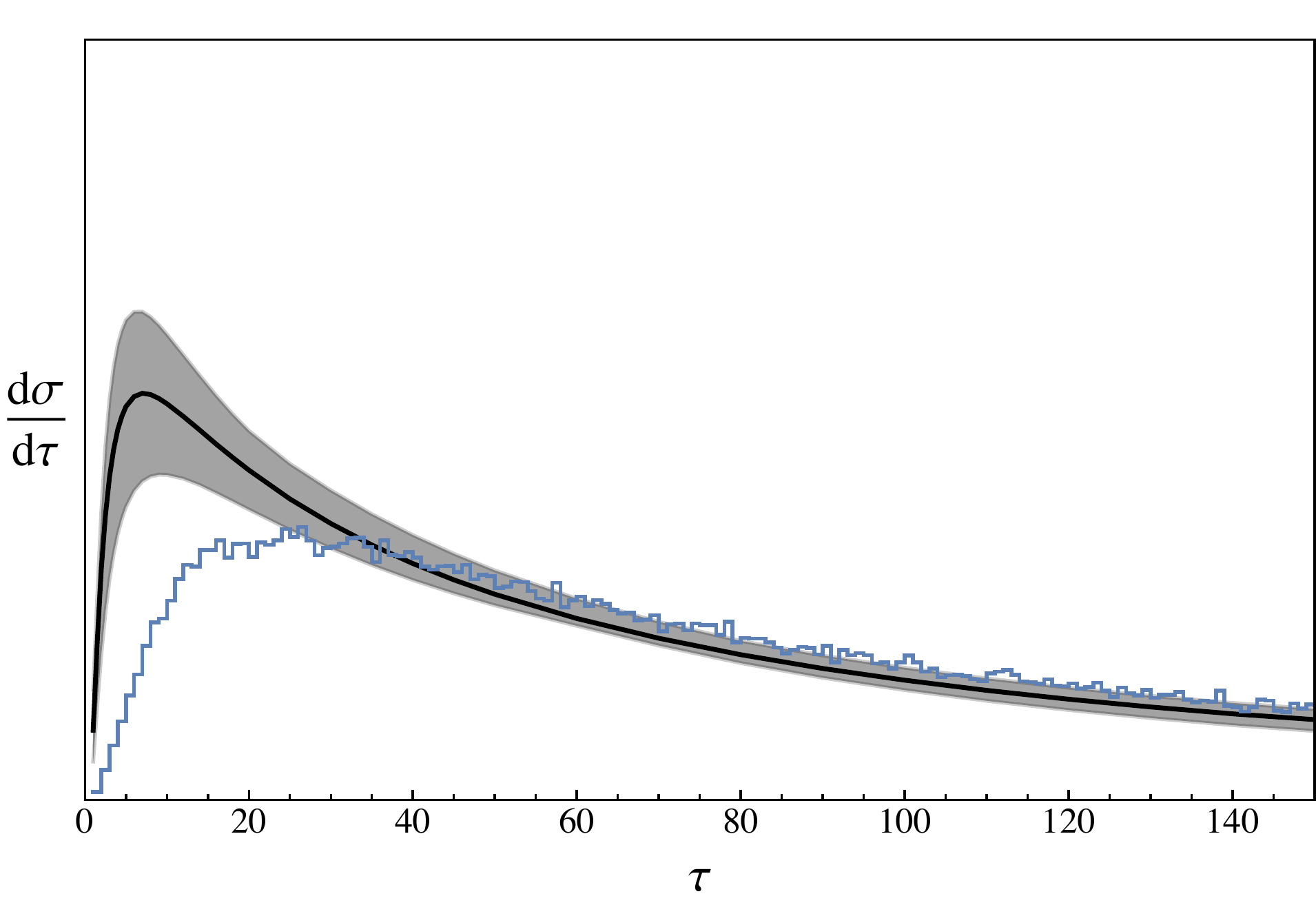} &
      \includegraphics[width=0.22\textwidth]{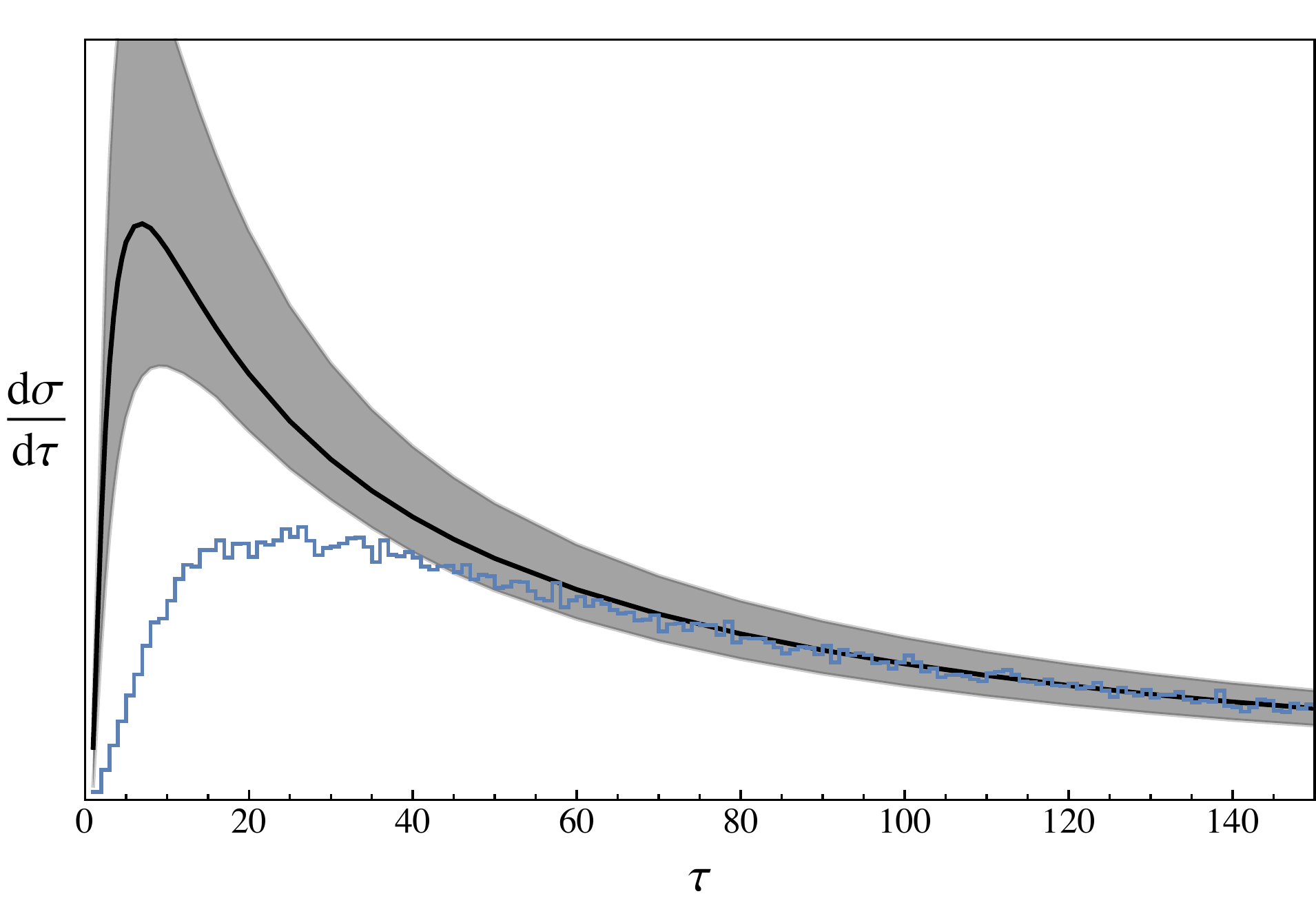} \\
      {} & \multicolumn{3}{c}{\includegraphics[width=0.66\textwidth]{Plots/Legend}}
    \end{tabular}
  \end{center}
  \caption{$\cT_4$ distributions in events with the angled phase space point in \Eq{scissor} for $e^+e^- \to u\bar{u} u\bar{u}$. Top row has the nearby momenta quarks color-connected, and bottom row has back-to-back partons color connected.}
  \label{fig:ee4jScissorsLargeEvent}
\end{figure}

\Fig{ee4jScissorsLargeEvent} shows the distributions for different choices of $\mu_h$ of $\cT_4$ in events with $e^+e^- \to u\bar{u} u \bar{u}$ at the parton level. The top row has the close $u$ and $\bar{u}$ ($p_1$ and $p_2$ in \Eq{scissor}) color-connected. The bottom row has the back-to-back partons, $p_1$ and $p_3$, color connected. As discussed in greater detail at the end of \Sec{2jetti}, the agreement between events showered with \Pythia and analytic distributions at individual phase space points is worse than in the case of \Fig{ee4jSquareEvent} due to the presence of additional logarithms that the analytic expression does not resum. However, these additional logarithms cancel out in the averaging over phase space.

\begin{figure}[h]
\begin{center}
\includegraphics[width=0.7\textwidth]{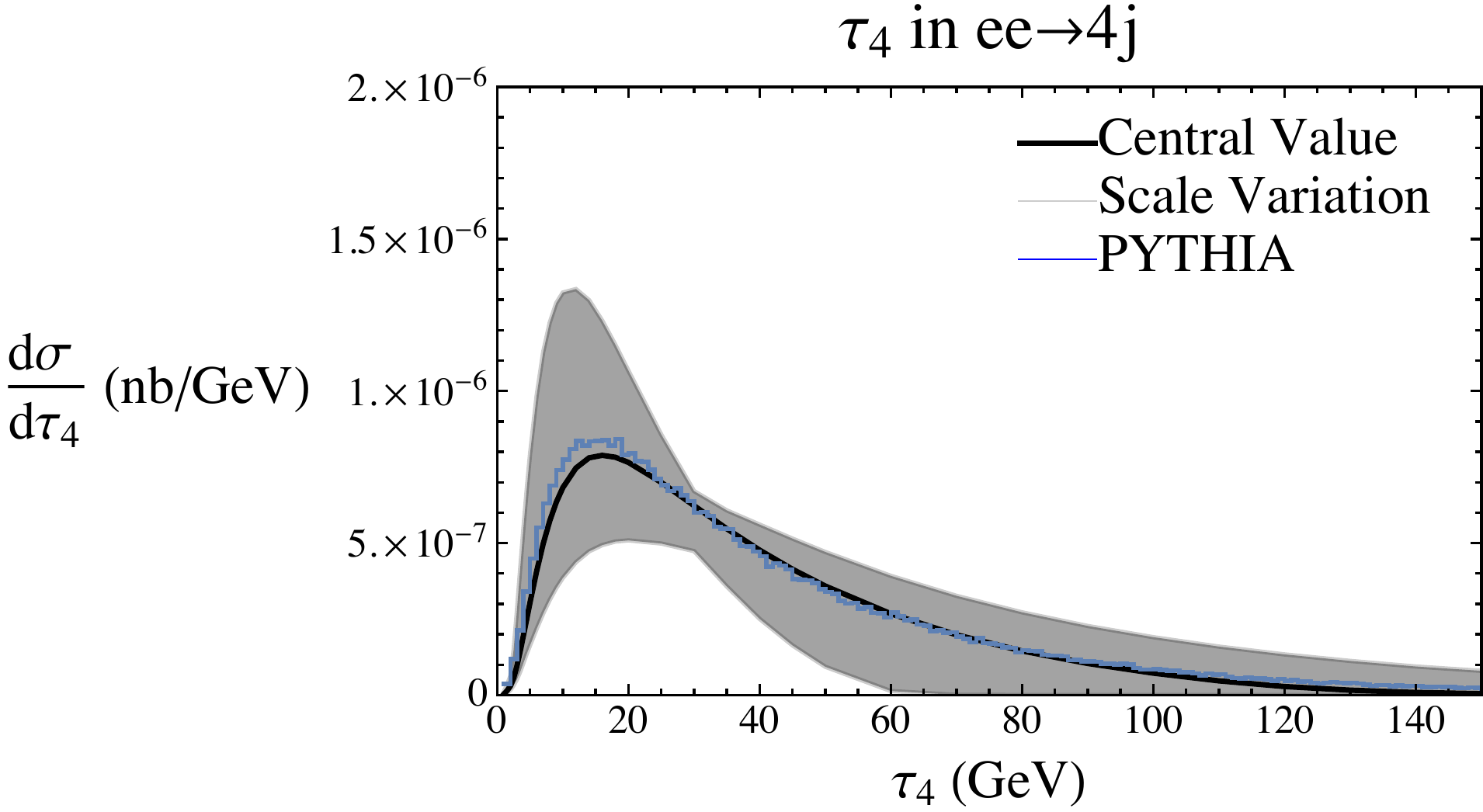}
\caption{Full distribution of $\cT_4$ (integrated over phase space, summed over channels) in $e^+e^-\to$ 4 jets at $\ecm = 500~\GeV$. Black line is the central value of the SCET/\MadGraph 
calculation with $x=1$ (so $\mu_h \approx 125~\GeV$). The gray band the union of the hard,
jet and soft scale variations. \Pythia is shown as the blue histogram.}
\label{fig:eejjjj}
\end{center}
\end{figure}

\begin{figure}[t]
  \begin{center}
    \begin{tabular}{lcccc}
      {} & $\mu_H = 400\GeV$ & $\mu_H = 600\GeV$ & $\mu_H = 800\GeV$ & $\mu_H = 1600\GeV$ \\
      \tikz{\node [rotate=90] (text) {$uu \to uu$};} &
      \includegraphics[width=0.2\textwidth]{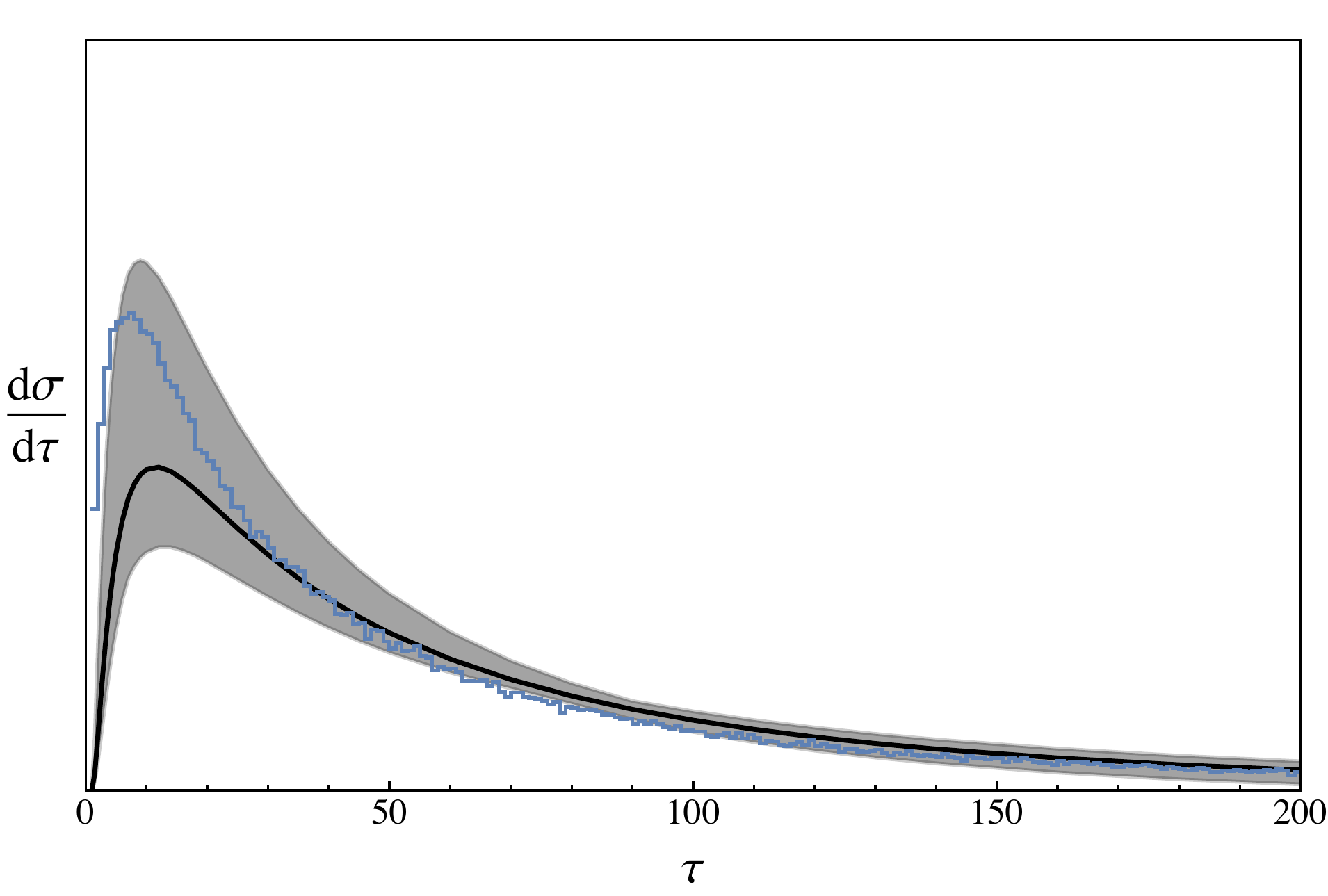} &
      \includegraphics[width=0.2\textwidth]{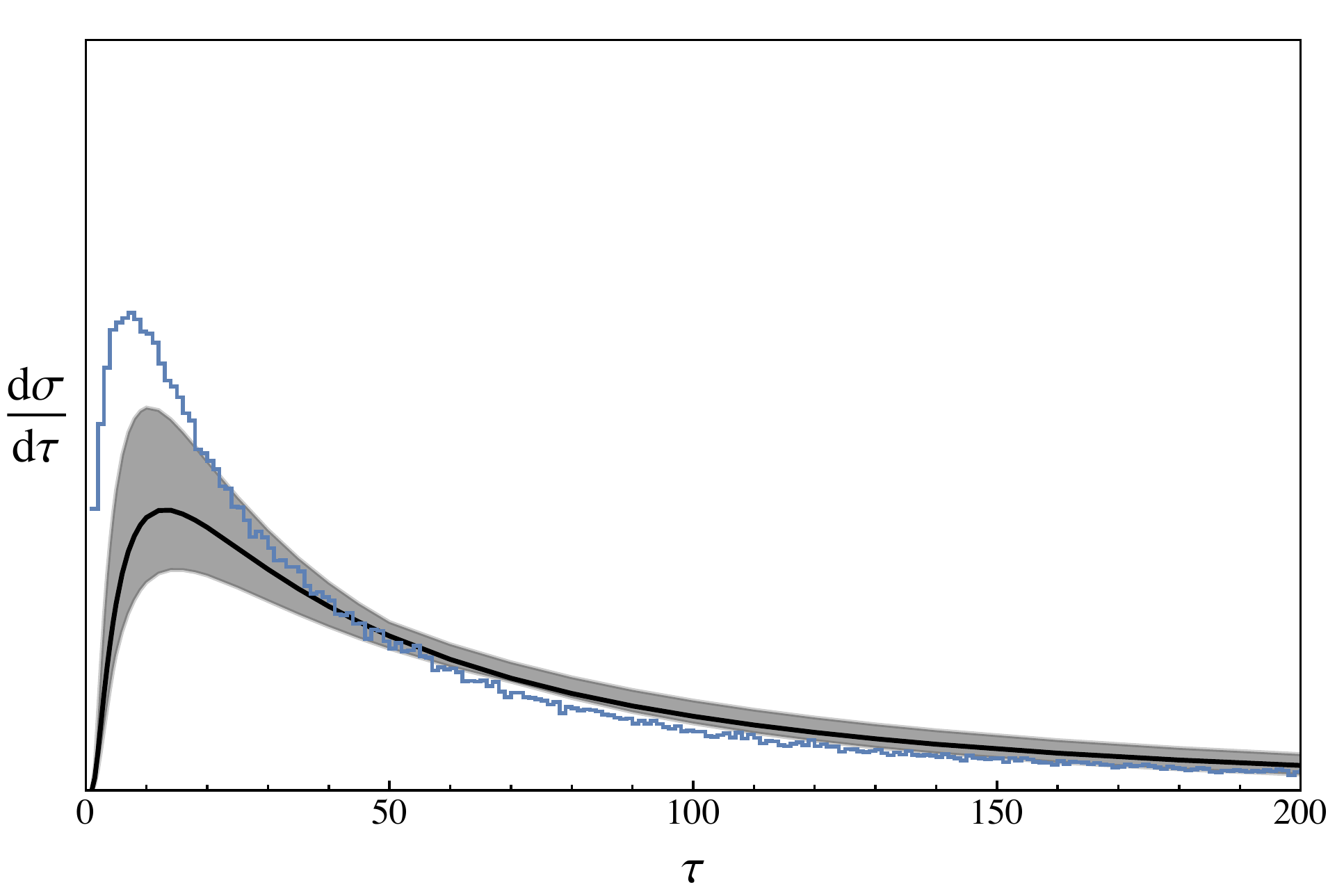} &
      \includegraphics[width=0.2\textwidth]{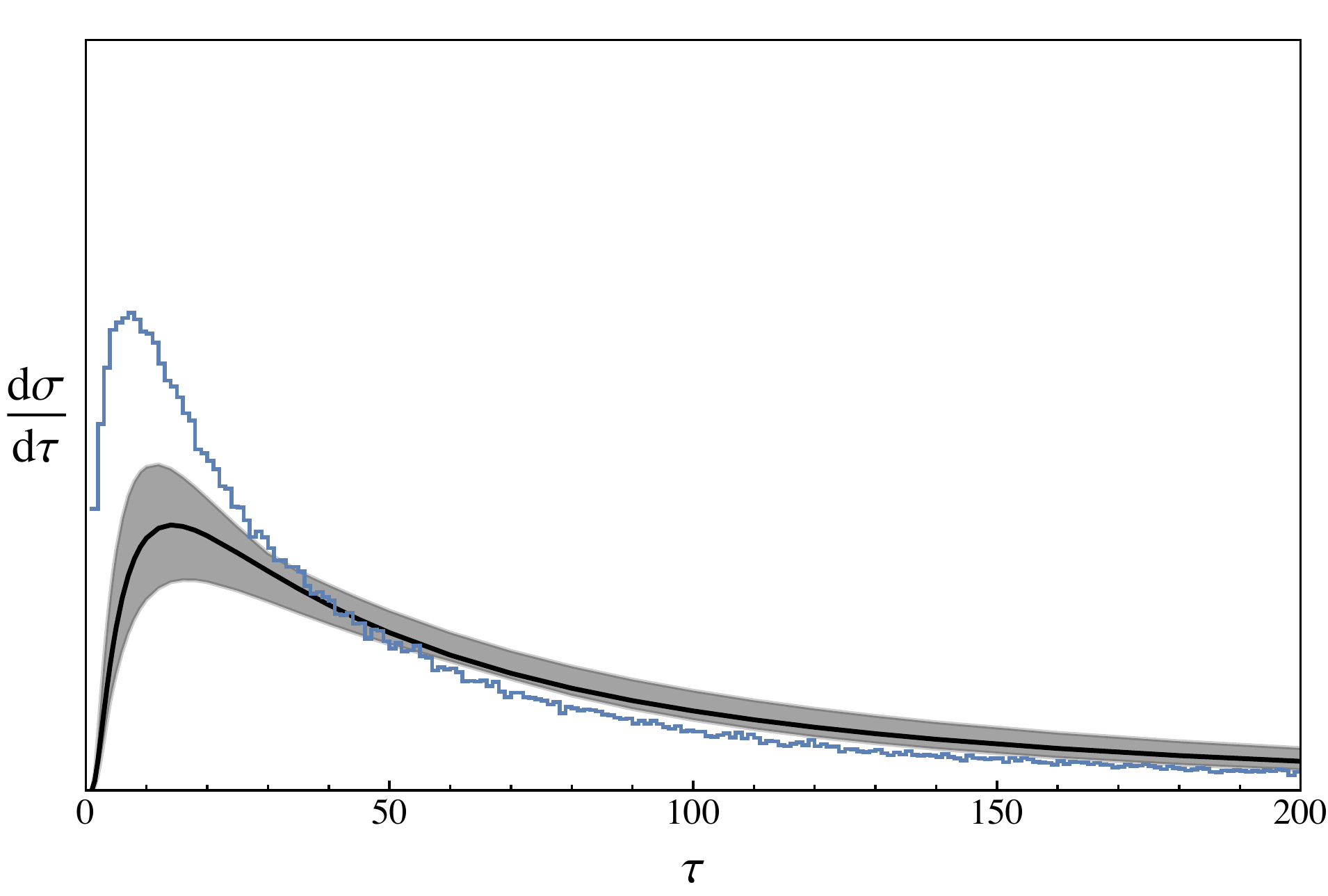} &
      \includegraphics[width=0.2\textwidth]{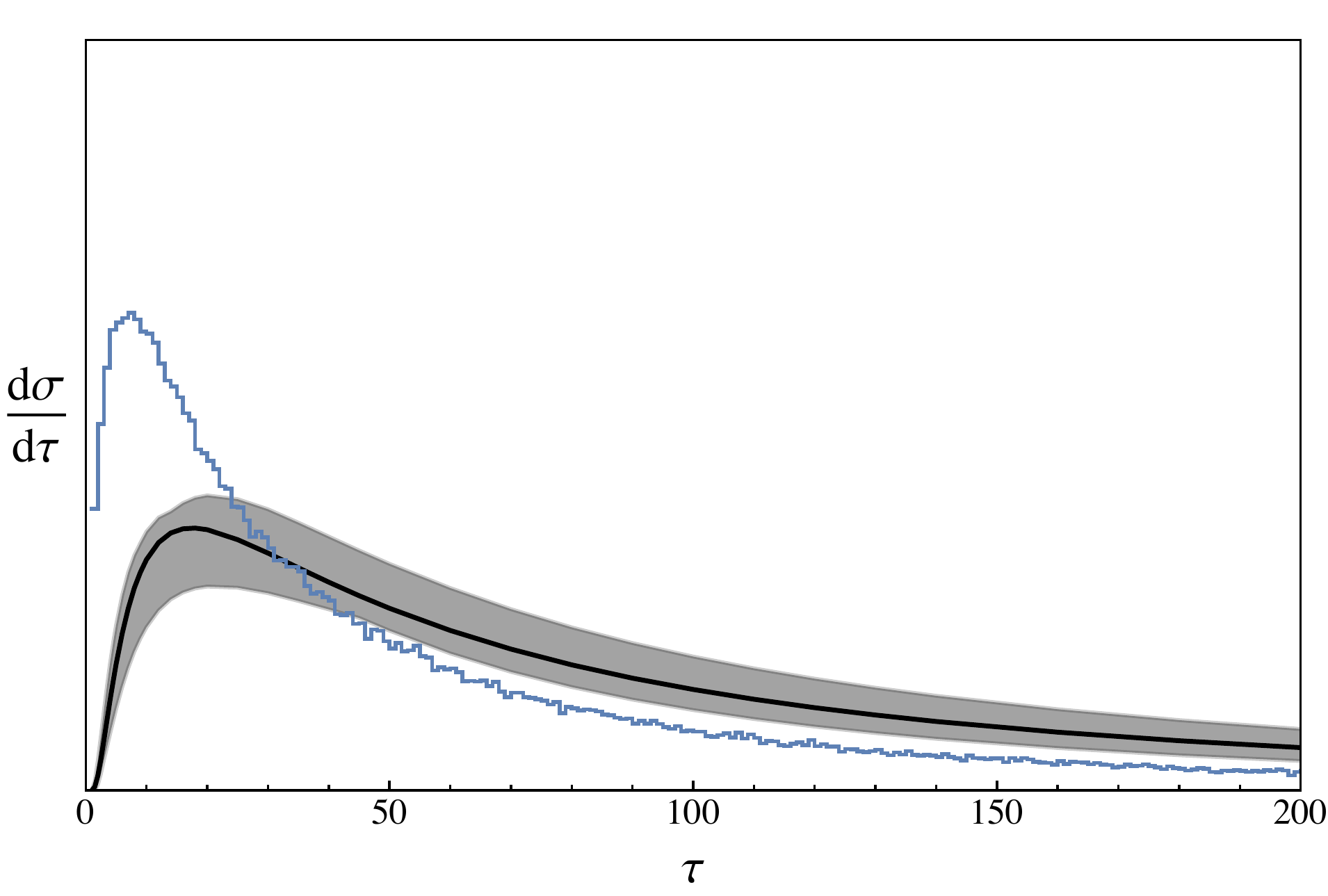} \\
      \tikz{\node [rotate=90] (text) {$u\bar{u} \to gg$};} &
      \includegraphics[width=0.2\textwidth]{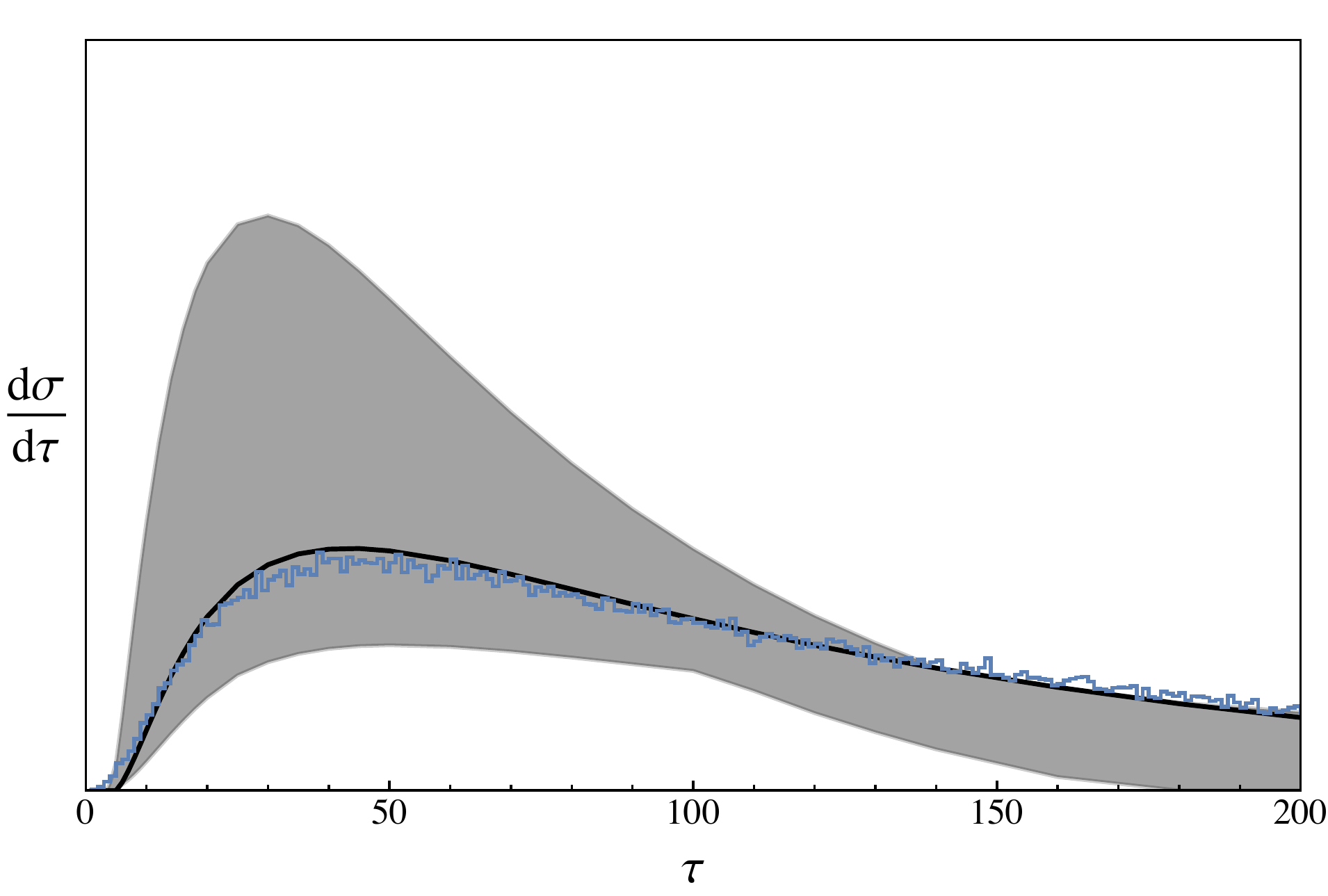} &
      \includegraphics[width=0.2\textwidth]{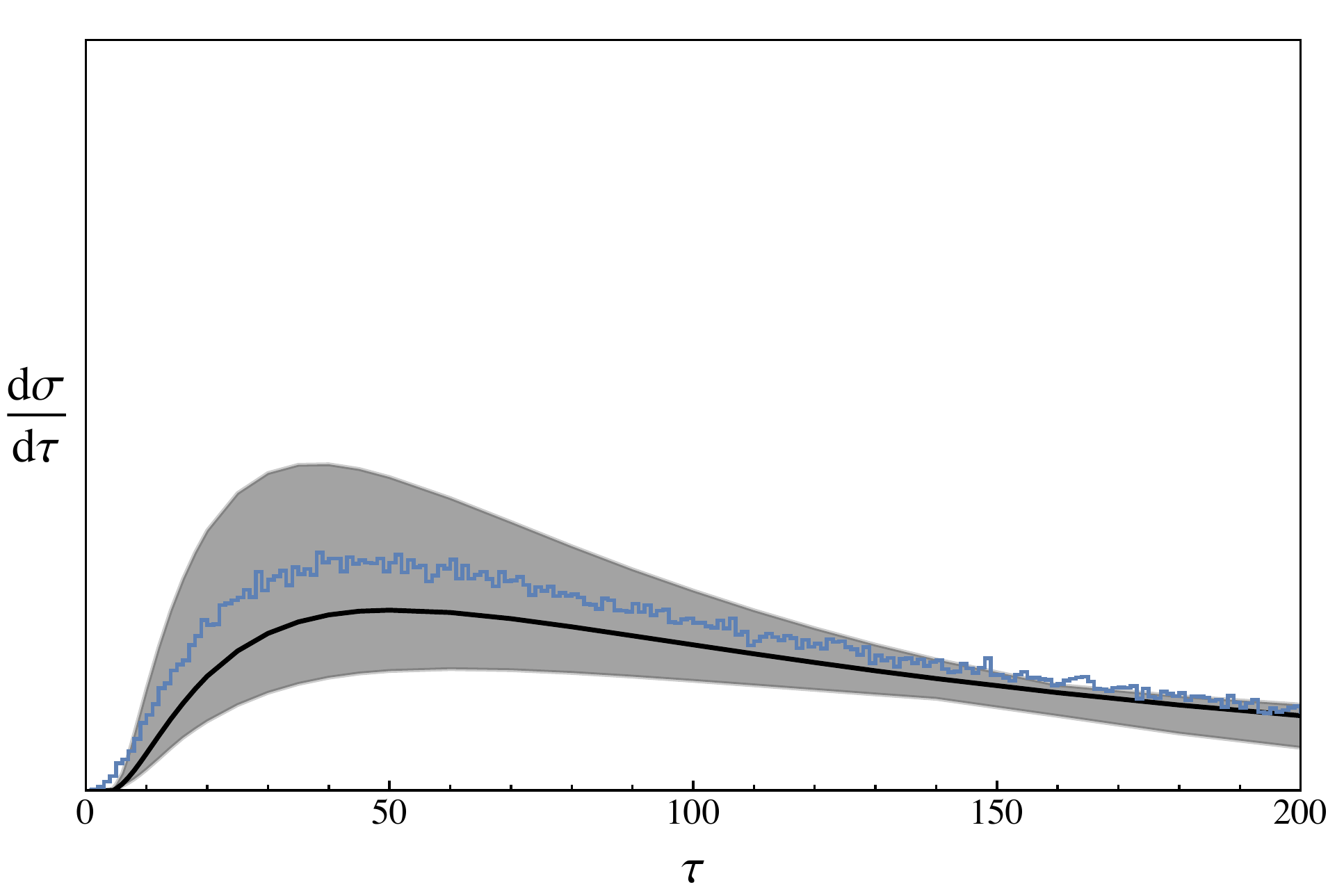} &
      \includegraphics[width=0.2\textwidth]{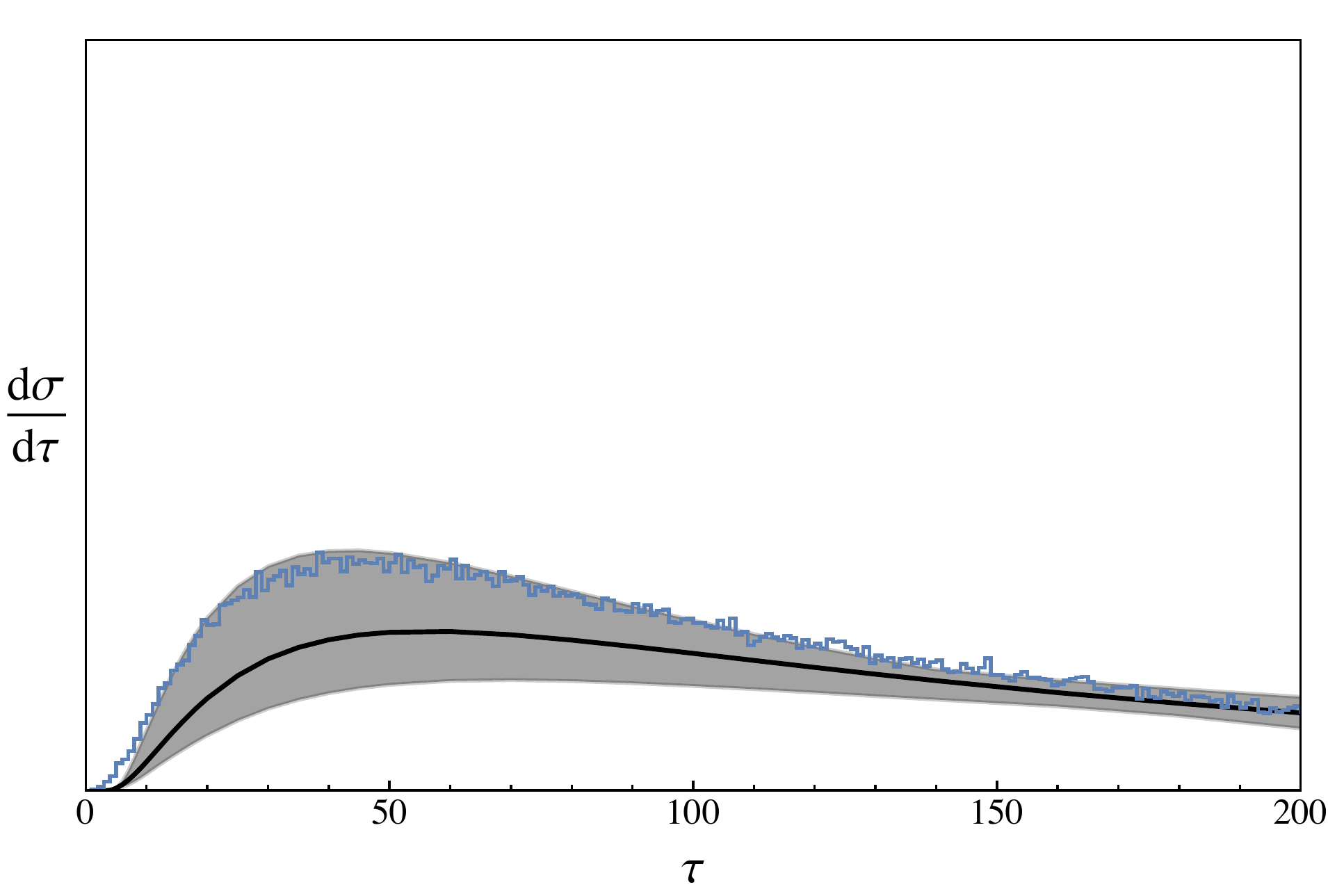} &
      \includegraphics[width=0.2\textwidth]{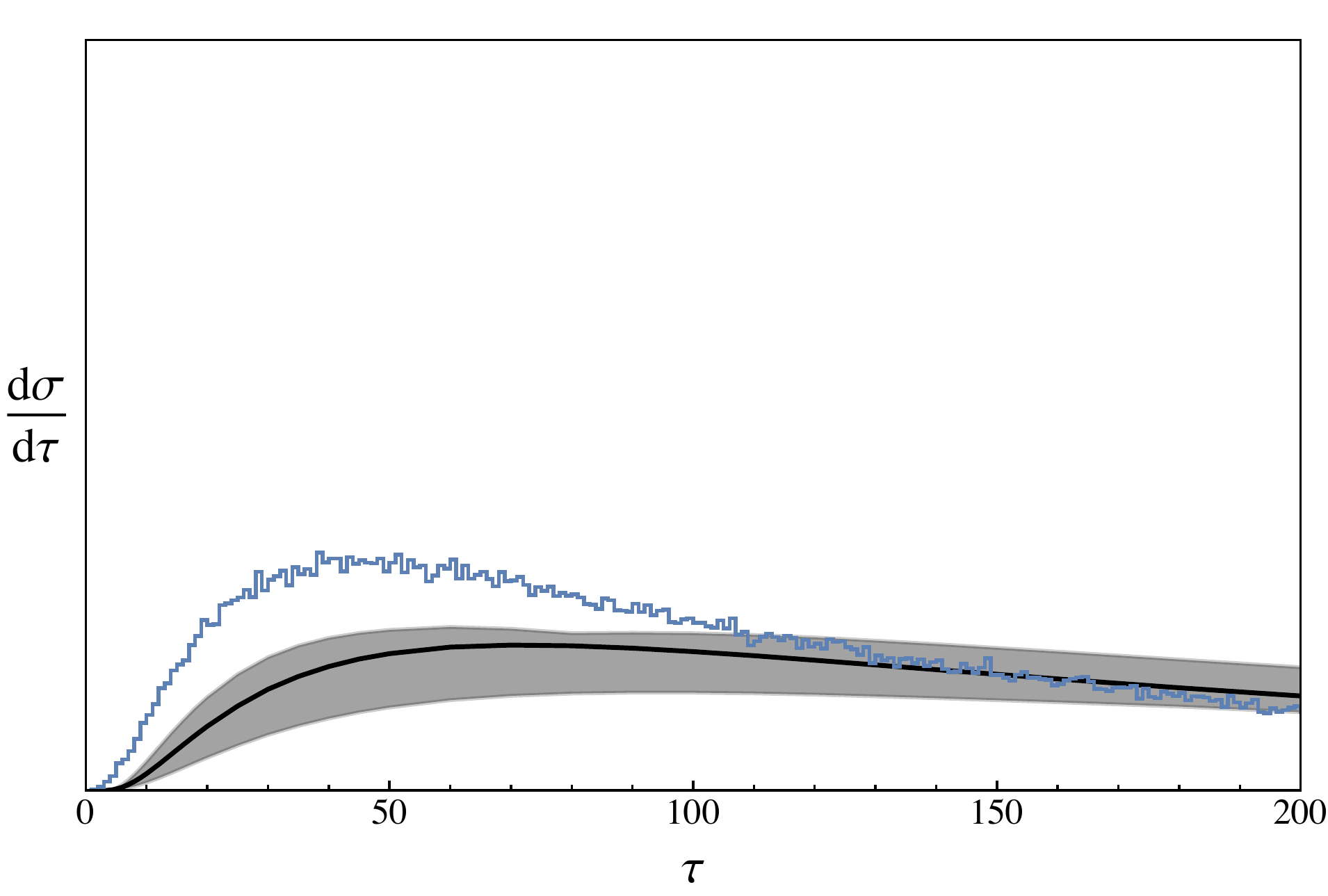} \\
      \tikz{\node [rotate=90] (text) {$ug \to ug$};} &
      \includegraphics[width=0.2\textwidth]{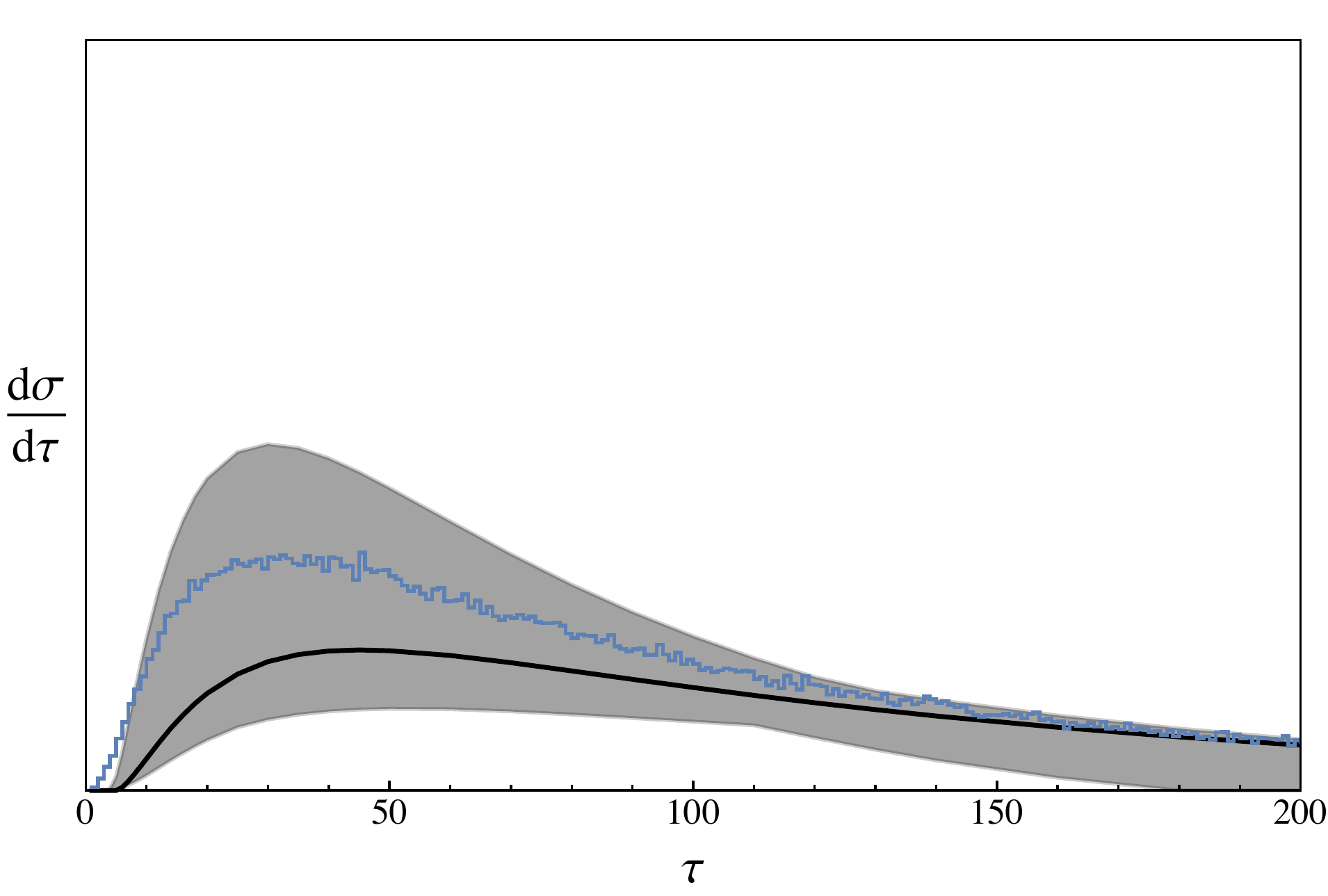} &
      \includegraphics[width=0.2\textwidth]{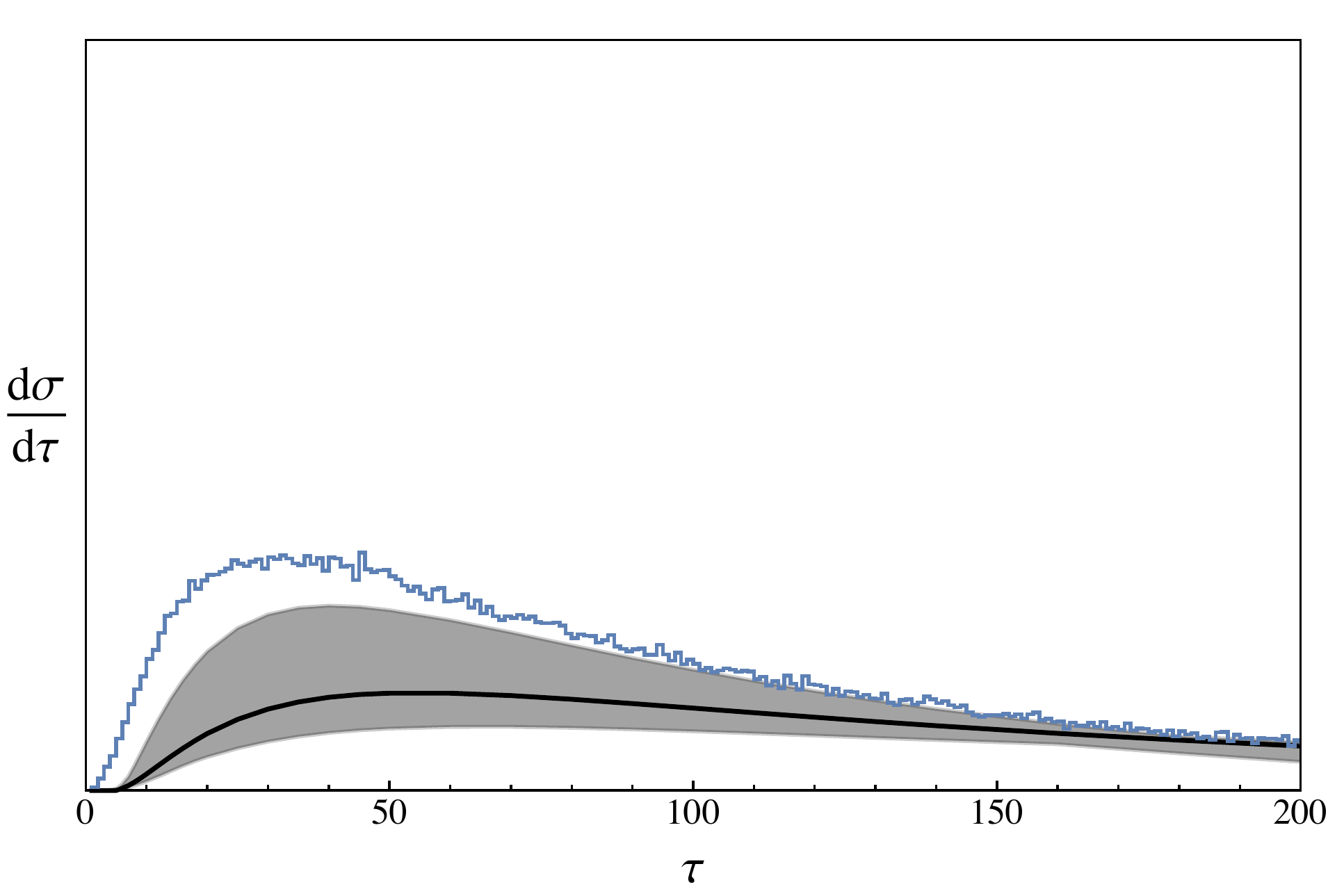} &
      \includegraphics[width=0.2\textwidth]{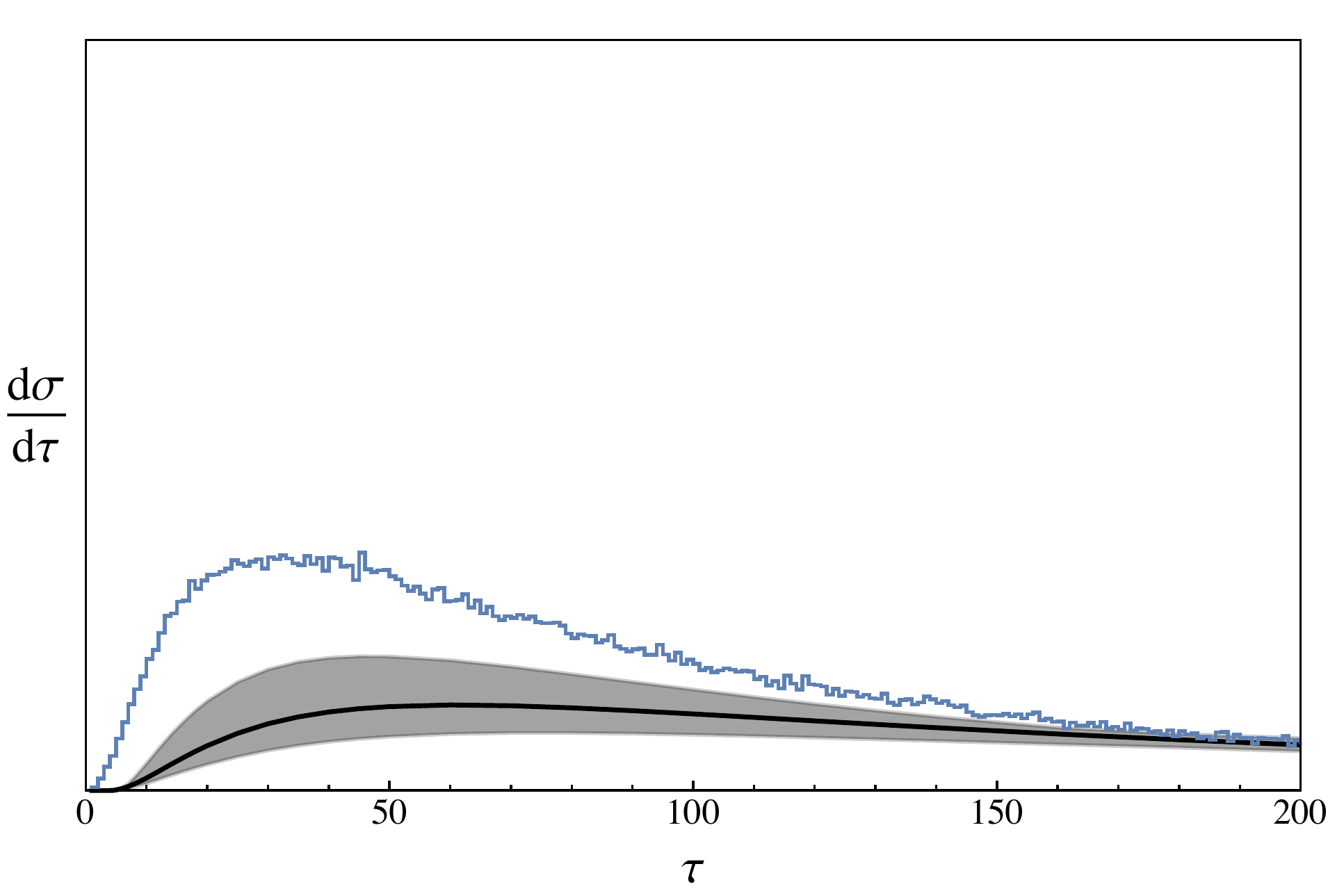} &
      \includegraphics[width=0.2\textwidth]{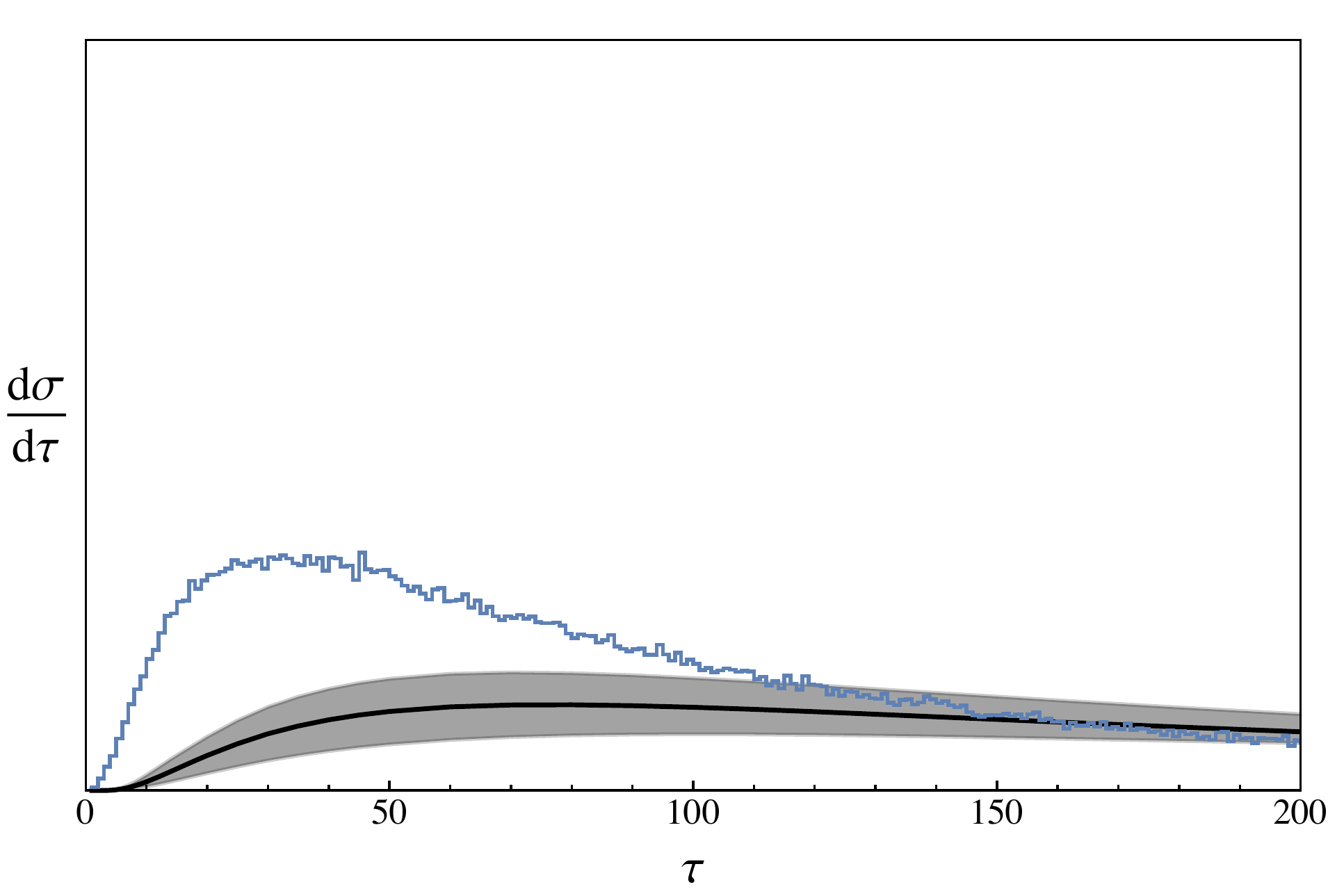} \\
      \tikz{\node [rotate=90] (text) {$gg \to u\bar{u}$};} &
      \includegraphics[width=0.2\textwidth]{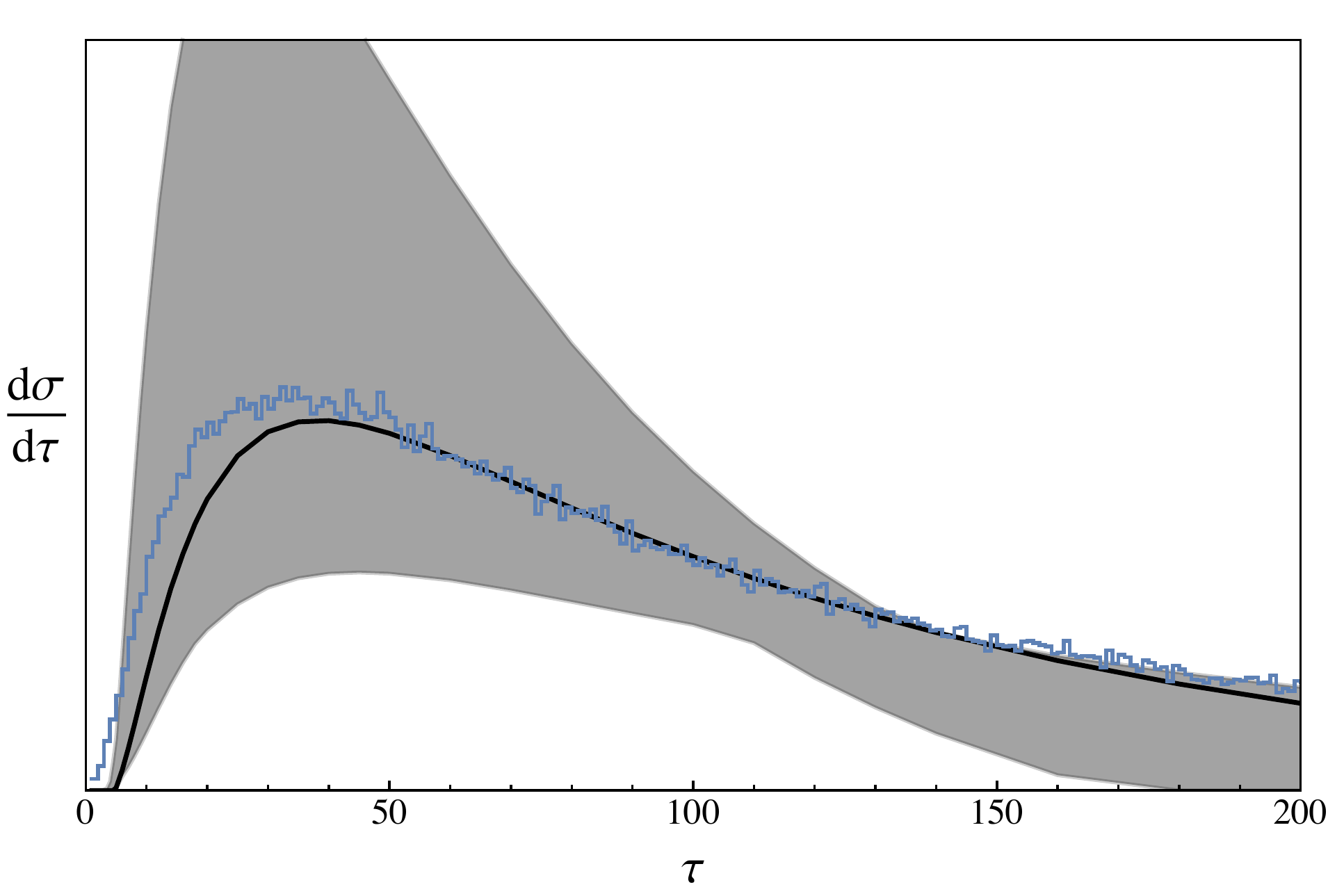} &
      \includegraphics[width=0.2\textwidth]{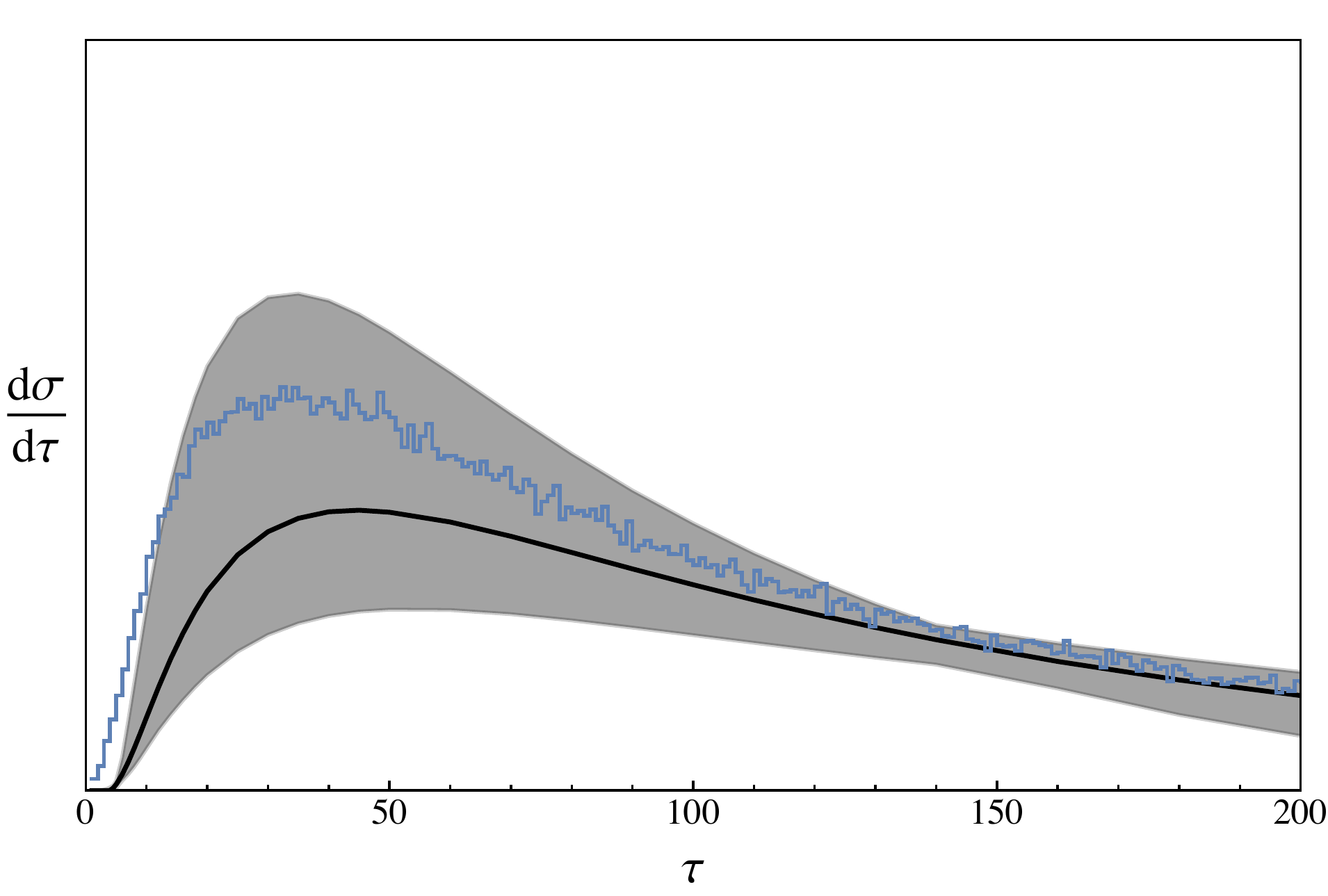} &
      \includegraphics[width=0.2\textwidth]{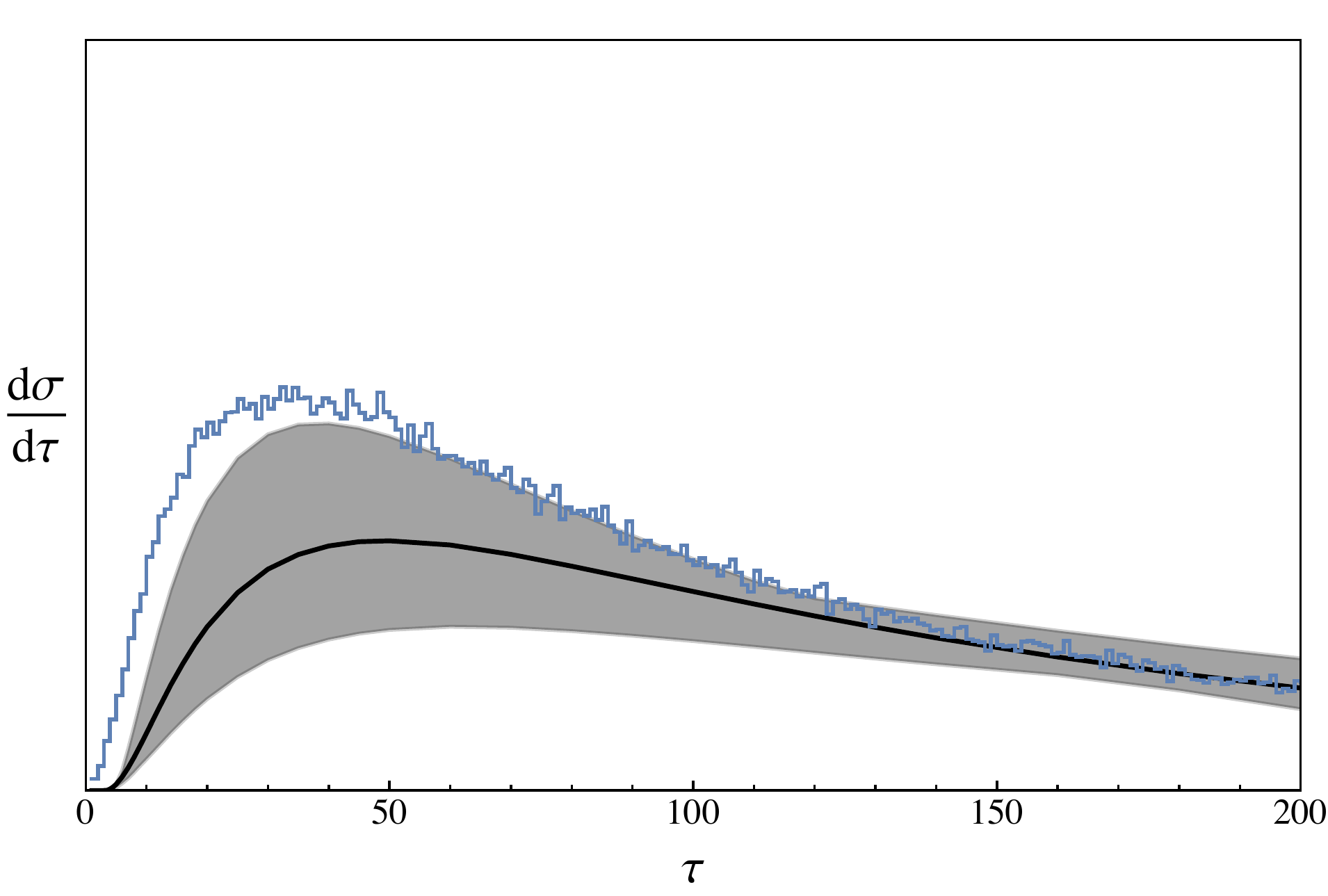} &
      \includegraphics[width=0.2\textwidth]{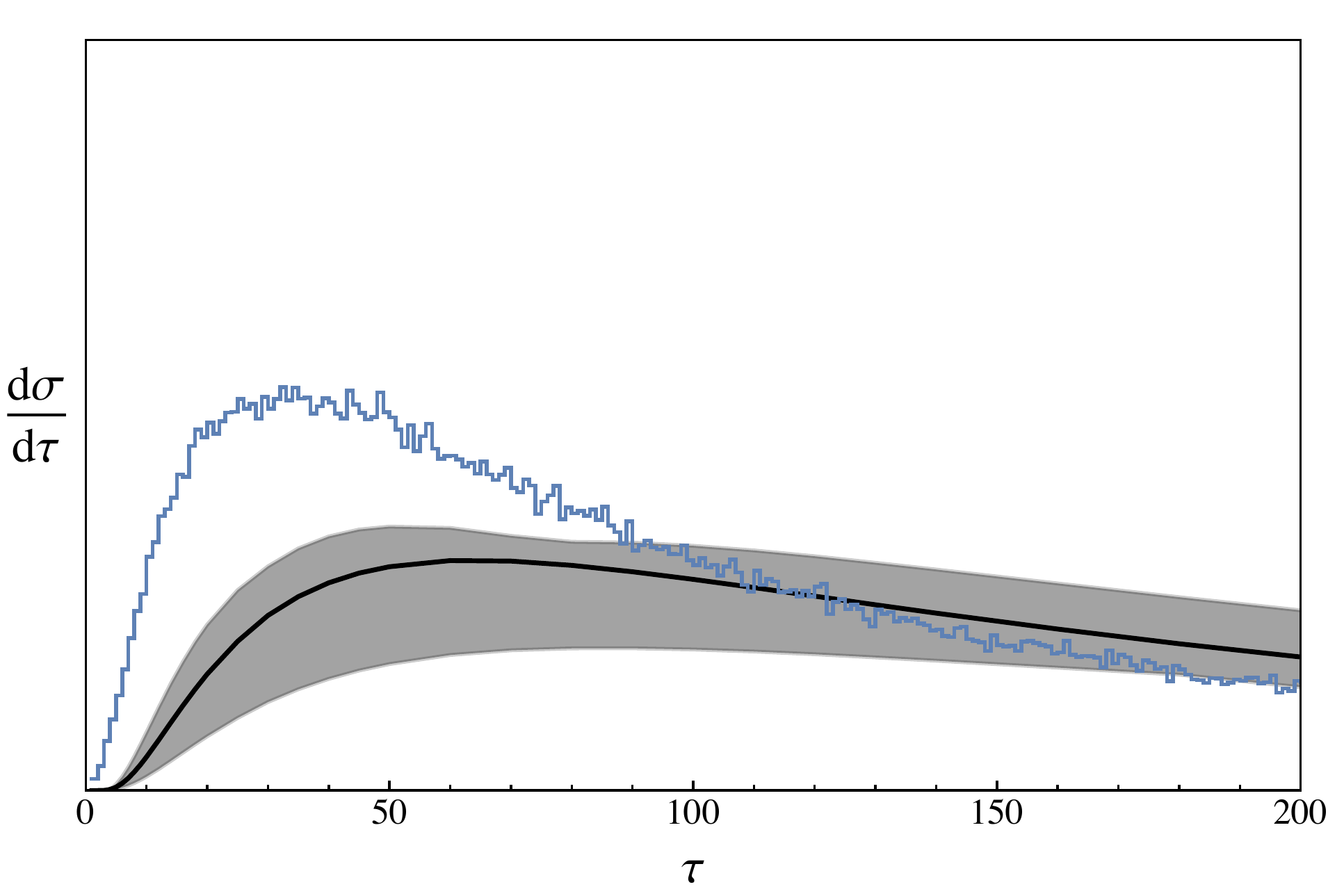} \\
      {} & \multicolumn{4}{c}{\includegraphics[width=0.66\textwidth]{Plots/Legend}}
    \end{tabular}
  \end{center}
  \caption{$\cT_2$ distribution in $pp$ collisions at the single symmetric phase space point in \Eq{symm} for different channels. Top row is $uu \to uu$, second row is $u\bar{u} \to gg$, third row is $ug \to ug$, and bottom row is $gg\to u\bar{u}$.}
  \label{fig:ppSquare}
\end{figure}

Finally, we show the complete result, integrated over phase space, summed over channels in \Fig{eejjjj}.

\subsection{\texorpdfstring{$pp \to \text{2 jets} + X$}{2 jets + X at hadron colliders}}
  \label{sec:2jetti}
Now we change processes, form $e^+e^-\to 4j$ to $pp \to 2j + \gamma$. This example will demonstrate the flexibility of our calculations, since most of the work from the previous process can be reused here, since the two processes $e^+e^-\to 4j$ and $pp\to 2j + \gamma$ have the same partonic channels. It will also demonstrate the changes needed to apply the calculations in a $pp$ collider with PDFs.

A powerful feature of the \MadGraph/SCET  framework is that it neatly separates the observable-dependent pieces from the partonic-channel dependent pieces and the phase-space dependent pieces. From the viewpoint of the jet and soft functions, the only difference between a $e^+e^-\to 4j$ event and a $pp\to 2j + \gamma$ event is a few factors of $i\pi$ from the crossing (see e.g. \cite{Kelley:2010fn}, Eq. (38)), and that the latter should use beam functions for the two incoming partons instead of jet functions. 

For $pp \to 2j + \gamma$, we will again look at $n$-jettiness type observables. The factorization theorem for $2$-jettiness is formally differential in the two jet masses $\cT^1$ and $\cT^2$ and the beam jettinesses $\cT^a$ and $\cT^b$. $2$-jettiness is the sum of these
\begin{equation}
\cT_2 = \cT^a + \cT^b + \cT^1 + \cT^2
\label{2jettinessDef}
\end{equation}
Thus, 2-jettiness in $pp$ collisions is a crossing of 4-jettiness in $e^+e^-$.

\Fig{ppSquare} shows the $\cT_2$ distribution starting with the symmetric partonic configuration in \Eq{symm}. We show in the different rows different partonic channels ($u\bar{u} \to u \bar{u}$, $u\bar{u} \to gg$, $ug \to ug$ and $gg \to uu$). $gg\to gg$ can be studied similarly, but we do not consider it in this paper (there is no 4 gluon channel in $e^+e^- \to 4j$ or in $pp \to 2j + X$ with $X = W^\pm, Z,\gamma$). We can see that for $\cT_2$, the agreement with \Pythia is not as good as for $\cT_4$ in $e^+e^-$ events.  This is to be expected, because of the way initial state radiation is treated in the two approaches. \Pythia combines initial and final state parton showers in an intricate way that avoids double counting; SCET handles initial-state radiation with beam functions and final state radiation with jet functions. Since these methods are different, we expect worse agreement than when only final-state radiation is present.

Now that the overall agreement is less perfect, it becomes clear that the scale choice of \Pythia is not the same as the natural scale choice of SCET. Take for example the bottom row ($gg\to u\bar u$). Here we see that there is a SCET scale choice, around 400 GeV, that matches \Pythia (as expected, since the tunes in \Pythia get it to mostly NLL accuracy). On the other hand, the scale choice that minimizes the SCET error bands is around 800-1600 GeV. Each of these criteria can be achieved separately, but there is no reason that the \Pythia scale choices should be the same as the SCET natural scale choices. This story is the same in each of the four rows. For $uu\to uu$ and $ug\to ug$, the scale choice that matches \Pythia are 200 GeV and 300 GeV respectively; although they are lower than 400 GeV and thus not shown, they match as well as the others.

\begin{figure}[t]
  \begin{center}
    \begin{tabular}{cccc}
      {} & $\mu_H = 400\GeV$ & $\mu_H = 800\GeV$ & $\mu_H = 1600\GeV$ \\
      \begin{tikzpicture}[thick,scale=1] %, every node/.style={scale=1}]
        \useasboundingbox (1,-1) rectangle (0,1);
        \tikzset{>=latex};
        \draw[->, line width=1,opacity=1] (1,0) -- (0.05,0) ;
        \draw[->, line width=1,opacity=1,rotate=30] (0.05,0) -- node[above]{}  (1,0) ;
        \draw[->, line width=1,opacity=1]  (-1,0) -- (-0.05,0)  ;
        \draw[->, line width=1,opacity=1,rotate=30] (-0.05,0) --  node[below]{} (-1,0) ;
        \draw[-,line width=1,red]   (-0.8,0.2) to[out=-10,in=-140] (0.6, 0.6);
        \draw[-,line width=1,blue]   (-0.6,-0.6) to[out=40,in=170] (0.8, -0.2);
      \end{tikzpicture} &
      \includegraphics[width=0.25\textwidth]{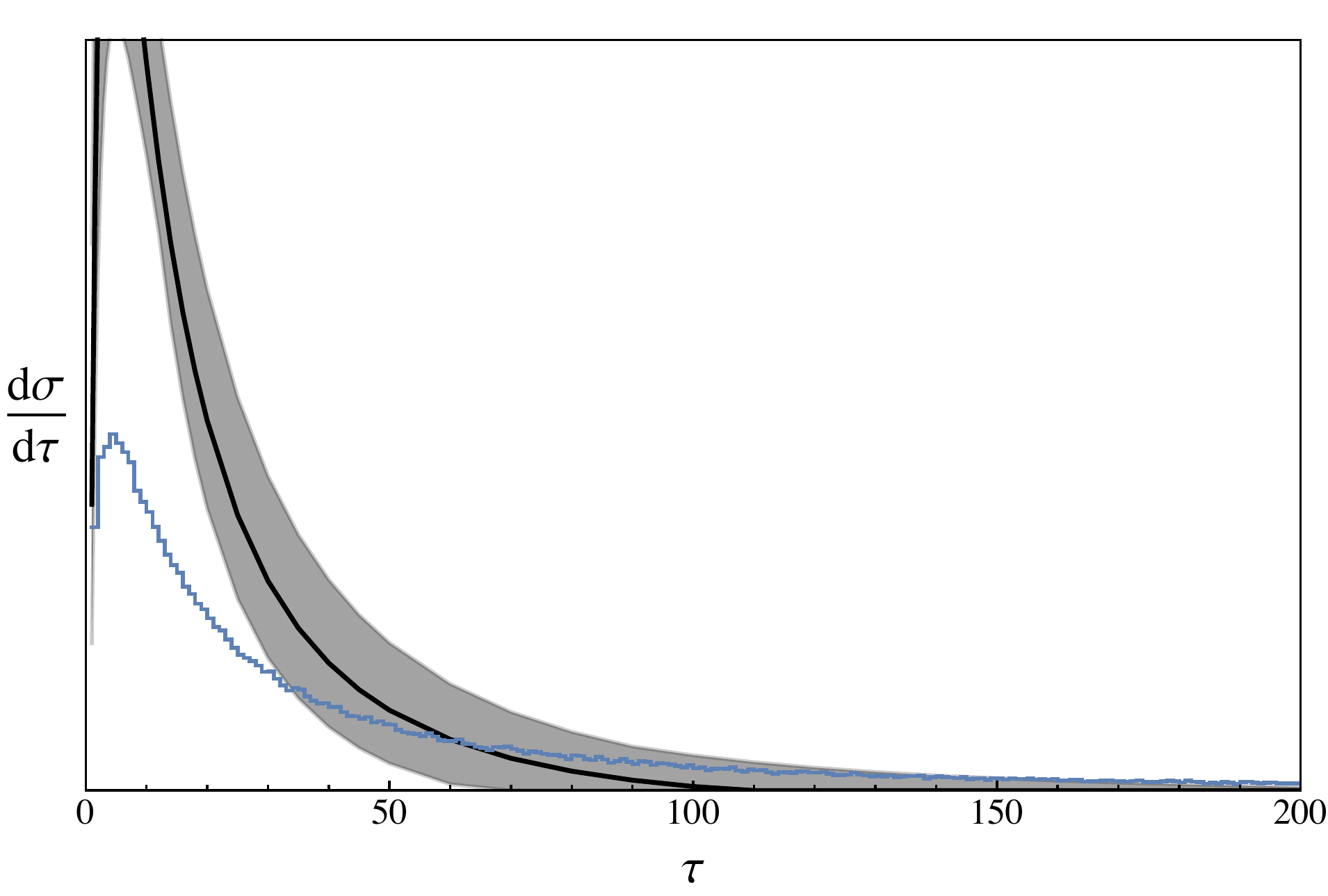} &
      \includegraphics[width=0.25\textwidth]{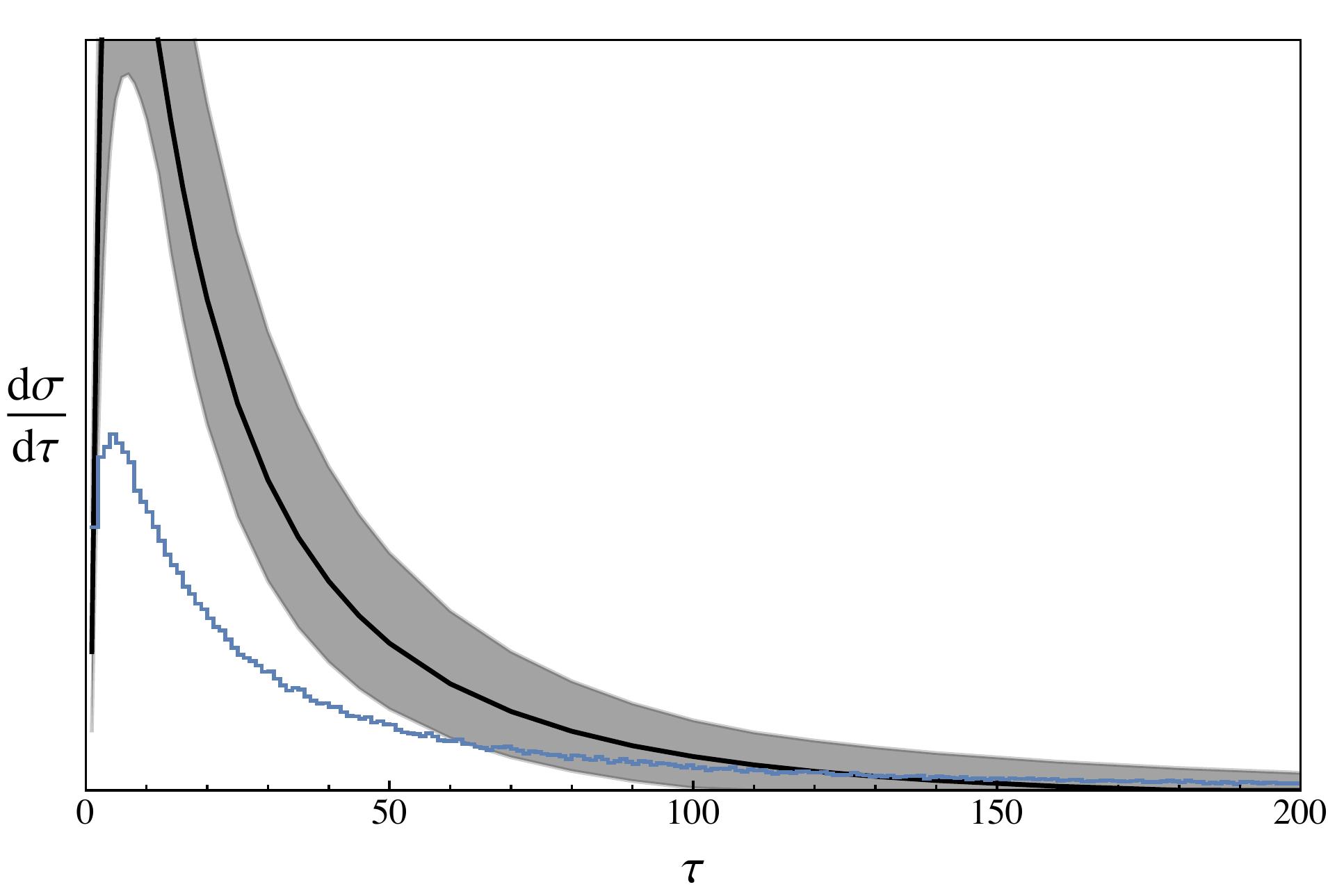} &
      \includegraphics[width=0.25\textwidth]{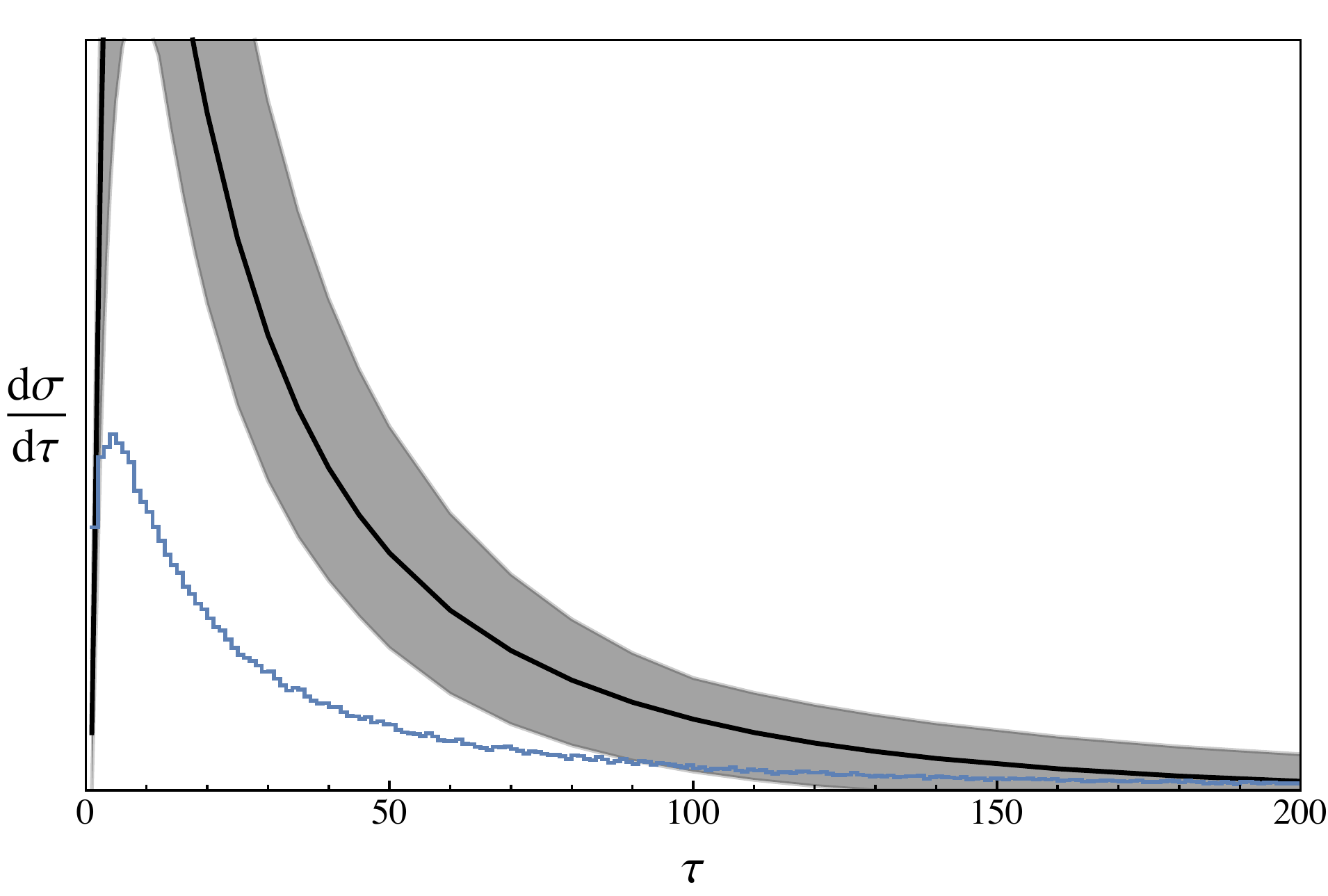} \\

      \begin{tikzpicture}[thick,scale=1] %, every node/.style={scale=1}]
        \tikzset{>=latex};
        \useasboundingbox (1,-1) rectangle (0,1);
        \draw[->, line width=1,opacity=1] (1,0) -- (0.05,0) ;
        \draw[->, line width=1,opacity=1,rotate=30] (0.05,0) -- node[above]{}  (1,0) ;
        \draw[->, line width=1,opacity=1]  (-1,0) -- (-0.05,0)  ;
        \draw[->, line width=1,opacity=1,rotate=30] (-0.05,0) --  node[below]{} (-1,0) ;
        \draw[-,line width=1,red]   (-0.8,-0.1) to[out=5,in=25,distance = 10] (-0.7, -0.3);
        \draw[-,line width=1,blue]   (0.8,0.1) to[out=185,in=205,distance = 10] (0.7, 0.3);
      \end{tikzpicture} &
      \includegraphics[width=0.25\textwidth]{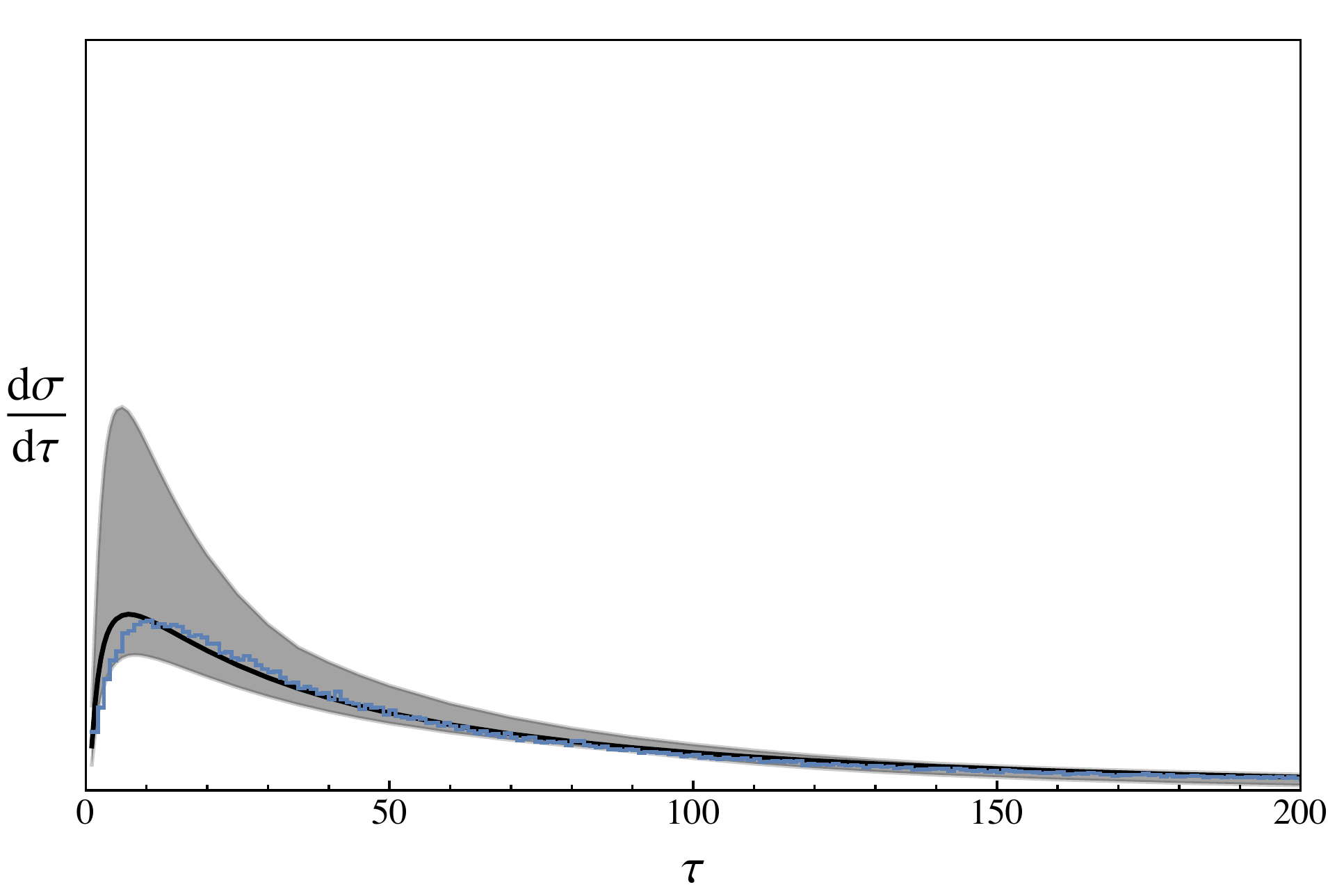} &
      \includegraphics[width=0.25\textwidth]{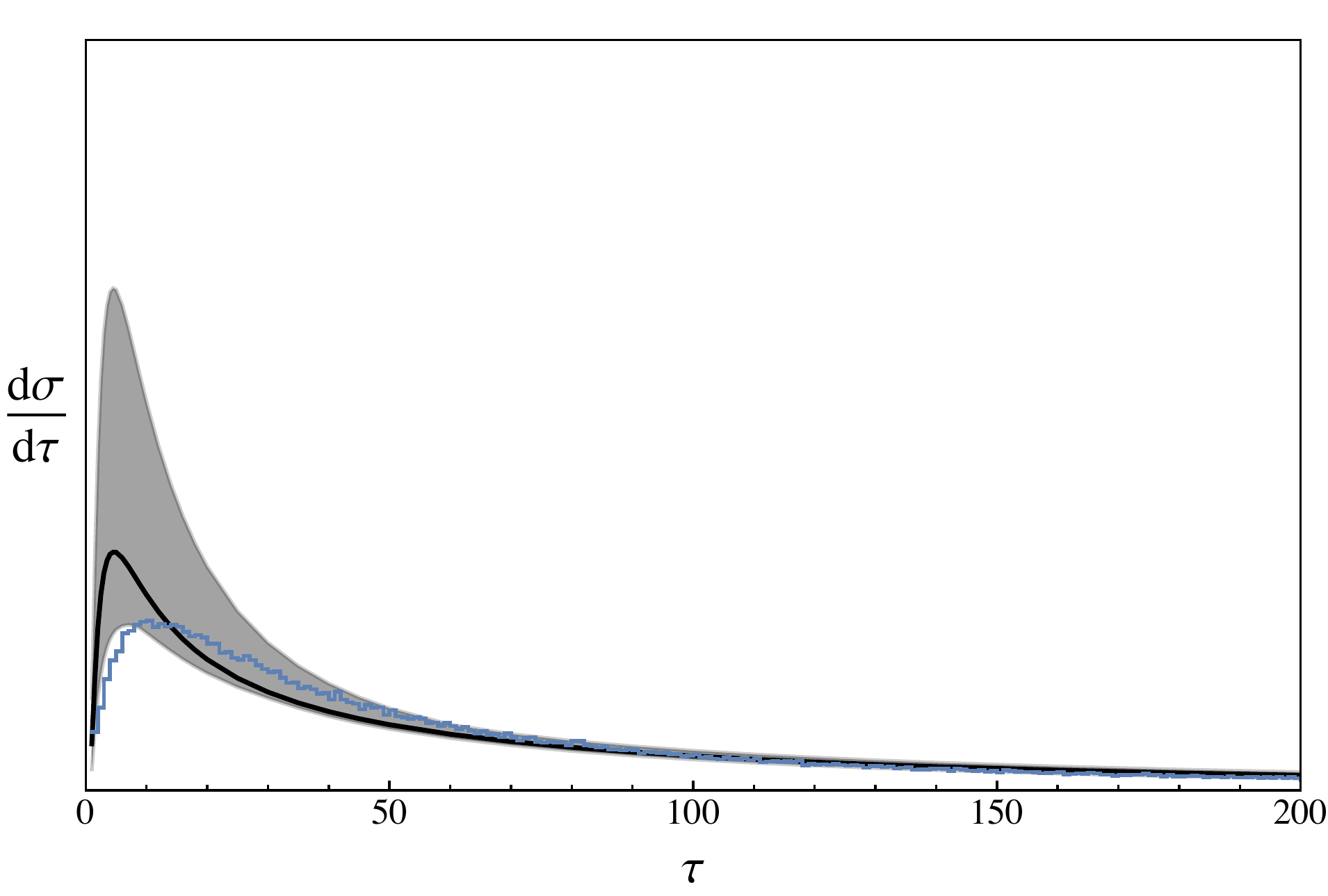} &
      \includegraphics[width=0.25\textwidth]{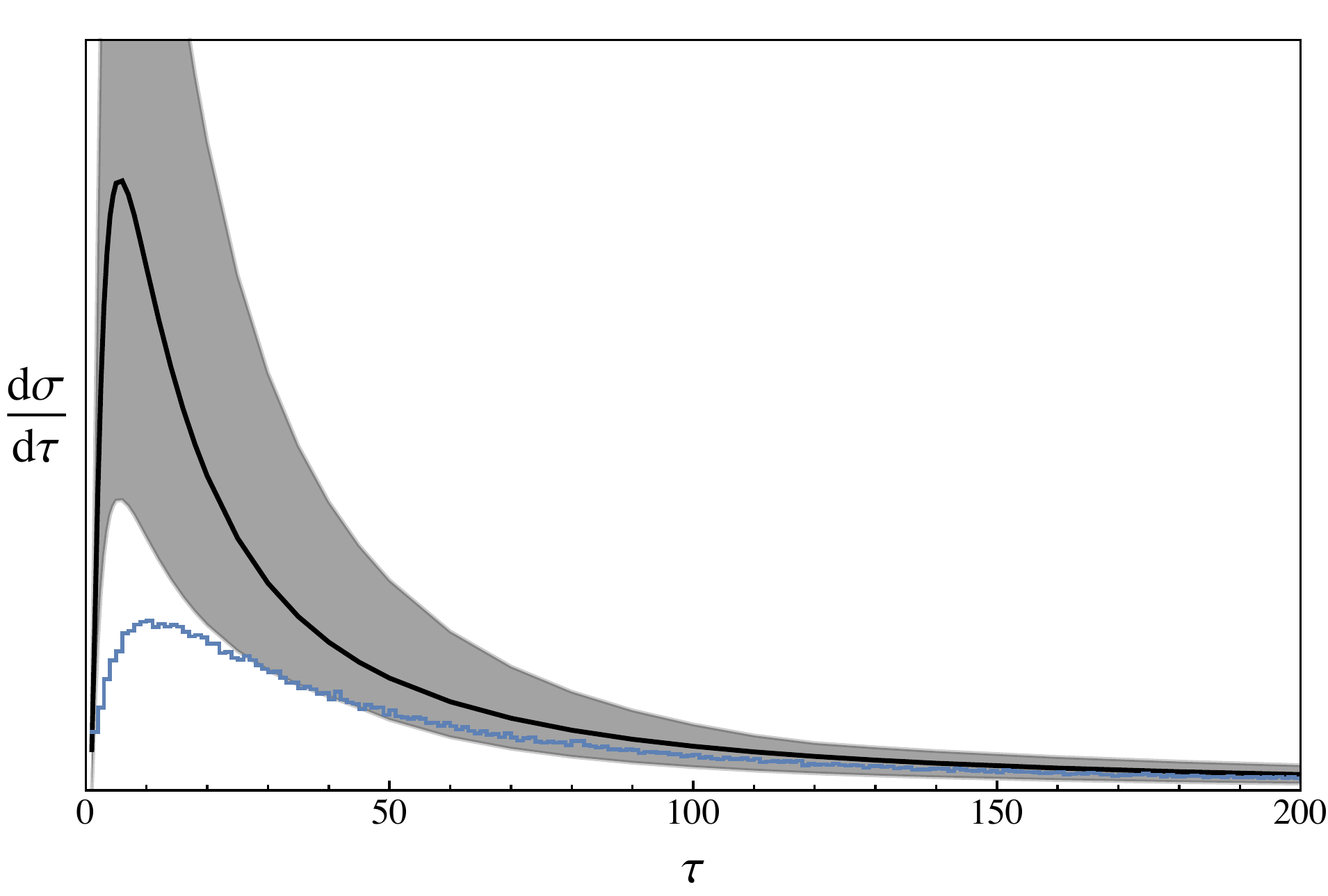} \\
      {} & \multicolumn{3}{c}{\includegraphics[width=0.66\textwidth]{Plots/Legend}}
    \end{tabular}
  \end{center}
  \caption{$\cT_2$ distributions in events with the angled phase space point in \Eq{scissor} for $uu \to uu$. Top row has the nearby momenta pairs color-connected (forward scattering) and bottom row has the larger-angled color connection.}
  \label{fig:ppjjTiltEventLarge}
\end{figure}

The analogous plots for the angled event with momenta in \Eq{scissor} are shown for the two $u\bar u\to u \bar u$  color connections in \Fig{ppjjTiltEventLarge}. We see that these angled events, in particular the top row where the color connection is associated with the forward scattering, produce uniformly worse agreement with \Pythia than the symmetric events.

\begin{figure}[h]
\begin{center}
\includegraphics[width=0.6\textwidth]{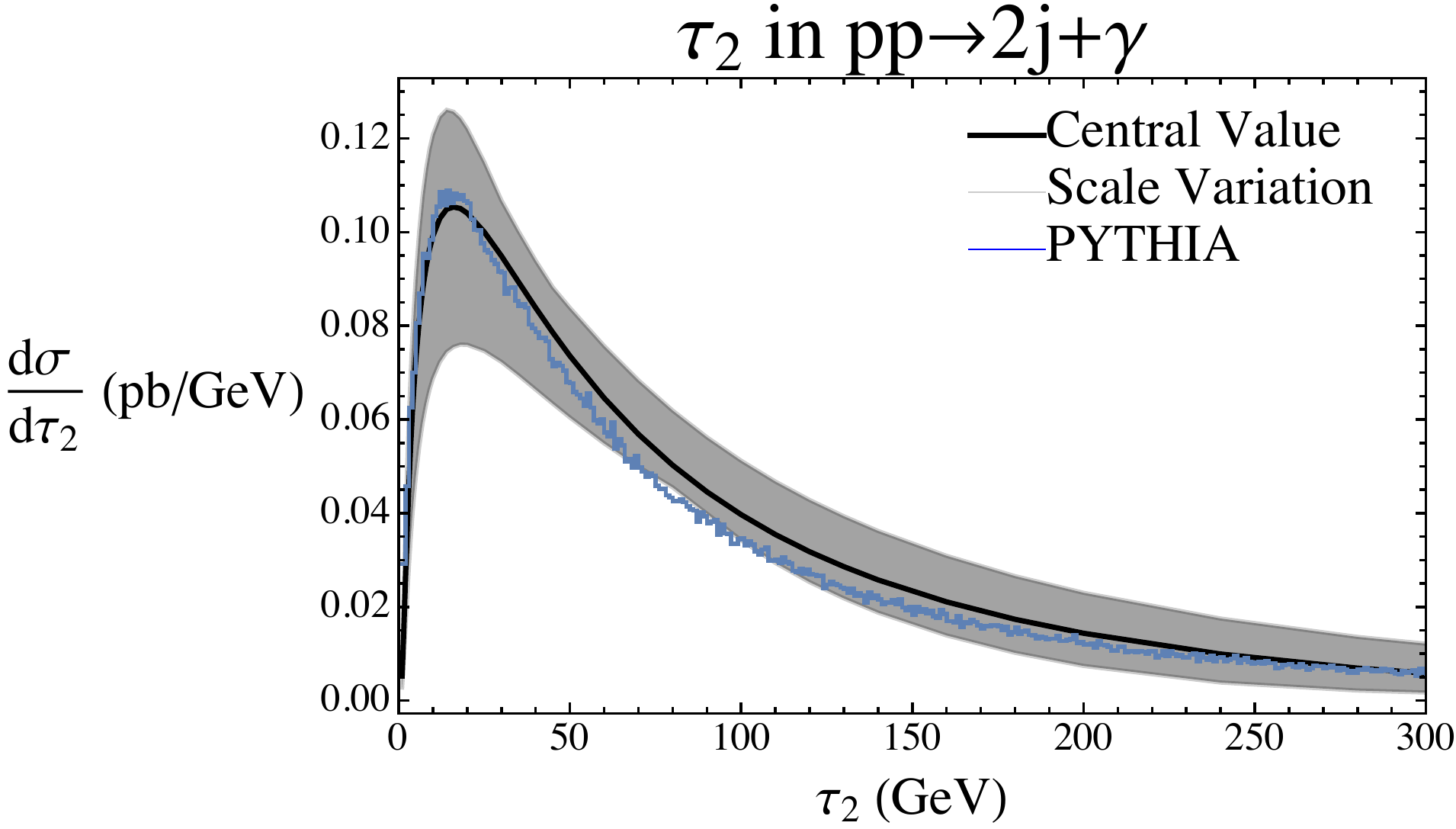}
\caption{Distribution of $\cT_2$ in $pp\to 2j + \gamma$ events using \MadGraph/SCET compared to \MadGraph/\Pythia.}
\label{fig:ppt2}
\end{center}
\end{figure}

After computing the cross section differential in the phase space, we can average the distributions over phase-space points to obtain the full cross section. In \Fig{ppt2}, we show the prediction for $\cT_2$ in $pp \to 2j + \gamma$ events using 50,000 phase-space points generated at $\sqrt{s} = 8\TeV$. The jets are constrained to have $p_T^j > 300\GeV$ and $|\eta_j| < 5$, with the equivalent cuts for photons at $p_T^\gamma > 500\GeV$ and $|\eta_\gamma| < 2.5$. To avoid collinear singularities and additional large logarithms, relatively stringent isolations cuts of $\Delta R_{jj} = \Delta R_{\gamma j} > 0.8$ are imposed. The grey band represents the \MadGraph/SCET NLL calculation and the blue histogram the \MadGraph/Pythia result. Note that agreement in the full cross section is actually better than for the individual phase-space points, as expected for an inclusive observable.

Rather than $\cT_2$, we can also look at masses of the two outgoing jets placing a cut on the mass of the incoming jets. A similar version of 1-jettiness was first studied in Ref.~\cite{Jouttenus:2013hs}. To be precise, recall that $\cT^a$ and $\cT^b$ are the jettiness of the two incoming jets, and $\cT^1$ and $\cT^2$ are those for the outgoing jets. Then, we consider
\begin{align}
  \tc = (\cT^1+\cT^2)\theta(\Tcut-\cT^a-\cT^b)
\end{align} 
$\tc$ is approximately the sum of the masses-squared of the jets divided by their energies: $\tc \sim \frac{m_1^2}{E_1} + \frac{m_2^2}{E_2}$, thus it is closely related to jet mass in dijet events. To make this change, we compute
\begin{align}
  F_{\tc}(\tc, \Tcut) = \int d^4\cT \frac{dF}{d^4\cT} \delta(\tc-\cT^1-\cT^2)\theta(\Tcut-\cT^a-\cT^b)
\end{align} 
This integral is easily done analytically. Thus using \MadGraph/SCET we compute the cross section for $\tc$ with minimal changes from $\cT_2$.

\begin{figure}[h]
  \begin{center}
   \subfigure[][]{
\includegraphics[width=0.4\textwidth]{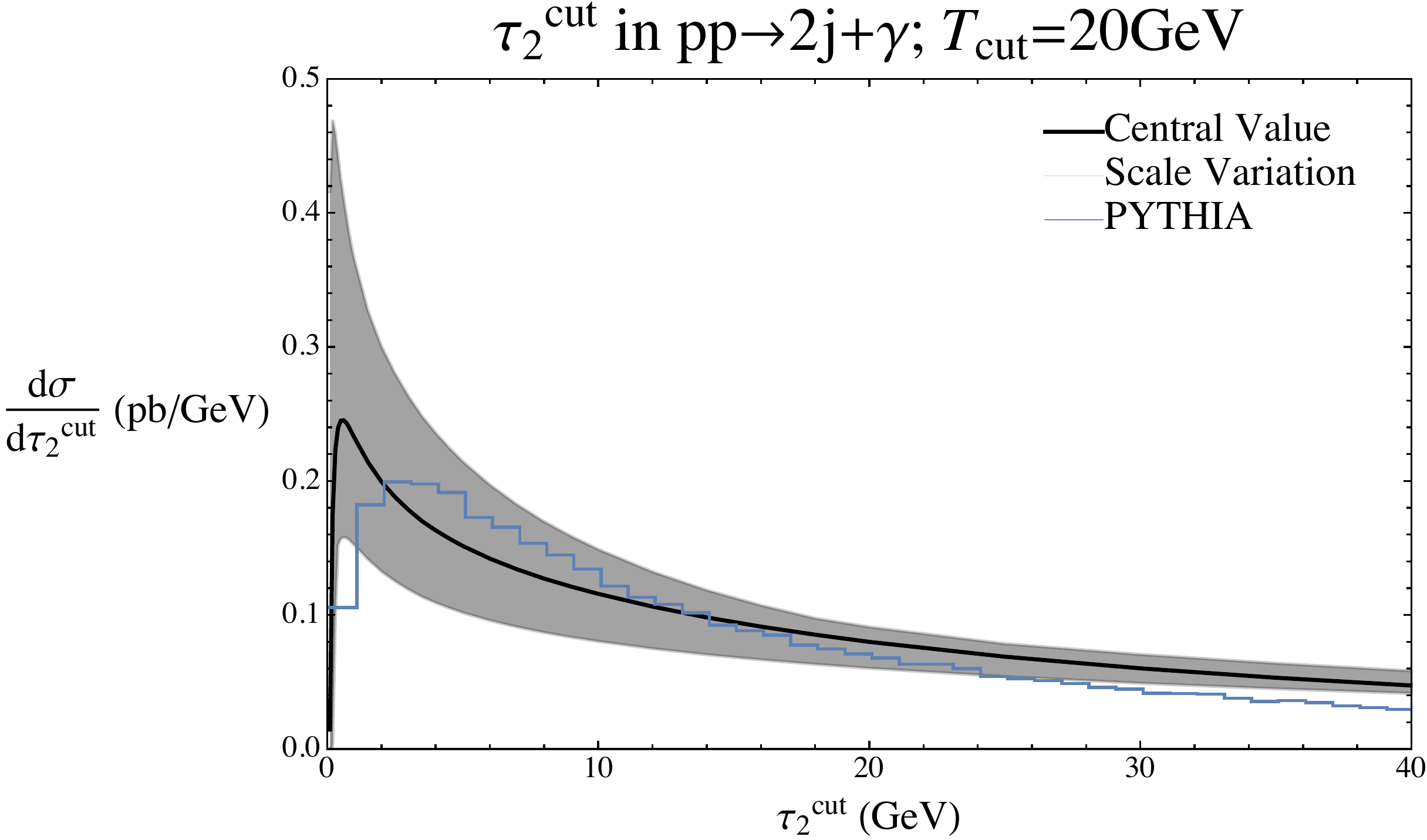}
}
  \subfigure[][]{
\includegraphics[width=0.4\textwidth]{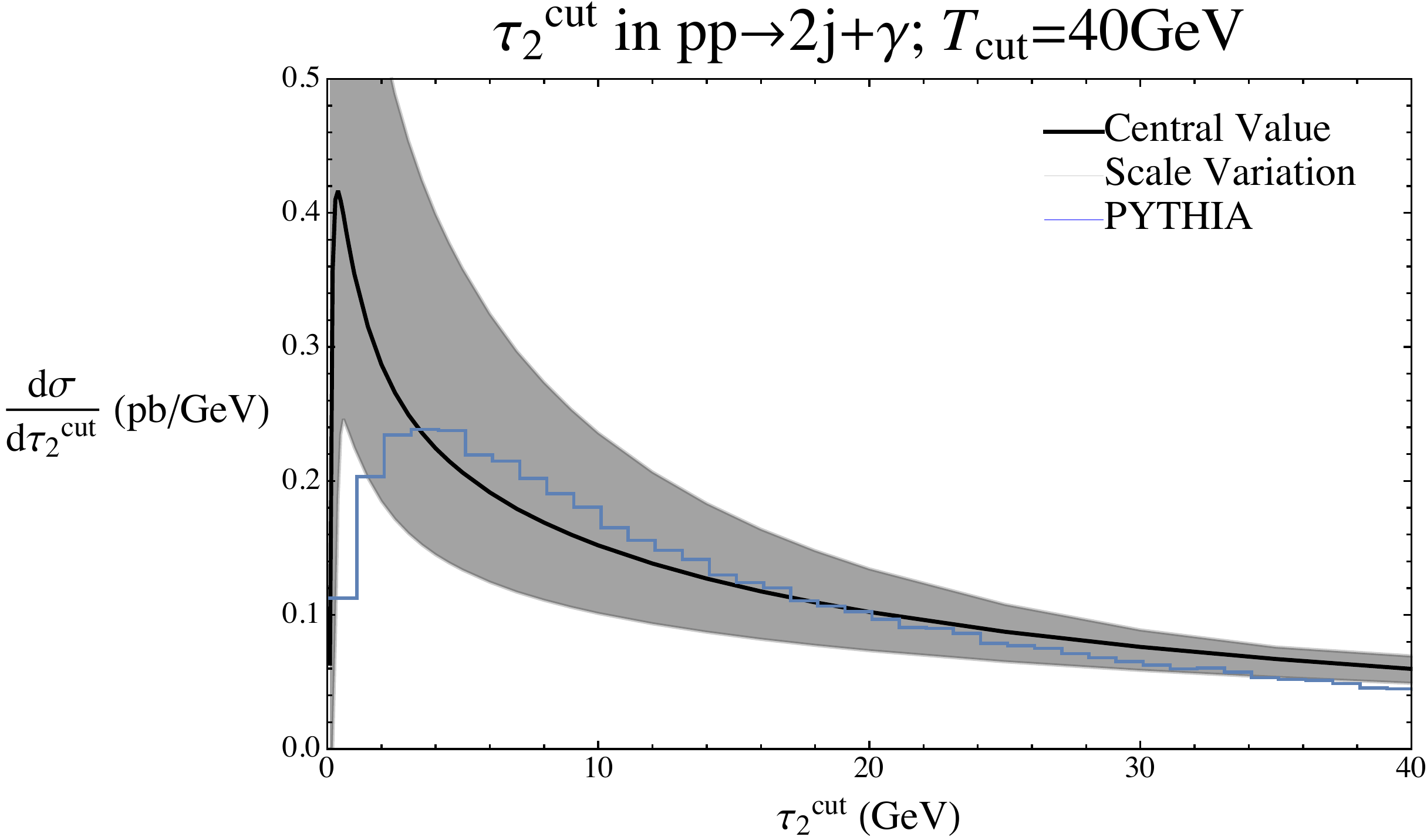}
}
\end{center}
\caption{Cumulative $\tc$ distributions, with comparisons to \Pythia. These are integrated over phase space, with (a)  $\Tcut=20$ GeV and (b) $\Tcut=40$ GeV.}
\label{fig:ppt2cutCumulative}
\end{figure}

\begin{figure}[h]
  \begin{center}
\includegraphics[width=0.65\textwidth]{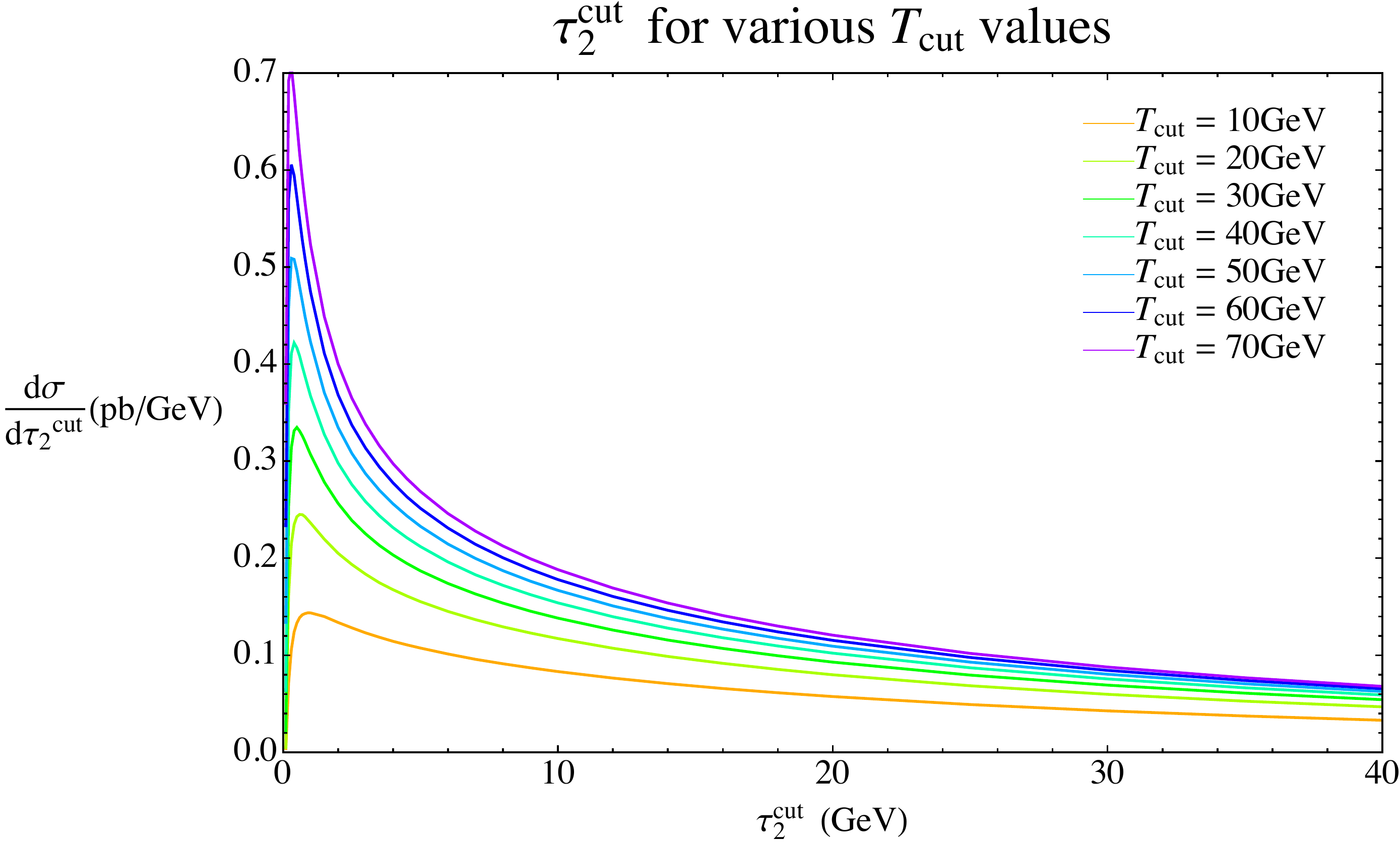}
\end{center}
\caption{The cross section for $\tc$ for various values of $\Tcut$.}
\label{fig:ppt2cutRescaling}
\end{figure}

Results for $\tc$ are shown in \Fig{ppt2cutCumulative}. As observed in Ref.~\cite{Jouttenus:2013hs} (see Fig.~15), \Pythia does not match the SCET distribution for $\tc$ nearly as well as it did for $\cT_2$. This is to be expected: while the $\cT_2$ distribution only depends on $\cT_2$, the $\tc$ distribution depends on both $\Tcut$ and $\tc$, so there can be large logs of their ratio. With $\tc$, either $\Tcut$ or $\tc$ can be used as the soft scale and choose $\mu_s=\sqrt{\Tcut \tc}$. This choice resums some large logarithms, but not others. Consistent with this explanation, we note that in the region where $\tc \approx \Tcut$, \MadGraph/SCET and \MadGraph/\Pythia match. We also show, in \Fig{ppt2cutRescaling}, the effect on the $\tc$ distribution of changing $\Tcut$. 

\begin{figure}[h]
  \begin{center}
   \subfigure[][ 1 event]{
\includegraphics[width=0.4\textwidth]{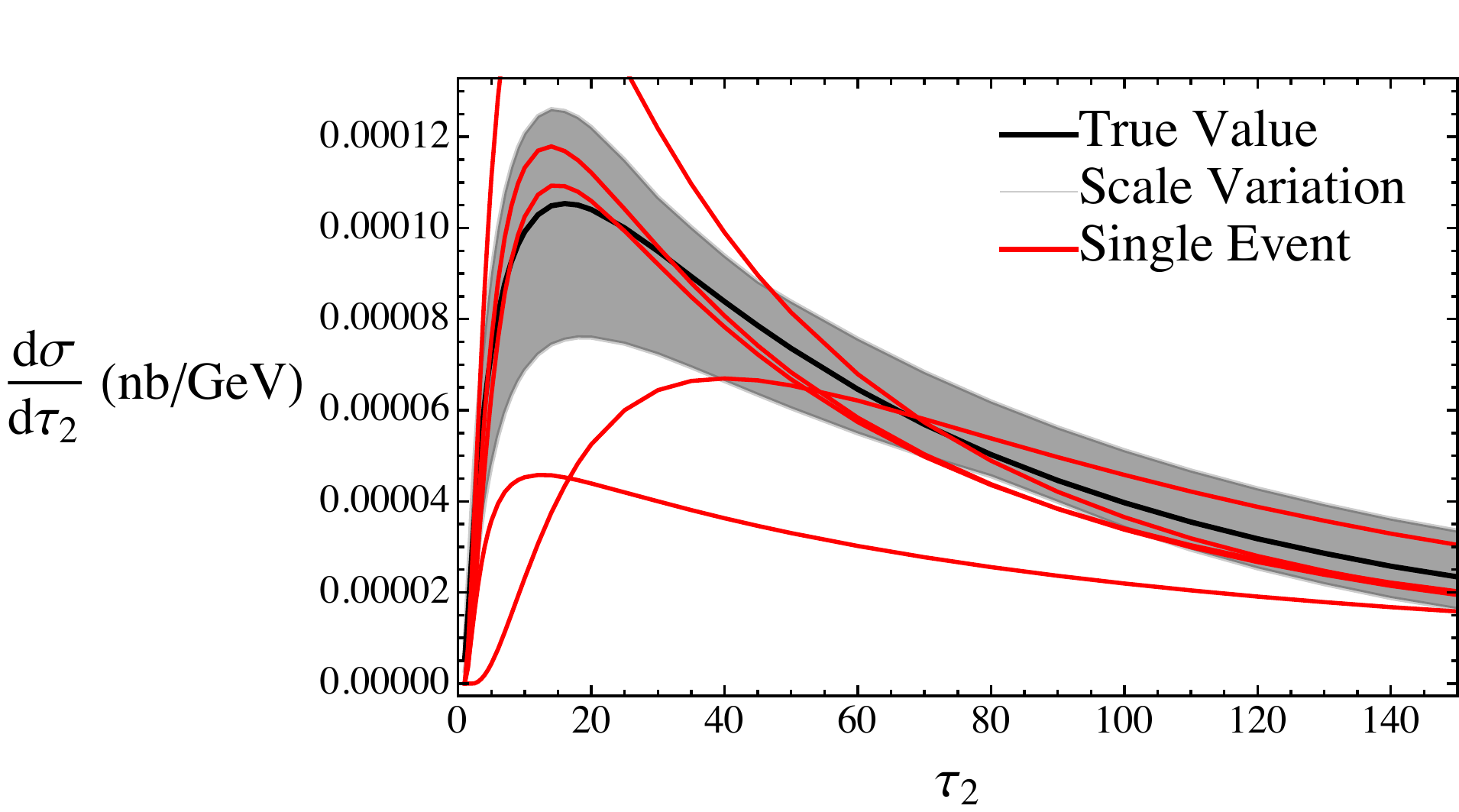}
\label{fig:convergence1}
}
  \subfigure[][ 10 events]{
\includegraphics[width=0.4\textwidth]{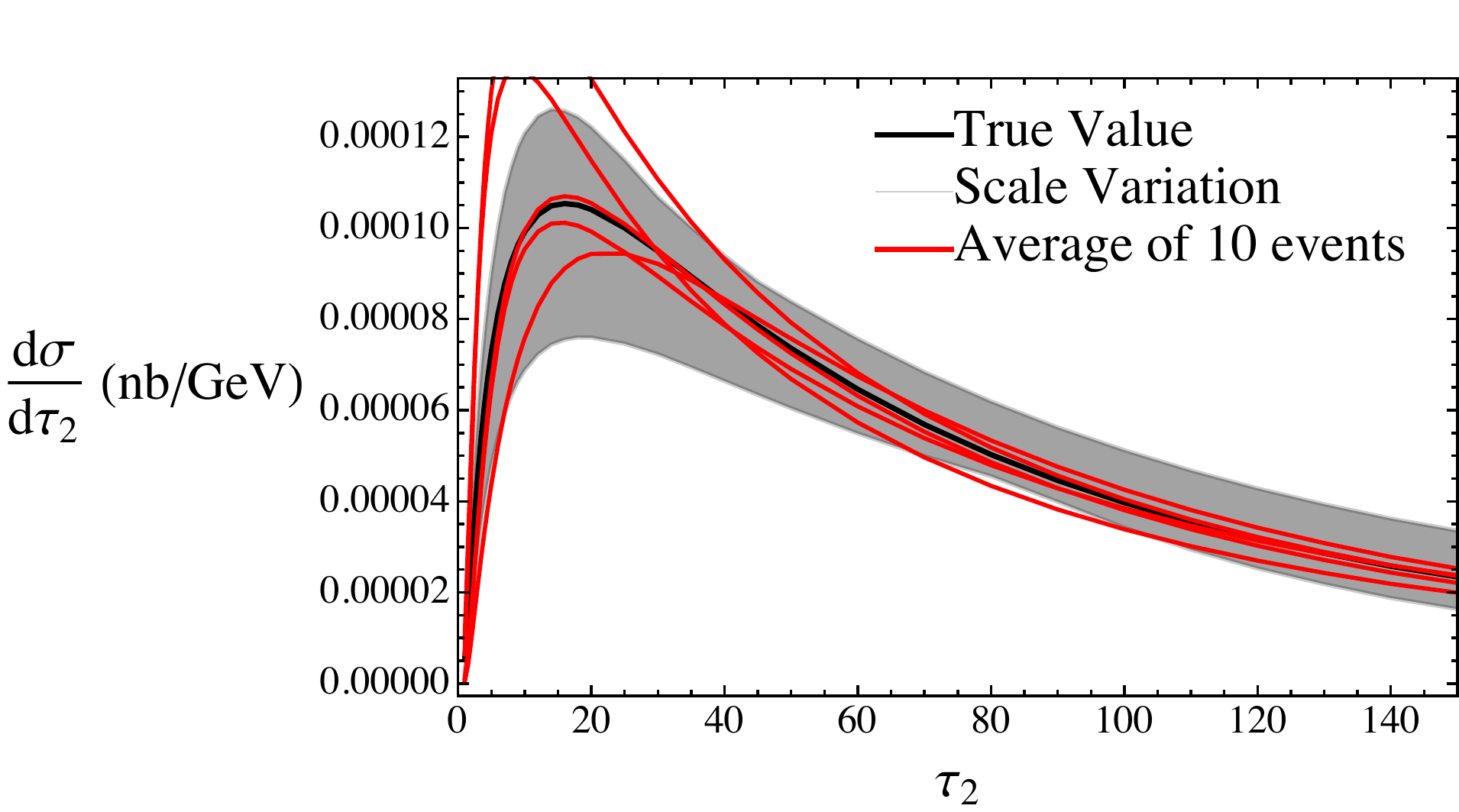}
\label{fig:convergence10}
}
  \subfigure[][ 100 events]{
\includegraphics[width=0.4\textwidth]{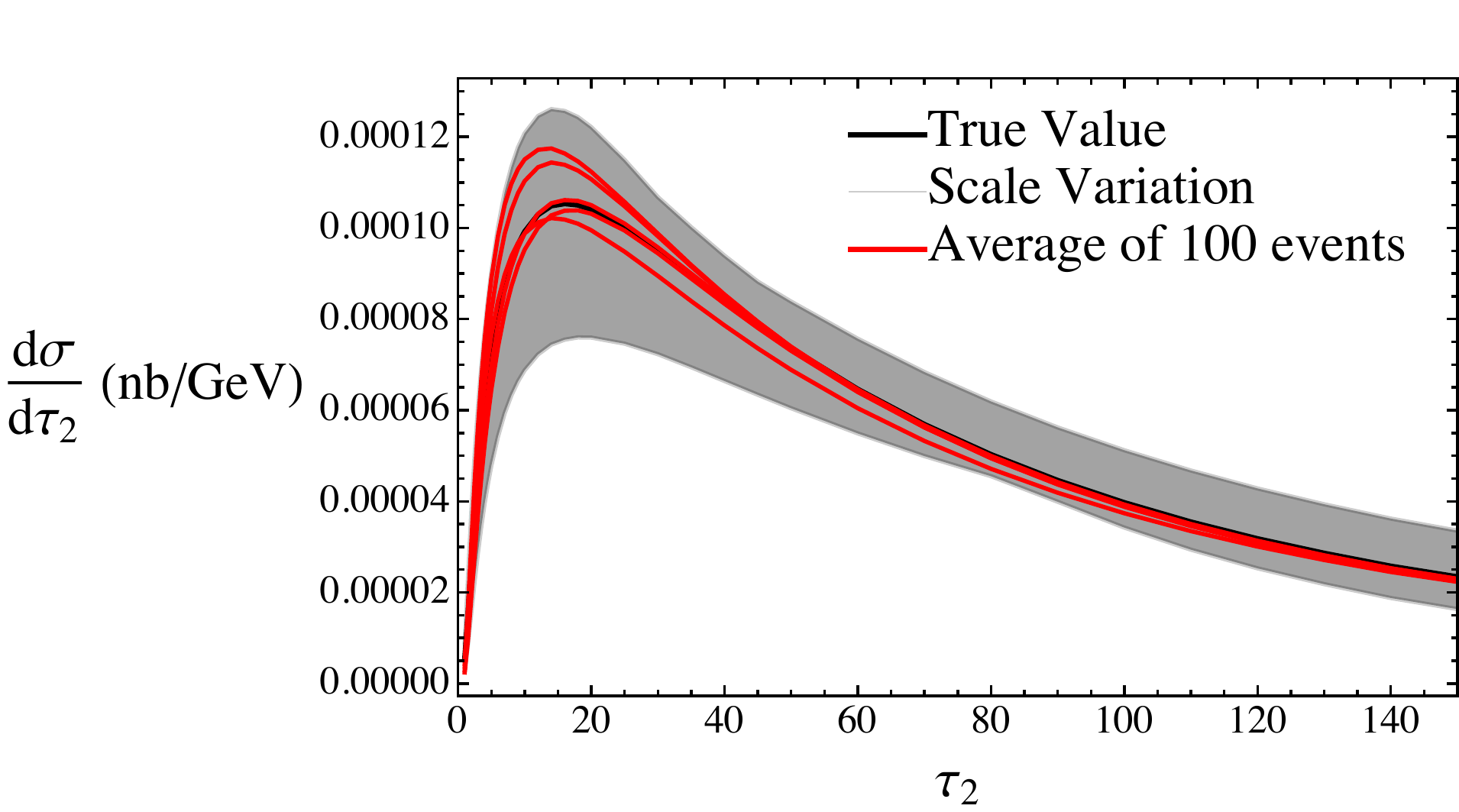}
\label{fig:convergence100}
} 
  \subfigure[][ 1000 events]{
\includegraphics[width=0.4\textwidth]{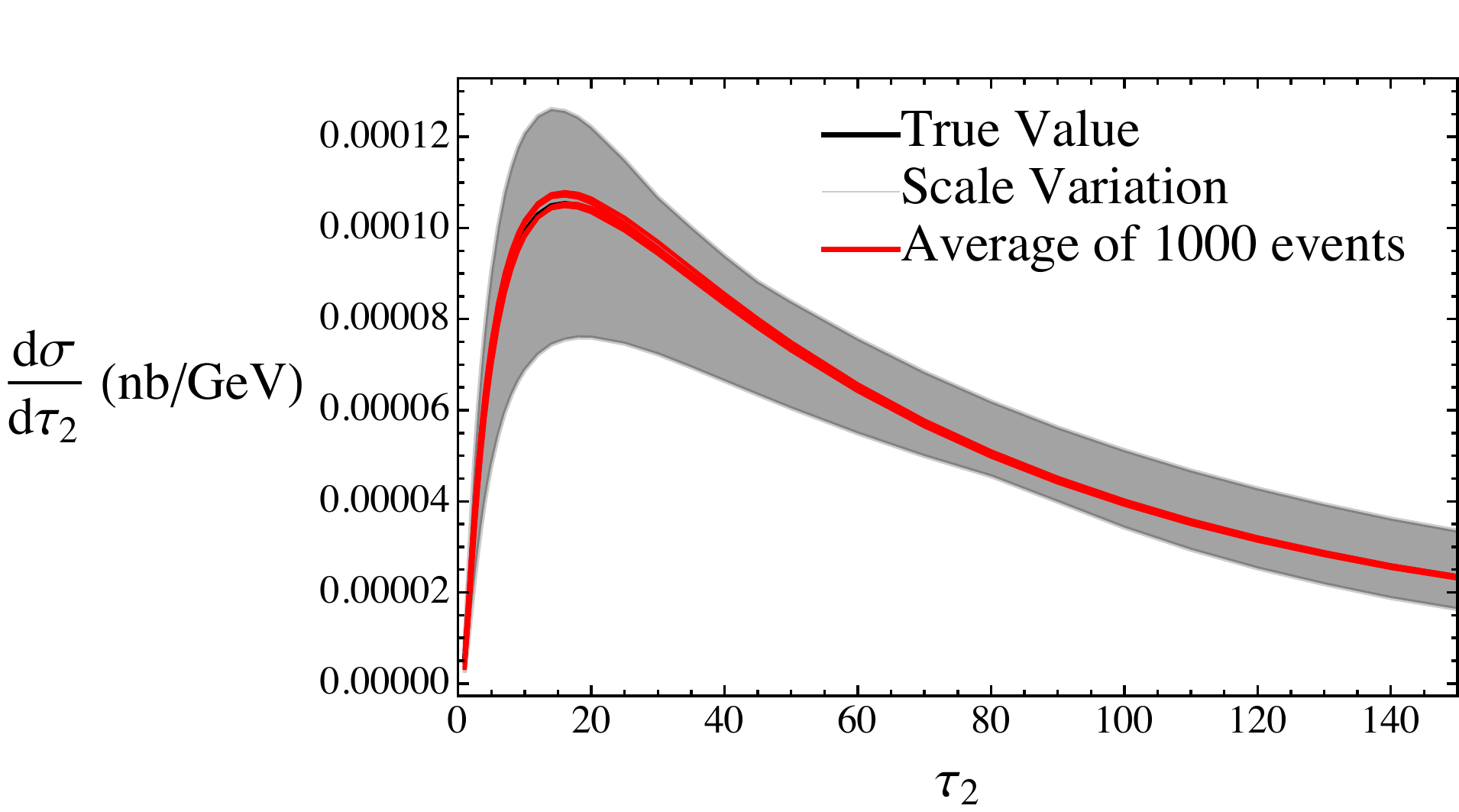}
\label{fig:convergence1000}
} 
\end{center}
\caption{Convergence of the MC integration. In each plot the red curves are several different runs with a fixed number of \MadGraph events in each run, as indicated. 
 The background black curve is the asymptotic answer, with 50,000 events. }
\label{fig:convergence}
\end{figure}

\subsection{Convergence of the phase space sum}
All the plots throughout this paper sum over many thousands of \MadGraph events (10,000 to 50,000) to ensure sufficient convergence. Such a large number of events can be computationally expensive, so one might wonder how many are actually needed. 

In \Fig{convergence} we show the $\cT_2$ distribution summing  $n$ phase-space points with $n=1,10,100$ or $1000$. In all four panels, the case with $n=50,000$ is displayed with uncertainty bands in grey. We see convincingly that, at least for 2-jettiness with our cuts, about 1000 phase-space points is enough given the scale uncertainties inherent to NLL resummation. Needless to say, using 1000 rather than 50,000 events greatly speeds up the calculation.

%%%%%%%%%%%%%%%%%%%%%%%%%%%%%%%%%%
%%%%%%%%%%%%%%%%%%%%%%%%%%%%%%%%%%
\section{Summary and conclusions}
%%%%%%%%%%%%%%%%%%%%%%%%%%%%%%%%%%
%%%%%%%%%%%%%%%%%%%%%%%%%%%%%%%%%%
Calculating a resummed cross section for a process like $pp\to 2j + W$ can be an arduous endeavor. For anything more complicated, such as $pp\to W + 3j$, performing the calculation completely by hand, \emph{i.e.}, with \Mathematica, is essentially impossible. Fortunately, the most numerically intensive and error-prone part of such an endeavor is largely automatable. This paper discusses a method of combining numerical calculations of the phase space integrals and tree-level hard functions using \MadGraph, with semi-analytic calculations of the jet, soft and beam functions using SCET and \Mathematica.
This method brings the calculation of many resummed  cross sections for the LHC within reach.

The method discussed in this paper uses \MadGraph to generate a set of events. The final-state partons in these events indicate the directions and energies of the jets, and the initial-state partons indicate the starting point for the evolution of beam functions (or simply PDFs for threshold resummation). We call the \MadGraph output $H^\colorindex(\Phi)$; it is equivalent to the tree-level hard function in SCET. For a given event, we can then compute the jet and soft functions, evolve the beam functions, and correct the hard function from tree level to the appropriate order (for NLL resummation, this means adding in logarithms). We call this $F^\colorindex(\Phi,\obs)$. For each event, we then have a distribution of the observable $H^\colorindex(\Phi)F^\colorindex(\Phi,\obs)$. Finally, we average these according to the weights from \MadGraph into the final distribution, $\frac{d\sigma}{d\obs} = \int d\Phi H^\colorindex(\Phi)F^\colorindex(\Phi,\obs)$.

We first tested this method on the thrust distribution at $e^+e^-$ collisions, where analytical results from SCET are reproduced exactly. Next, we examined the 4-jettiness event shape $\cT_4$ in $e^+e^- \to $ 4 jets. Here we compared to \Pythia, finding agreement within errors. We also explored how the distribution changes for different phase space points. As expected for very symmetric phase space points, where the partons all have the same energy and no angles are small, the agreement with \Pythia is very good. For less symmetric phase space points, the agreement is worse. This can be understood because there are logarithms of ratios of kinematic quantities which \Pythia resums at the LL level, but which SCET does not resum at all. When the phase space is integrated over, the agreement with \Pythia is very good, comparable to the agreement in the symmetric phase space points.

Next, we applied the method to the calculation of 2-jettiness $\cT_2$ in $pp \to 2j + \gamma$ events. The same calculation can be used for $pp \to 2j + W$ or $pp \to 2j + Z$. We found in this case again good agreement with \Pythia, but somewhat worse agreement than for $e^+e^-$ events. Again, this can be understood because the treatment of initial state radiation, using beam functions or the parton shower, are more dissimilar than the treatment of final-state radiation with the two methods. We also looked at the sum of the two jet masses in $pp \to 2j + \gamma$ with a cut $\Tcut$ on the beam thrust. This can be thought of as a generalization
of the jet mass calculation of~\cite{Jouttenus:2013hs} for 1-jet events to the dijet case.

Besides being straightforward to implement, one of the main features of this method is that its components are readily recyclable. After calculating $F^\colorindex(\Phi, \obs)$ and integrating against $H^\colorindex(\Phi)$ for some process (\emph{e.g.}, $\cT_2$ in $pp\to 2j + \gamma$), one can then calculate the same observable in the same process with different cuts, or even a different hard process (\emph{e.g.}, 2-jettiness in $pp\to 2j + W$), by simply generating a new MC sample. One can even make larger changes (changing the observable, or performing crossings on the incoming/outgoing particles), with relative ease, provided the jet, soft, and beam functions necessary remain the same. A new calculation only needs to be done when a new channel opens up (\emph{e.g.}, going from $pp \to 2j + W$ to $pp\to 2j$ opens up the $gg\to gg$ channel) or switching  to an observable sensitive to a jet shape (\emph{e.g.}, going from observables sensitive to jet mass to ones sensitive to jet broadening).

The \MadGraph/SCET method we describe is equivalent to a calculation in SCET, but with an efficient numerical computation of the hard function. Because the method is just SCET it is systematically improvable in the usual manner. The anomalous dimensions of the various components can be computed at any logarithmic order, and the fixed-order pieces computed to the corresponding order in $\alpha_s$. To go to NNLL resummation would require in addition the finite parts of the 1-loop hard, jet and soft functions. The jet functions can usually be computed to 1-loop without too much effort, and the soft functions can be computed to 1-loop numerically~\cite{Bauer:2011hj}. For the hard function, one can interface to an NLO matrix-element calculator, such as using \MadGraph at NLO~\cite{Alwall:2014hca}. In fact, to compute the NNLL distribution, one does not have to worry about matching the NLO calculation to parton showers: one simply needs the dimensionally regulated virtual contribution (the NLO Wilson coefficient). Indeed an NNLL calculation has already been performed~\cite{Becher:2014aya} exploiting \MadGraph at NLO for the hard function, albeit for an inclusive color-singlet final state. Thus the technology is already available to compute nearly any factorizable observable at NNLL using \MadGraph for the computationally intensive hard-function calculation.

\acknowledgments
MF thanks the Center for Future High Energy Physics, Bejing, China for its hospitality while this work was being completed. The authors were supported in part by the US Department of Energy, under grant DE-SC0013607. The work of MF was also supported in part by the National Science Foundation, under grant PHY-1258729.

\appendix

%%%%%%%%%%%%%%%%%%%%%%%%%%%%%%%%%%
%%%%%%%%%%%%%%%%%%%%%%%%%%%%%%%%%%
\section{Appendix: Details of the SCET calculation}
  \label{app:maincalculationappendix}
%%%%%%%%%%%%%%%%%%%%%%%%%%%%%%%%%%
%%%%%%%%%%%%%%%%%%%%%%%%%%%%%%%%%%
Throughout this appendix, color-structure indices $I,J$ on $H^{IJ}$, $S^{IJ}$ and $F^{IJ}$ are left implicit.

\subsection{Analytic piece}
  \label{app:HJSfuncs}
In this appendix we describe the modifications necessary to reproduce our results from those available in the literature and give explicit expressions for the functions $F^\colorindex(\Phi,\obs)$ for the observables we compute. 

Taking the hard-function anomalous dimensions for $2 \to 2$ QCD processes from Ref.~\cite{Kelley:2010fn}, the primary modification is the redefinition of the kinematic variables in order to make them applicable for arbitrary kinematics. Since we are deviating from 4-particle kinematics, we must go back to their Eq.~(69) and rederive the particular form for each channel. Moreover, the Mandelstam $t$ is no longer well defined, since $s_{13}\neq s_{24}$. But by a fortuitous property of the 4-parton channels, rederiving $M_{IJ}$ from scratch (from Eq. (69) of~\cite{Kelley:2010fn}) indicates that we can reuse their 4-particle results if we use
\begin{equation}
  \label{eq:stureplacement}
  \begin{split}
    s &\to \sqrt{s_{12}s_{34}}, \\
    u &\to \sqrt{s_{14}s_{23}}, \\
    t &\to \sqrt{s_{13}s_{24}},
  \end{split}
\end{equation}
The above replacements have only one caveat for the $qqqq$ and $qqgg$ channels we consider. The anomalous dimensions of \cite{Kelley:2010fn} contain a term $\sum_iC_iL(-t)$. In the $qqqq$ channel this is unchanged (modulo the replacement of $t$). In the $qqgg$ channel, however, this changes slightly:
\begin{equation}
  \label{eq:f}
  \sum_i C_i L(-t) = 2 C_F L(-t) + 2C_AL(-t)  \to 2 C_F L \left(-t\sqrt{\frac{s_{qq}}{s_{gg}}}\right)
                                                  + 2 C_A L \left(-t\sqrt{\frac{s_{gg}}{s_{qq}}}\right)
\end{equation}
We write this more concisely as $\sum_iC_iL(-t)\to \sum_iC_iL(-tf_i)$, where $f_q=1$ for the $qqqq$ channel and $f_q=f_g^{-1}=\sqrt{\frac{s_{qq}}{s_{gg}}}$ for the $qqgg$ channel.

The anomalous dimension of the jet function we get from Ref.~\cite{Jouttenus:2011wh}, while the anomalous dimension of the soft function can be extracted from the condition that scale dependence should cancel order-by-oder between the hard, jet, and soft contributions, $\mu\,d\sigma/d\mu=0$. For the fixed-order soft and jet functions, we take advantage of the fact that the anomalous dimensions of each are known to NNLL, and extract the desired fixed-order piece needed for NLL precision by expanding the anomalous dimensions as
\begin{equation}
  f(\mu_0) = f_{\text{tree}}+\int^{\mu_f} d\ln\mu \frac{d f(\mu)}{d\ln\mu}\, ,
\end{equation}
where $f$ is soft or jet function and $\mu_f$ is the natural scale, respectively, $\mu_s$ or $\mu_j$. This will be ambiguous up to a constant, \emph{i.e.}, $\mu_f$-independent, piece, but the ambiguity occurs at higher order than that at which we are working.

The fixed-order functions can be combined with the running to get the Laplace transforms of the full jet, beam, and soft functions.\footnote{We must take care combining different pieces from the literature since \cite{Kelley:2010fn} and \cite{Jouttenus:2011wh} define the jet anomalous dimension $\gamma_J^0$ with differing factors of -2.}
\begin{align}
  \label{eq:fulljet}
  \tilde{J}_i\!\left(\frac{\omega_i}{E_i},\mu\right)
    &= \left. e^{-4C_iS(\mu_j^i,\mu)+2A_J^i(\mu_j^i,\mu)}
       \left[1 + \frac{1}{2}C_i\Gc\partial_{\eta_i}^2 - \gamma_J^{0i}\partial_{\eta_i}\right]
       (\omega_i m_J^{i})^{\eta_i}\right|_{\eta_i={-2C_iA_\Gamma(\mu_j^i,\mu)}}  \\
  \begin{split}
    \label{eq:fullbeam}
    \tilde{B}_i\!\left(\frac{\omega_i}{E_i},x,\mu\right)
     &= f(x,\mu_B^i) e^{-4C_iS(\mu_B^i,\mu)+2A_J^i(\mu_B^i,\mu)}
         \left[1 + \frac{1}{2}C_i\Gc\partial_{\eta_i}^2 - \gamma_J^{0i}\partial_{\eta_i}\right. \\
     &\qquad + \left.\left.\frac{\alpha}{2\pi}\frac{1}{f(\mu_B^i, x)}
         \int_x^1 \frac{d\xi}{\xi}P_{ij}\left(\frac{x}{\xi}\right)f_j(\xi, \mu^i_B)\right]
         (\omega_i m_J^{i})^{\eta_i}\right|_{\eta_i={-2C_iA_\Gamma(\mu_j^i,\mu)}}
  \end{split} \\
  \begin{split}
    \tilde S (\mu)
      &= \prod_i (\abs{t_n}f_i^n)^{C_iA_\Gamma(\mu_s,\mu)}
           e^{A_\Gamma(\mu_s,\mu)M_{IJ}^\dagger} \left[\tilde S_{\text{tree}}\right. \\
      &\qquad \left.- \Gc \sum_i \left( \left[\frac{M^\dagger_{IJ}\tilde S_{\text{tree}}
           +\tilde S_{\text{tree}}M_{IJ}}{4}+C_i\tilde S_{\text{tree}}\ln(\abs{t_n}f_i^n)\right] \partial_{\theta_i} 
           + C_i\tilde S_{\text{tree}}\partial_{\theta_i}^2\right)\right]  \\
      &\qquad \left. e^{A_\Gamma(\mu_s,\mu)M_{IJ}}
           \prod_i\left( (\omega_i m_S^{i})^{\theta_i} e^{2C_i S(\mu_s,\mu)} \right)
           \right|_{\theta_i={2C_iA_\Gamma(\mu_s,\mu)}}
  \end{split}
\end{align}
In the above expression, the tilde represents Laplace transform, the index $i$ indicates a parton which might be a quark or a gluon, $C_i$ is either $C_F$ (if $i$ is a quark) or $C_A$ (if it's a gluon), $\Gamma$ is the cusp anomalous dimension (defined without a factor of $C_F$ or $C_A$), $S$ and $A$ are defined from $\Gamma$ as in \cite{Kelley:2010fn}, $m_S = \mu_se^{\gamma_E}$, and $m_J^i = \frac{\mu_j^2}{E_i}e^{\gamma_E}$. $t_n$ is the Mandelstam variable $t$ applied to the directions $n_i$ only; $t_n\equiv n_{13}n_{24}$, $f_i$ is the correction factor defined in \Eq{f}, and $M_{IJ}$ is the hard anomalous dimension matrix taken from \cite{Kelley:2010fn}.

Note that $\tilde S_{\text{tree}}$, $M_{IJ}$, and $M^\dagger_{IJ}$ are all matrices in color space which may not be simultaneously diagonalizable, so keeping track of their order is important. We now present two concrete examples of the functions $F^a(\Phi,\obs)$ used in our calculations.

\subsection{$F$ for 4-jettiness at $e^+e^-$}
  \label{app:4jetti}
Putting together the separate pieces in \App{HJSfuncs} and taking inverse Laplace transforms, the function $F(\Phi,\cT_4)$, fully differential in the separate $\cT^i$ contributions is
\begin{align}
  \label{eq:dFd4T}
\frac{d^4 F}{d \cT^1d \cT^2 d \cT^3 d \cT^4}(\mu, \{\cT^i\}) &= 
e^{A_\Gamma(\mu_s,\mu)M^\dagger} \tilde S_{\text{tree}}^{1/2}
G(d_{s1i}, d_{s2i}, d_{j1i}, d_{j2i})S_{\text{tree}}^{1/2}e^{A_\Gamma(\mu_s,\mu)M}
\times\\\nonumber&\times
\prod_i
\left(\abs{t_n}f_i^n\right)^{C_iA_\Gamma(\mu_s,\mu)}
e^{2C_i S(\mu_s,\mu)-4C_iS(\mu_j^i,\mu)+2A_J^i(\mu_j^i,\mu)}
\times\\\nonumber&\times\left.
\prod_i
\frac{1}{\cT^i}\frac{1}{\Gamma(\theta_i+\eta_i)}
\left(\frac{\cT^i}{m_S^{i}}\right)^{\theta_i}\left(\frac{\cT^i}{m_J^{i}}\right)^{\eta_i}\right|_{\scriptsize \begin{array}{l}
\theta_i={-2C_iA_\Gamma(\mu_s,\mu)}\\\eta_i={2C_iA_\Gamma(\mu_j^i,\mu)}\end{array}}
\end{align}
Here, $G$ is defined through
\begin{align}
  \begin{split}
   \tilde{G}(s_{1i},s_{2i},j_{1i},j_{2i})
     &\equiv 1 + \sum_i \left( \left[\frac{\tilde S_{\text{tree}}^{-1/2}M^\dagger\tilde S_{\text{tree}}^{1/2}
       +\tilde S_{\text{tree}}^{1/2} M\tilde S_{\text{tree}}^{-1/2}}{4}+C_i \ln(\abs{t_n}f_i^n)\right]\Gc s_{1i}\right. \\
     &\qquad - \left. C_i\Gc s_{2i}+ \frac{1}{2}C_i\Gc j_{2i} + \gamma_J^{0i}j_{1i} \right)
   \end{split} \\
  \intertext{and after performing the inverse Laplace transform, $G$ depends on}
  \label{eq:defined}
  \begin{split}
    d_{s1i} &\equiv \ln\left(\frac{\cT^i}{m_S^i}\right)-\psi(\eta_i+\theta_i) \\
    d_{s2i} &\equiv \left(\ln\left(\frac{\cT^i}{m_S^i}\right)
                    - \psi(\eta_i+\theta_i)\right)^2-\psi'(\eta_i+\theta_i) \\
    d_{j1i} &\equiv \ln\left(\frac{\cT^i}{m_J^i}\right)-\psi(\eta_i+\theta_i) \\
    d_{j2i} &\equiv \left(\ln\left(\frac{\cT^i}{m_J^i}\right)
                    - \psi(\eta_i+\theta_i)\right)^2-\psi'(\eta_i+\theta_i)
  \end{split}
\end{align}

To get $F$ itself from $d^4F/d\cT^i$ we integrate against a $\delta$ function defining our observable as a function of the $\cT^i$. Since 
\begin{equation}
\frac{d^4 F}{d \cT^1d \cT^2 d \cT^3 d \cT^4}
 = \mathcal{N} \prod_i(\cT^i)^{\eta_i+\theta_i-1},
\end{equation}
for 4-jettiness the integral is
\begin{equation}
  \begin{split}
    F(\cT_4) &= \mathcal{N} \int d\cT^1 d\cT^2 d \cT^3 d \cT^4 \delta(\cT_4-\cT^1-\cT^2-\cT^3-\cT^4)  
      \prod_i(\cT^i)^{\eta_i+\theta_i-1}\\
             &= \frac{\mathcal{N}}{\cT_4\Gamma \left(\sum_i\eta_i+\theta_i\right)}
                \prod_i\left(\frac{\cT_4}{m_J^i}\right)^{\eta_i}\left(\frac{\cT_4}{m_S^i}\right)^{\theta_i}
  \end{split}
\end{equation}
We define the following functions for a general observable $\mathcal{O}$, given by $f(\cT^i)$,
\begin{equation}
  \label{eq:Idef}
  I_0 \equiv \int d^4\cT^i\, \delta(\mathcal{O}-f(\cT^i)) \prod_i(\cT^i)^{\eta_i+\theta_i-1}
    \qquad
  I(d, O) \equiv \frac{1}{I_0} \int d^4\cT^i\, \delta(O-f(\cT^i)) d \prod_i(\cT^i)^{\eta_i+\theta_i-1},
\end{equation}
so that in integrating \Eq{dFd4T}, we need to simply replace $\prod_i(\cT^i)^{\eta_i+\theta_i-1}$ with $I_0$ and replace each $d$ with $I(d)$.

\subsection{$F$ for 2-jettiness at $pp$}
  \label{app:2jetti}
For the $pp$ observables considered in this paper, the above discussion is modified by replacing two of the jet functions with beam functions \cite{Stewart:2009yx}. The beam functions have the same RG running as the jet functions, so in practice using beam functions or jet functions changes the calculation very little. The only subtlety is that we must divide the beam functions by the PDFs, since \MadGraph will include PDFs in calculating the hard function, as discussed in \Sec{SCET}.

Putting together all the pieces, the full function $F(\Phi, \cT_2)$ is:
\begin{equation}
  \label{eq:t4}
  \begin{split}
  F(\Phi, \cT_2) &= \left(\frac{\alpha_s(\mu_h)^2}{\alpha_s(\mu_{MG})^2} \frac{f(x_a, \mu_{J}^a)}{f(x_a, \mu_{MG})}\frac{f(x_b, \mu_{J}^b)}{f(x_b, \mu_{MG})}\frac{H(\mu_h)}{H(\mu_{MG})}\right) \\
                 &\qquad e^{A_\Gamma(\mu_s,\mu_h)M^\dagger} \tilde S_{\text{tree}}^{1/2} G\left[I(d_{s1i}), I(d_{s2i}), I(d_{j1i}), I(d_{j2i})\right]\tilde S_{\text{tree}}^{1/2}e^{A_\Gamma(\mu_s,\mu_h)M} \\
                 &\qquad \prod_i \left(\abs{t_n}f_i^n\right)^{C_iA_\Gamma(\mu_s,\mu)} \left(\frac{\mu^2_H}{\abs{t}f_i}\right)^{C_iA_{\Gc}(\mu_h,\mu)} \\
                 &\qquad \prod_i \exp\left[2C_iS(\mu_h,\mu)+2C_i S(\mu_s,\mu)-4C_iS(\mu_j^i,\mu)+2A_J^i(\mu_j^i,\mu_h)\right] \\
                 & \qquad \frac{1}{\cT_2}\prod_i \frac{1}{\Gamma[2C_iA_\Gamma(\mu_j^i,\mu_s)]} \left(\frac{\cT_2}{m_S}\right)^{-2C_iA_\Gamma(\mu_s,\mu)}\left(\frac{\cT_2}{m_J^{i}}\right)^{2C_iA_\Gamma(\mu_j^i,\mu)},
  \end{split}
\end{equation}
where $\mu_s$, $\mu_h$, $\mu_j^i$ are the natural scales, $\mu_{R}$ is the renormalization/factorization scale used by \MadGraph and $\mu$ is arbitrary  (the independence of the above formula of $\mu$ is a nontrivial cross-check). $G$, $I$, and $d$ have the same definitions as in \App{4jetti}, with the appropriate redefinition for $\mathcal{O} = \cT_2$.

\section{Appendix: Color matrix structures in \MadGraph}
  \label{app:MGappendix}

In this appendix we briefly summarize the conventions \MadGraph uses to construct its color matrices. We confine ourselves to processes with only quarks and gluons, although \MadGraph can handle more general representations of $SU(3)$ as well. In \MadGraph particles are referred to by the order in which they are entered at the generation step, and for the purpose of color decomposition, we can treat the color of the initial state as its crossing-symmetric analogue. That is to say, the color decomposition of $e^+e^- \to \bar{q}qgg$ is identical to $q\bar{q} \to gg$, since the latter is related to $0 \to \bar{q}qgg$ by crossing symmetry while keeping the order intact. We first describe the simpler cases of pure gluons and pure quarks, which make understanding the more involved mixed case more straightforward.

For pure gluons, \MadGraph uses the conventional (overcomplete) single-trace operator basis, with cyclic rotations modded out, so the first gluon always retains its position. The basis takes the form, for $n$ gluons,
\begin{equation}
  \label{eq:allgbasis}
  \mathcal{B}_i = \{ \Tr(T^a T^{\sigma_i(b)} T^{\sigma_i(c)} \dotsm)
                     A_1^{a\mu} A_2^{b\nu} A_3^{c\rho} \dotsm \}.
\end{equation}
Here $\sigma_i$ represents the $i$th permutation of $(n-1)$ gluons, and runs over all $(n-1)!$ possible permutations. Due to the symmetries of the resulting amplitudes, every ordering of gluons will give the exact same squared matrix element when summed over all permutations, but as it helps clarify the other cases, we specify \MadGraph's convention here. The ordering in generated recursively, with the first term having all gluons in order, and all subsequent terms generated considering all permutations of the last $m$ generators $T^a$ before incrementing the $(n-m)$-th one by one and repeating until the 2nd position has been occupied by all $(n-1)$ gluons in the permutation. For example, with 4 gluons, the order of permutations generated is $\sigma = \{234,243,324,342,423,432\}$. In the RG evolution involved in resummation, these operators will mix with multi-trace terms, and therefore the basis provided by any tree-level MC generator will have to be enlarged.

For fermion-only operators, \MadGraph takes the dipole basis of all possible fermion color-factor contractions into bilinears, $q_i \bar{q}^i$. Crucially, the ordering of pairs in the color matrix can be of importance here. The convention \MadGraph adopts is to have the first entry order the quarks in the bilinears as they are defined in the process. The antiquarks are then contracted in the order they appear, and shuffled around in the same order as the gluons in the all-gluon case. As another illustrative example, the basis for $d_1 u_2 \to u_3 d_4 \bar{u}_5 u_6$ is
\begin{equation}
  \label{eq:6qbasis}
  \begin{split}
    \mathcal{B}_i = \{&(u_3\bar{d}_1)(d_4\bar{u}_2)(u_6\bar{u}_5),\
                       (u_3\bar{d}_1)(d_4\bar{u}_5)(u_6\bar{u}_2),\
                       (u_3\bar{u}_2)(d_4\bar{d}_1)(u_6\bar{u}_5), \\
                      &(u_3\bar{u}_2)(d_4\bar{u}_5)(u_6\bar{d}_1),\
                       (u_3\bar{u}_5)(d_4\bar{d}_1)(u_6\bar{u}_2),\
                       (u_3\bar{u}_6)(d_4\bar{u}_2)(u_6\bar{d}_1)\}.
  \end{split}
\end{equation}

Finally, mixed quark-gluon processes are significantly more complicated in generality. They involve a combination of permutations over the orderings of both the quark and anti-quark labels and partitions of adjoint operators over the quark bilinears. This is done with a combination of overcomplete enumerations of operators and vetoes of operators that have already been enumerated. At the same time, most mixed quark-gluon channels are of such high multiplicity that it is difficult to imagine them being phenomenologically relevant in the near future. We therefore content ourselves with listing the bases for all matrix elements with up to 6 partons, which will be sufficient for all processes up to $pp \to 4j (+ X)$. The processes $q\bar{q} gg$, $q\bar{q}ggg$, and $q\bar{q}gggg$ are handled exactly the same as the all-gluon case, except that the position of the first gluon's adjoint generator is not held fixed, the permutations generated being among all the gluons. For processes with multiple quark bilinears, \MadGraph tries to keep the adjoint generators as far back as possible at every step. This means that after shuffling through the antiquarks, the basis must shuffle through the quarks as well to generate every permutation. For $u_1\bar{u}_2 u_3\bar{u}_4 g_5$, this yields
\begin{equation}
  \label{eq:4qgbasis}
  \mathcal{B}_i = \{(u_1\bar{u}_2)(u_3 T^a \bar{u}_4) A^{a\mu}_5,\
                    (u_1\bar{u}_4)(u_3 T^a \bar{u}_2) A^{a\mu}_5,\
                    (u_3\bar{u}_2)(u_1 T^a \bar{u}_4) A^{a\mu}_5,\
                    (u_3\bar{u}_4)(u_1 T^a \bar{u}_2) A^{a\mu}_5\}.
\end{equation}
Adding another gluon for $u_1\bar{u}_2 u_3\bar{u}_4 g_5 g_6$ expands this basis to
\begin{equation}
  \begin{split}
  \label{eq:4q2gbasis}
  \mathcal{B}_i = \{&(u_1\bar{u}_2)(u_3 T^a T^b \bar{u}_4) A^{a\mu}_5 A^{b\nu}_6,\
                     (u_1\bar{u}_2)(u_3 T^b T^a \bar{u}_4) A^{a\mu}_5 A^{b\nu}_6,\
                     (u_1\bar{u}_4)(u_3 T^a T^b \bar{u}_2) A^{a\mu}_5 A^{b\nu}_6, \\
                    &(u_1\bar{u}_4)(u_3 T^b T^a \bar{u}_2) A^{a\mu}_5 A^{b\nu}_6,\
                     (u_3\bar{u}_2)(u_1 T^a T^b \bar{u}_4) A^{a\mu}_5 A^{b\nu}_6,\
                     (u_3\bar{u}_2)(u_1 T^b T^a \bar{u}_4) A^{a\mu}_5 A^{b\nu}_6, \\
                    &(u_3\bar{u}_4)(u_1 T^a T^b \bar{u}_2) A^{a\mu}_5 A^{b\nu}_6,\
                     (u_3\bar{u}_4)(u_1 T^b T^a \bar{u}_2) A^{a\mu}_5 A^{b\nu}_6,\
                     (u_1 T^a \bar{u}_2)(u_3 T^b \bar{u}_4) A^{a\mu}_5 A^{b\nu}_6, \\
                    &(u_1 T^a \bar{u}_4)(u_3 T^b \bar{u}_2) A^{a\mu}_5 A^{b\nu}_6,\
                     (u_3 T^a \bar{u}_2)(u_1 T^b \bar{u}_4) A^{a\mu}_5 A^{b\nu}_6,\
                     (u_3 T^a \bar{u}_4)(u_1 T^b \bar{u}_2) A^{a\mu}_5 A^{b\nu}_6\},
  \end{split}
\end{equation}
completing the listing of all bases for processes with up to 6 partons.

\bibliography{MadGraphSCET}
\bibliographystyle{jhep}
\end{document}